\documentclass[twocolumn, twocolappendix]{aastex631}

\usepackage{graphicx}	% Including figure files
\usepackage{amsmath}	% Advanced maths commands
\usepackage{ulem}

\newcommand{\kpc}{\mbox{$\>{\rm kpc}$}}
\newcommand{\kmskpc}{\mbox{$\>{\rm km/s/kpc}$}}
\newcommand{\Gyr}{\mbox{$\>{\rm Gyr}$}}
\newcommand{\Omphi}{\mbox{$\>\Omega_\varphi$}}
\newcommand{\OmphiR}{\mbox{$\>(\Omphi - \OmP)/\Omega_R$}}
\newcommand{\OmP}{\mbox{$\>\Omega_\mathrm{P}$}}
\newcommand{\Msun}{\>{\rm M_{\odot}}}
\newcommand\degrees{^\circ}
\newcommand{\ommax}{\mbox{$\>\omega_\mathrm{max}$}}

\DeclareMathOperator{\sign}{sign}

\newcommand{\Omphiz}{\mbox{$\>(\Omphi - \OmP)/\Omega_z$}}
\newcommand{\OmzR}{\mbox{$\>\Omega_z/\Omega_R$}}
\newcommand{\EJ}{\mbox{$\>E_{\rm J}$}}
\newcommand{\ecc}{\mbox{$\>{\rm ecc}$}}
\newcommand{\zmax}{\mbox{$z_{\rm max}$}}
\newcommand{\abar}{\mbox{$a_{\rm bar}$}}
\newcommand{\poteff}{\mbox{$\>\Phi_{\rm eff}$}}

 \definecolor{ForestGreen}{RGB}{34,139,34}
 \definecolor{bostonred}{rgb}{0.8, 0.0, 0.0}

\defcitealias{Anderson2022}{Paper~I}
\defcitealias{BT}{BT08}

\begin{document}

\title[Orbits and evolution of flat bars]{Orbital support and evolution of flat profiles of bars (shoulders)}

\author[0000-0002-0740-1507]{Leandro {Beraldo e Silva}}
\affiliation{Department of Astronomy, University of Michigan, 1085 S. University Ave., Ann Arbor, MI, 48109 USA}
\affiliation{Jeremiah Horrocks Institute, University of Central Lancashire, Preston PR1 2HE, UK}
\author[0000-0001-7902-0116]{Victor P. Debattista}
\affiliation{Jeremiah Horrocks Institute, University of Central Lancashire, Preston PR1 2HE, UK}
\author[0000-0002-6379-5053]{Stuart Robert Anderson}
\affiliation{Jeremiah Horrocks Institute, University of Central Lancashire, Preston PR1 2HE, UK}
\author[0000-0002-6257-2341]{Monica Valluri}
\affiliation{Department of Astronomy and Astrophysics, University of Michigan, Ann Arbor, MI, USA}
\author[0000-0003-4588-9555]{Peter Erwin}
\affiliation{Max Planck Institut f\"ur Extraterrestrische Physik, Giessenbachstrasse, D-85748 Garching, Germany}
\author[0000-0003-2594-8052]{Kathryne J. Daniel}
\affiliation{Department of Astronomy \& Steward Observatory, University of Arizona,
Tucson, AZ 85721, USA}
\affiliation{Center for Computational Astrophysics, Flatiron Institute, New York, NY 10010, USA}
\author[0000-0003-3523-7633]{Nathan Deg}
\affiliation{Department of Physics, Engineering Physics, and Astronomy, Queen's University, Kingston, ON, K7L 3N6, Canada}

\correspondingauthor{Leandro {Beraldo e Silva}}
\email{lberaldo@umich.edu, lberaldoesilva@gmail.com}

%% Mark off the abstract in the ``abstract'' environment. 
\begin{abstract}
Many barred galaxies exhibit upturns (shoulders) in their bar major-axis density profile. Simulation studies have suggested that shoulders are supported by looped $x_1$ orbits, occur in growing bars, and can appear after bar-buckling. We investigate the orbital support and evolution of shoulders via frequency analyses of orbits in simulations. We confirm that looped orbits are shoulder-supporting, and can remain so, to a lesser extent, after being vertically thickened. We show that looped orbits appear at the resonance $\OmphiR=1/2$ (analogous to the classical Inner Lindblad Resonance, and here called ILR) with vertical-to-radial frequency ratios $1 \lesssim\OmzR \lesssim 3/2$ (vertically {\it warm} orbits). {\it Cool} orbits at the ILR (those with $\OmzR > 3/2$) are vertically thin and have no loops, contributing negligibly to shoulders. As bars slow and thicken, either secularly or by buckling, they populate warm orbits at the ILR. Further thickening carries these orbits towards crossing the vertical ILR [vILR, $\Omphiz=1/2$], where they convert in-plane to vertical motion, become chaotic, kinematically hotter and less shoulder-supporting. Hence, persistent shoulders require bars to trap new stars, consistent with the need for a growing bar. Since buckling speeds up trapping on warm orbits at the ILR, it can be followed by shoulder formation, as seen in simulations. This sequence supports the recent observational finding that shoulders likely precede the emergence of BP-bulges. The python module for the frequency analysis, \texttt{naif}, is made available.

\end{abstract}

%% Keywords should appear after the \end{abstract} command. 
%% The AAS Journals now uses Unified Astronomy Thesaurus concepts:
%% https://astrothesaurus.org
%% You will be asked to selected these concepts during the submission process
%% but this old "keyword" functionality is maintained in case authors want
%% to include these concepts in their preprints.
\keywords{Galactic dynamics --- Galactic bars --- Frequency analysis}

%% From the front matter, we move on to the body of the paper.
%% Sections are demarcated by \section and \subsection, respectively.
%% Observe the use of the LaTeX \label
%% command after the \subsection to give a symbolic KEY to the
%% subsection for cross-referencing in a \ref command.
%% You can use LaTeX's \ref and \label commands to keep track of
%% cross-references to sections, equations, tables, and figures.
%% That way, if you change the order of any elements, LaTeX will
%% automatically renumber them.
%%
%% We recommend that authors also use the natbib \citep
%% and \citet commands to identify citations.  The citations are
%% tied to the reference list via symbolic KEYs. The KEY corresponds
%% to the KEY in the \bibitem in the reference list below. 

\section{Introduction}
\label{sec:intro}
Stellar bars are present in $\approx 60-70\%$ of nearby spiral galaxies \citep[][]{Eskridge_2000, Menendez-Delmestre_2007, Sheth_2008, Nair_2010, Erwin_2018}. Barred galaxies have surface-brightness profiles along the bar major axis which are traditionally classified as either exponential or as having, on top of an exponential profile, a nearly-flat part in the bar outskirts (on both sides) -- hereafter the ``shoulders'' \citep[e.g.][but see \cite{Erwin_2023} for an updated classification scheme]{Elmegreen_1985, Carvalho_1987, Elmegreen_1996, Prieto_2001, Gadotti2007}.

\cite{Elmegreen_1996}, \cite{Regan_1997} and \cite{Elmegreen_2011} found that shoulders are generally present in profiles of early-type barred spiral galaxies, while late-type galaxies normally have bars with exponential profiles. In a sample of 144 barred galaxies from the Spitzer Survey of Stellar Structure in Galaxies \citep[$S^4G$,][]{Sheth_2010}, \cite{Kim_2015} found that massive and bulge-dominated galaxies tend to have flat bar profiles, while bulgeless galaxies tend to have exponential profiles, in accordance with early predictions based on $N$-body simulations \citep[][]{Combes_1993}. The same trends were found by \cite{Kruk_2018} in a sample of 3461 barred galaxies from SDSS, while \cite{Erwin_2023} find that the presence of shoulders is more strongly correlated with stellar mass. Shoulders in the bar major axis profile have also been observed in M31 \citep[][]{Athanassoula_2006} and in several simulation studies \citep[e.g.][hereafter {\bf Paper~I}]{Athanassoula_Misiriotis_2002, Bureau_2005, Anderson2022}.

\begin{figure}
    \centering
% 	% Allowable file formats are eps or ps if compiling using latex
% 	% or pdf, png, jpg if compiling using pdflatex
	\includegraphics[scale=0.35]{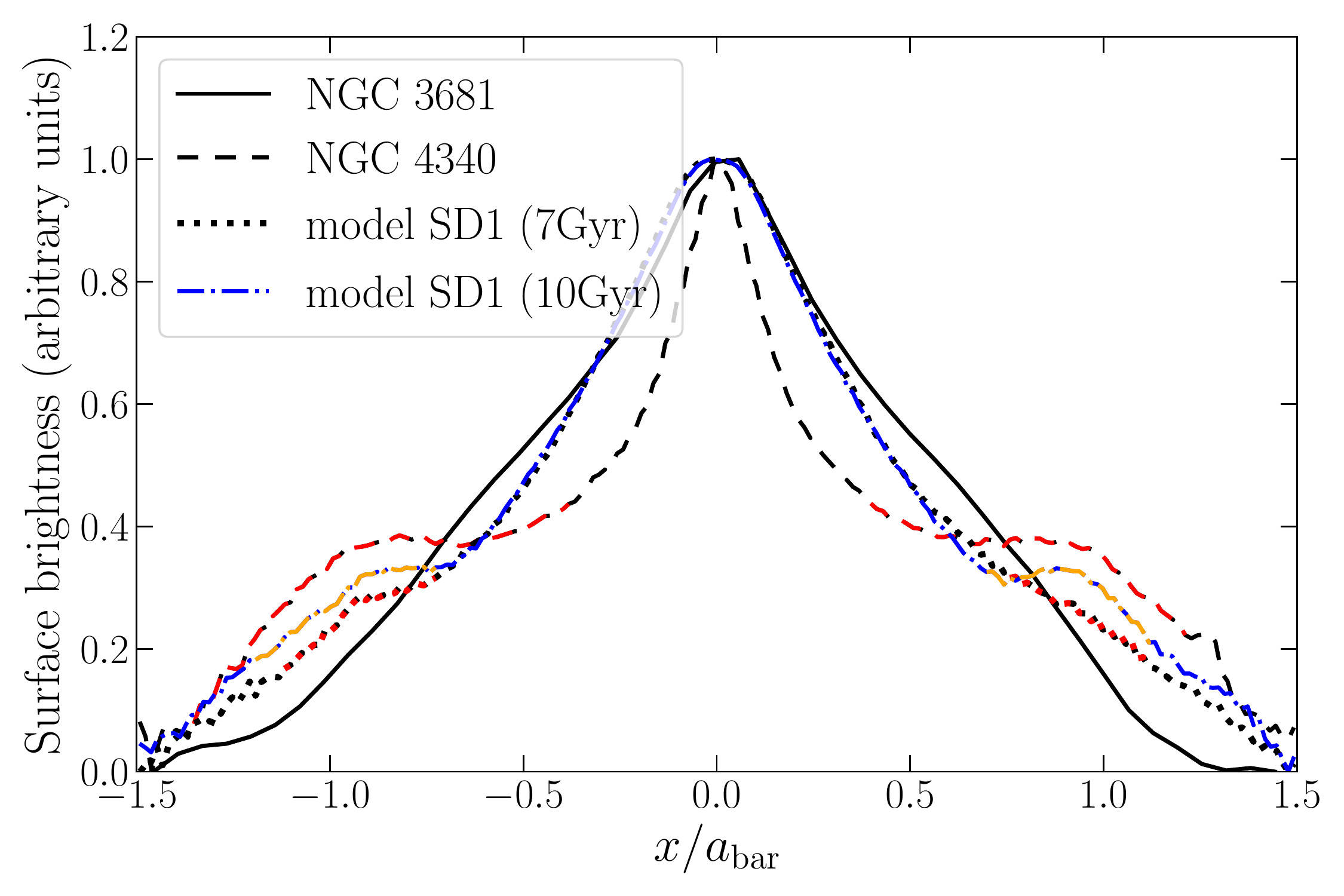}
    \caption{Normalised observed bar-major-axis brightness profiles for the \textit{Spitzer} $3.6\micron$ band for NGC~3681 and NGC~4340, in arbitrary units \citep[S$^4$G,][]{Sheth_2010}. Also shown the bar-major-axis surface density profiles for simulation SD1 at $7\Gyr$ (dotted black) and at $10\Gyr$ (dash-dotted blue) -- see Sec.~\ref{sec:sims} and Fig.~\ref{fig:Sigma_xy_xz_all}. NGC 3681 has a pure-exponential profile, while NGC 4340 and model SD1 have shoulders (red and orange), as detected in \citetalias{Anderson2022}.}
    \label{fig:Sigma_1D_examples}
\end{figure}

Fig.~\ref{fig:Sigma_1D_examples} shows examples of bar major-axis profiles for the galaxies NGC 3681 and NGC 4340 from $S^4G$ \citep[][]{Sheth_2010} and for the simulation SD1 (described in Sec.~\ref{sec:sims}) at 7 Gyr (dotted black/red) and at 10Gyr (dash-dotted blue/orange). The bars are aligned with the $x$-axis and for the simulation we select all star-particles within $|y|/\abar<0.145$, where $\abar$ is the bar length -- see Sec.~\ref{sec:sims} for details. For the observed galaxies, we show one-pixel-wide profiles along the bar major axis from \textit{Spitzer} 3.6\micron{} images. NGC 3681 has a pure-exponential profile, while NGC 4340 and the model SD1 have profiles with shoulders (red), as detected by \citetalias{Anderson2022}.

The galaxy NGC 4340 was chosen as an example of a bar profile with particularly pronounced shoulders, and these look much more pronounced than the ones in the model SD1 at 7Gyr. The shoulder strength, estimated as the excess mass of the shoulder over a purely exponential profile, normalized by the total mass within the shoulder region, typically grows over time, as shown in \citetalias{Anderson2022}. In fact, we see that the shoulders of SD1 get stronger at 10Gyr (blue/orange dash-dotted lines), although its length (in units of $\abar$) is approximately constant and smaller than in the observed galaxy NGC 4340.

In principle, profiles with shoulders might simply signal the presence of a strong and long bar. However, exploring a diverse set of simulations, \citetalias{Anderson2022} found only a weak dependence on bar strength or bar length, while \cite{Erwin_2023} found no evidence of such correlation in their observational sample. Importantly, \citetalias{Anderson2022} demonstrated that shoulders tend to be present in galaxies with growing bars, although their results suggest that this is a necessary but not a sufficient condition. Furthermore, \citetalias{Anderson2022} found that in secularly evolving (non-buckling) bars, once formed, shoulders tend to persist. On the other hand, in bars that buckle, shoulders tend to appear right after the buckling, and to be erased (at least temporarily) if a second buckling happens. Finally, \citetalias{Anderson2022} suggested that shoulders are mostly supported by looped $x_1$ orbits.

In this paper, we further investigate the shoulders in bar major-axis profiles of galaxies, focusing on their orbital support and the time evolution of these orbits. The main questions we aim to answer are: how and why do some bars develop shoulders? Which orbits support the shoulders? Why do bars need to grow in order to retain the shoulders once formed? What is behind the role of buckling events in creating or erasing shoulders?

We answer these questions by selecting particles from $N$-body simulations, integrating their orbits in the ``frozen'' potentials of different snapshots, and performing frequency analyses of these orbits. The $N$-body simulations are described in Sec.~\ref{sec:sims}. In Sec.~\ref{sec:orbits}, we explain the procedures for orbit integration and frequency analyses. The resulting frequency maps and stacked density profiles for different orbital groups, as well as their time evolution, are presented in Sec.~\ref{sec:results}. We demonstrate the somewhat surprising result that, once shoulders have formed, vertical resonances act as the main culprit for erasing them, shedding light on the suggestion of \citetalias{Anderson2022} that, in order for shoulders to be present for long times, the bar needs to be growing and trapping new stars. In Sec.~\ref{sec:discussions} we discuss our results, and the conclusions are summarized in Sec.~\ref{sec:conclusions}.

\section{Simulations}
\label{sec:sims}

Here we describe the $N$-body simulations explored in this paper. These simulations are chosen to represent a diverse subset of the models analysed in \citetalias{Anderson2022}. Briefly, one model (SD1, our fiducial model) has clear and persistent shoulders appearing around 4Gyr; one model (SD1S) does not develop shoulders; model HG1 is a fully self-consistent star-forming $N$-body+SPH simulation with weak shoulders; and Model 4 has a bar that buckles around $3.8 \Gyr$, developing persistent shoulders after that. These simulations are described in detail in \citetalias{Anderson2022}, but we summarise them here -- see Table \ref{tab:sims}.

Model SD1 is a pure $N-$body model, built using a modified version of \textsc{GalactICS} \citep{kuijken_dubinski95, widrow_dubinski05} that generates discs using a S\'ersic surface density profile:
\begin{equation}
 \Sigma_{d}=\frac{M_{d}}{2\pi n R_{d}^{2}\Gamma(2n) }e^{-(R/R_{d})^{1/n}} \mathrm{sech}^2(z/z_d)~,
\end{equation}
where $M_{d}$ is the disc mass, $R_{d}$ is the scale length, $n$ is the S\'ersic index, and $\Gamma$ is the gamma function. The S\'{e}rsic disc has $M_{d}=5.74\times10^{10} \Msun$, $R_{d}=0.265\kpc$, $z_{d}=0.25\kpc$ and $n=2.05$. The model has a Hernquist dark matter (DM) halo, defined with \textsc{GalactICS} parameters $\sigma_h=550$ kms$^{-1}$, $R_h=30$ kpc, $\alpha=1$ and $\beta=4$. It has $5\times10^{6}$ DM particles and $4.4\times10^{6}$ disc particles. In this, and in the other pure $N$-body simulations described below, the softening length is $\epsilon=50$ pc and $\epsilon=100$ pc for star- and DM-particles, respectively.

Model SD1S suppresses most secular bar growth by setting the halo of model SD1 in full prograde rotation \citep{Debattista_2000, Long_2014, Collier_2018}. This results in a large halo spin ($\lambda \simeq 0.091$), rare in cosmological simulations \citep{bullock_angmom+01}. Our goal is to contrast this model with its counterpart SD1 whose bar is growing, so we are not concerned by its fully rotating halo being unrealistic. 

HG1 is a star-forming simulation described in \citet{Cole_2014}, \citet{Gardner_2014} and \citet{Debattista_2017}. It starts out with a gas corona but no stars, hence the evolution is totally self-consistent. It is evolved with the smooth particle hydrodynamics code \textsc{GASOLINE} \citep{gasoline}. The simulation uses high resolution ($\epsilon=50$ pc for the gas, and gas particles with initial mass $2.7\times 10^4 \Msun$) and after being evolved for 10 Gyr has $\sim1.1\times10^7$ stellar particles, and a total stellar mass of $\sim6.5\times10^{10}\mathrm{M}\odot$. It uses the gas cooling, star formation and stellar feedback prescriptions of \citet{stinson+06}. The DM halo consists of $5\times 10^6$ particles with $\epsilon=103$ pc, where $4.5\times 10^6$ of them have mass $8.5\times 10^4\Msun$, and $1.7\times 10^6\Msun$ for the remainder.

Finally, Model 4 is a disc $N$-body model (no gas or star formation) -- see also \citet{Debattista_2020}. It is a baryon-dominated disc Milky Way-like model, again set up using \textsc{GalactICS}. It has a Navarro-Frenk-White DM halo \citep{navarro+1996}, and an exponential disc with an isothermal vertical profile. The disc has a central radial velocity dispersion of 128 kms$^{-1}$, a disc scale radius of 2.4 kpc and a scale height of 150 pc. It has $6\times10^6$ particles in the disc and $4\times10^6$ in the halo. 

All $N$-body models were evolved using \textsc{PKDGRAV} \citep{pkdgrav} for 10 Gyr. In what follows, we take model SD1 as our fiducial model, because of its clear trends illustrating the main results of this paper, although model HG1 may be more realistic.

\subsection{Simulation global properties}
\label{sec:sims_properties}
We use {\sc pynbody} \citep[][]{Pontzen_2013} to read, center  and align the simulations such that, at each snapshot analysed, the bar is initially along the $x$-axis. We estimate the gravitational potentials $\Phi$ and integrate orbits using  \textsc{agama} \citep{Vasiliev_2019}. Smooth potentials due to star- and gas-particles (when present) are separately modelled as triaxial distributions and estimated with the ``cylspline'' potential type, while the halo contribution is assumed to be axisymmetric and is estimated with the ``multipole'' type. For each snapshot analysed, we then estimate the total potential.

\begin{figure}
    \centering
	\includegraphics[scale=0.325]{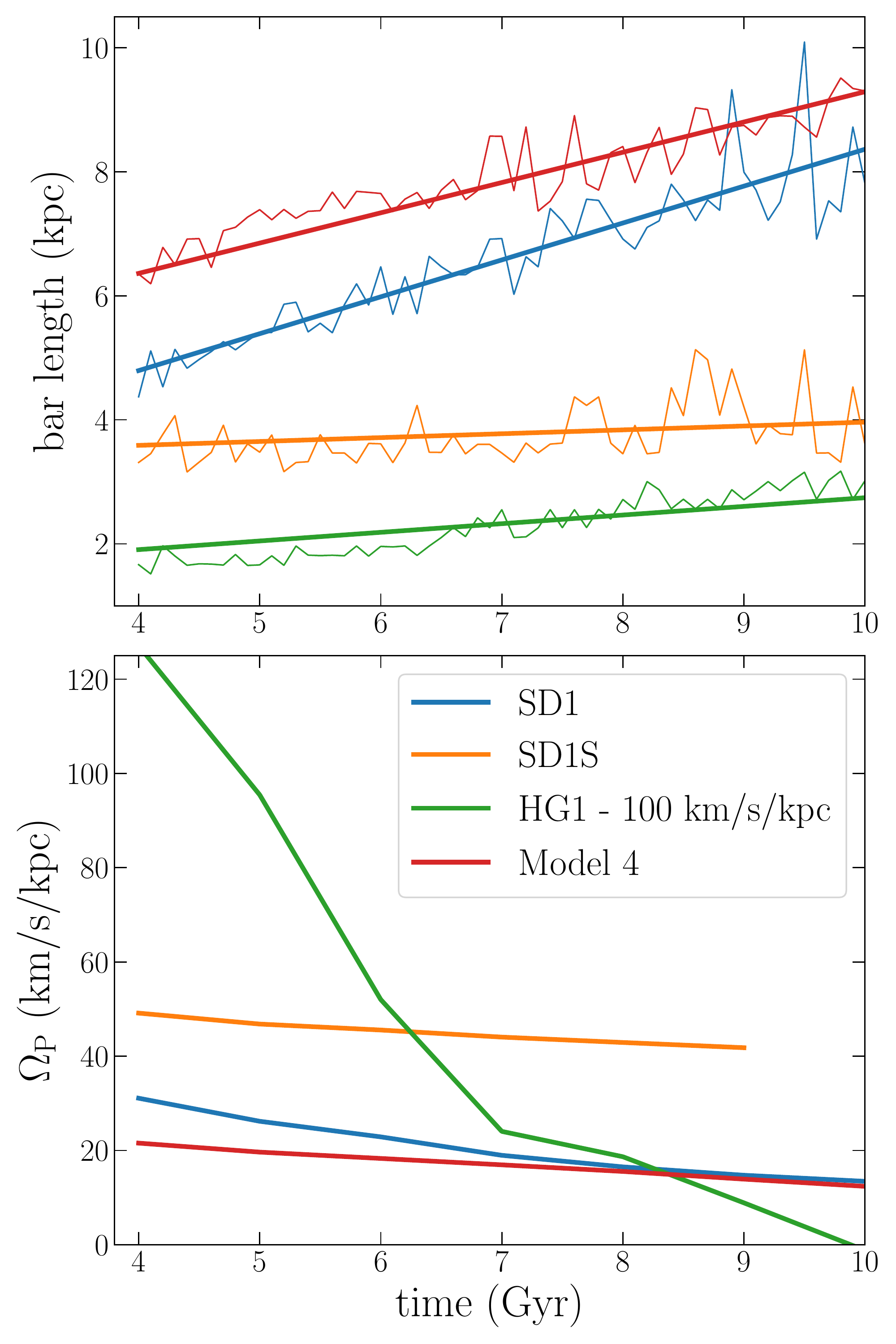}
    \caption{Evolution of bar lengths (top) and pattern speeds  (bottom) for our simulations. In the upper panel, thick lines show the linear fit to the bar length evolution (thin), which provides $\abar$ used in our analysis. In all models the bars slow down (but less significantly for SD1S and Model 4), and grow (but less significantly for models SD1S and HG1). For HG1, the values are shifted down by $100 \kmskpc$.}
    \label{fig:Om_p_all_sims}
\end{figure}

We compute the bar length (radius) $\abar$ as the mean of the cylindrical radius at which the amplitude of the $m=2$ Fourier moment of the $(x, y)$-plane surface density distribution reaches half its maximum value after its peak, and the radius at which the phase of the $m=2$ component deviates from a constant by more than $10\degrees$ -- see \citetalias{Anderson2022} for further details. We fit a straight line to the bar length as a function of time and use the fitted function to represent $\abar$ at each snapshot. Both estimated and fitted bar lengths are shown in the upper panel of Fig.~\ref{fig:Om_p_all_sims} as thin and thick lines, respectively, and we see that a straight line is a reasonable fit.

The bar pattern speeds $\OmP$ are computed as the time derivative of the phase angle of the $m=2$ Fourier mode of star-particles, as done by \cite{Debattista_2000}. We use 21 snapshots spaced by 5 Myr around the snapshots of interest. We then fit a straight line to the bar phase angle as a function of time in each of these time intervals, where the phase is assumed to increase monotonically with time. The pattern speeds are shown in Fig.~\ref{fig:Om_p_all_sims} (bottom panel) as functions of time. As often observed in simulations \citep[e.g.][]{Hernquist_1992, Debattista_1998, Oneill_2003, Athanassoula_2003}, the bars slow down significantly over the time span of the simulation -- see also \cite{Fragkoudi_2021} for a recent study of cosmological simulations and \cite{Hamilton_2022} for a recent theoretical treatment.

The bar in the Model SD1 starts with a slightly higher $\OmP$ in comparison to Model 4, but the pattern speeds in both models evolve to very similar, almost constant, values towards the end of the simulations. The model HG1 has a very centrally concentrated mass distribution, with a steep rotation curve in the central parts. Thus it has a very fast bar, which decelerates quickly without increasing in length significantly. However, these differences do not seem to alter the general trends of the shoulders we investigate.

\begin{figure*}
    \centering
	\includegraphics[scale=0.3]{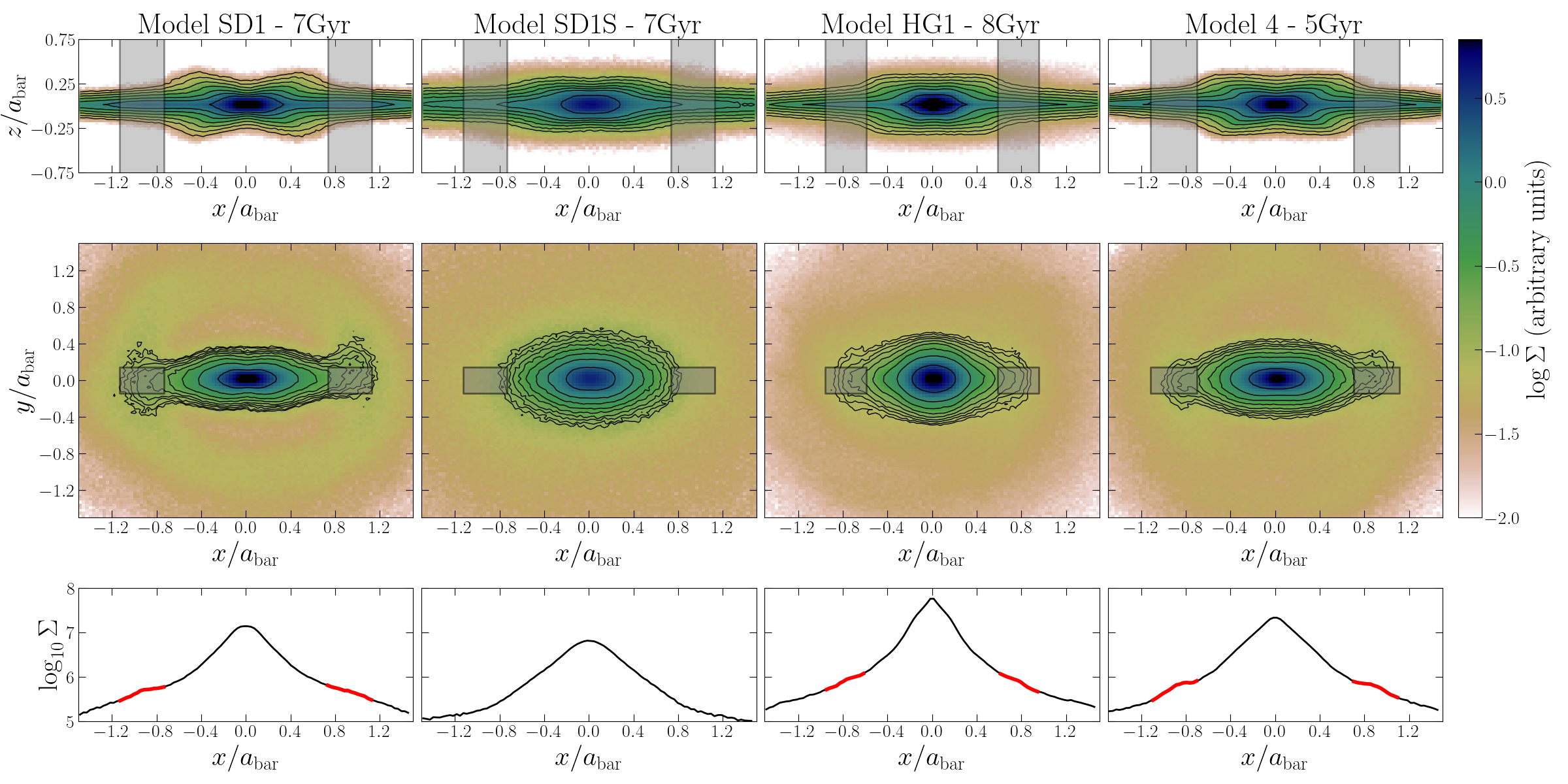}
    \caption{Edge-on (top row) and face-on (middle row) normalised surface density maps for the four simulations at the selection times. Shaded rectangles show the shoulder regions $|y|/\abar<0.145$ and $x$-edges identified in \citetalias{Anderson2022} (except for SD1S which does not have detected shoulders), where particles are selected. The bottom row shows the surface density along the stripe $|y|/\abar<0.145$, with the red snippets showing the shoulders (where detected).}
    \label{fig:Sigma_xy_xz_all}
\end{figure*}

Fig.~\ref{fig:Sigma_xy_xz_all} shows edge-on (top row) and face-on (middle row) views of the four models at the times when particles are selected for orbit integration. The coordinates are normalized by the respective bar lengths. Colors represent surface densities in log-scale. The $x-y$ ($x-z$) contours show 9 (7) equally-spaced percentiles, from 85 to 99 (84 to 99). The edge-on view shows clear signs of a BP-bulge in all models, although the ``X'' shape is weaker in models SD1S and HG1.

In the face-on view, we see the bars along the $x$-axis, and hints of density excess at their outskirts (except for SD1S, and barely for HG1) shown by the contours. The shaded rectangles represent the shoulder region, defined as $|y|/\abar<0.145$ and $x$-edges detected with the algorithm of \citetalias{Anderson2022}. Model SD1S has no shoulders, and we define its shoulder region using the same fractions of the bar length as those of SD1. In all models with shoulders, these are detected in the transition region between the thick and thin parts of the bar. The bottom row shows the surface density along the stripe defined by $|y|/\abar<0.145$ in each model, with red snippets showing where shoulders are detected.

Table~\ref{tab:sims} shows the number of star-particles selected in the shoulder region, the bar length $\abar$ and bar speed parameter $\mathcal{R} = R_\mathrm{CR}/\abar$, where $R_\mathrm{CR}$ is the co-rotation radius \citep[][]{Debattista_2000} at the selection times, and whether the model has gas or a strongly buckling bar (and when it buckles). Models SD1S and HG1 have fast bars ($\mathcal{R}\lesssim 1.4$) at these snapshots, while SD1 and Model 4 have slow bars ($\mathcal{R}\gtrsim 1.4$).

\begin{table}
    \begin{center}
     \caption{Selection times ($t_s$), bar speed parameter $\mathcal{R} = R_\mathrm{CR}/\abar$, bar length $\abar$, number of selected star-particles and whether the model has gas or a strongly buckling bar. Model SD1 and model 4 have reasonably strong shoulders. Model HG1 has weak shoulders and SD1S has a barely evolving bar and no shoulders. The simulation naming convention is the same as in \citetalias{Anderson2022}.}
    \label{tab:sims}
    \begin{tabular}{| l | l | l | l | l | l | l |}
    \hline
    Model & $t_s$ & $\mathcal{R}$ & $\abar/\kpc$ & $N$  & gas? & Buckling? \\ \hline
    SD1 & $7 \Gyr$ & 1.99 & 6.58 & 106032 & N & N \\ \hline
    SD1S & $7 \Gyr$ & 1.31 & 3.78 & 46973 & N & N \\ \hline
    HG1 & $8 \Gyr$ & 1.15 & 2.46 & 177426 & Y & N \\ \hline
    4 & $5 \Gyr$ & 1.60 & 6.85 & 136488 & N & Y, 3.8 Gyr\\
    \hline
    \end{tabular}
    \end{center}
\end{table}

\subsection{Epicyclic frequencies, resonance names and selection region}

\begin{figure*}
    \centering
% 	% Allowable file formats are eps or ps if compiling using latex
% 	% or pdf, png, jpg if compiling using pdflatex
	\includegraphics[scale=0.4]{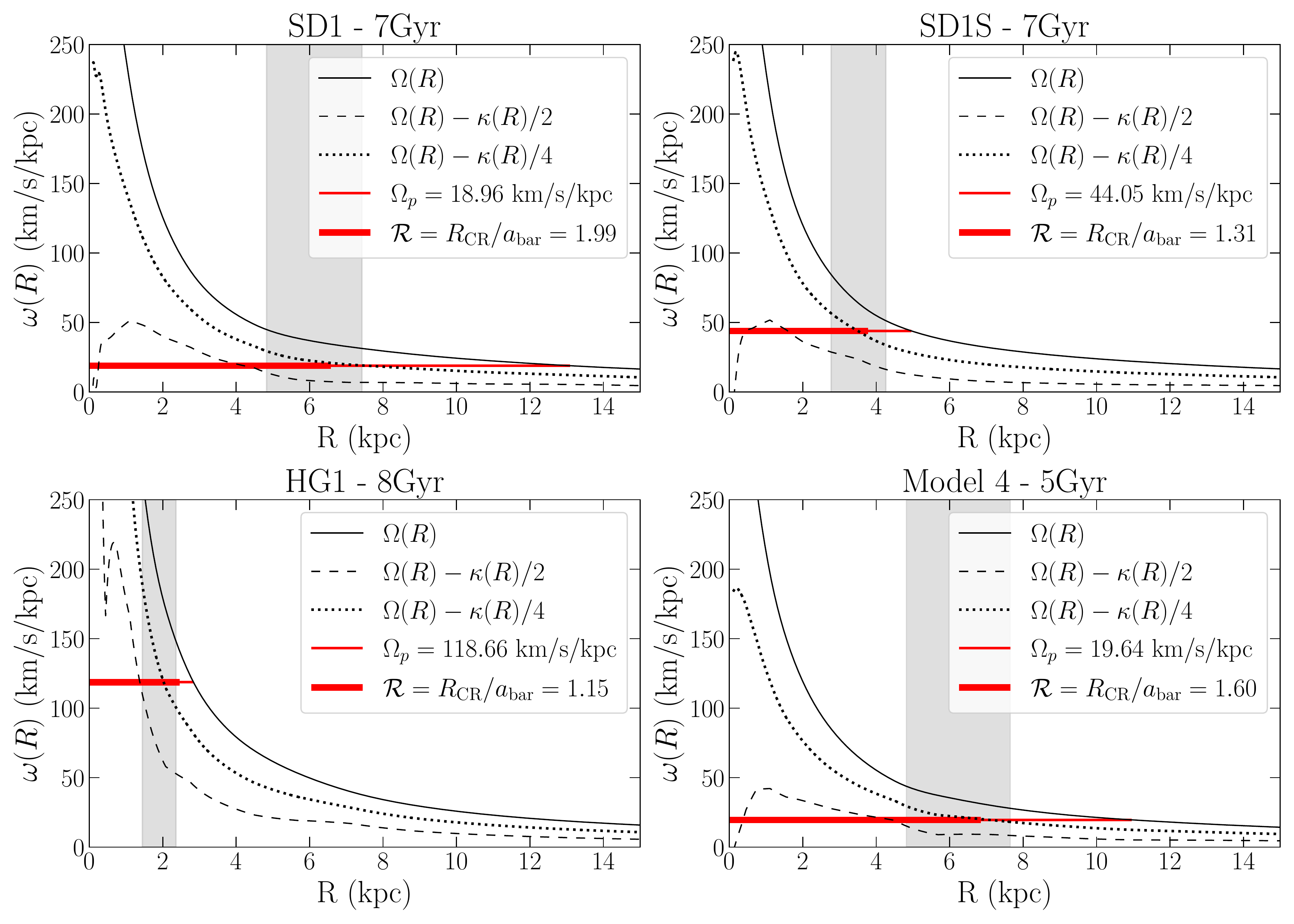}
    \caption{Rotation curves $\Omega(R)$ (solid black), $\Omega(R)-\kappa(R)/2$ (dashed) and $\Omega(R)-\kappa(R)/4$ (dotted) for the different simulations at the time when star-particles are selected. Gray areas show the average radial interval where shoulders are detected (SD1S has no shoulders, and we define this region using the same fractions of the bar length as those of SD1), within which we select particles in the simulations. Horizontal red lines show the bar pattern speeds, with the thick part representing the bar lengths, at the respective times. Model SD1 and Model 4 have slow bars ($\mathcal{R}\gtrsim 1.4$), while models SD1S and HG1 have fast bars. The shoulder region is around the Ultra Harmonic radius where $\Omega - \kappa/4 = \OmP$.}
    \label{fig:Omega_vs_R}
\end{figure*}

Fig.~\ref{fig:Omega_vs_R} shows the rotation curve $\Omega(R)$ (solid black), and the curves ${\Omega(R) - \kappa(R)/2}$ (dashed) and ${\Omega(R) - \kappa(R)/4}$ (dotted), where
\begin{equation}
\label{eq:Omega}
    \Omega(R)\equiv \sqrt{\frac{1}{R}\frac{\partial \Phi}{\partial R}},
\end{equation}
and
\begin{equation}
\label{eq:kappa}
    \kappa(R)\equiv \sqrt{\frac{\partial^2\Phi}{\partial R^2} + \frac{3}{R}\frac{\partial\Phi}{\partial R}}
\end{equation}
is the radial epicyclic frequency \citep{BT}, evaluated along the bar major-axis, at the times when star-particles are selected. The times are chosen when clear and stable shoulders are present (except for SD1S which does not have shoulders), and which are far from buckling events (in Model 4).

The bar pattern speeds at these times are shown as thin red lines, with the thick part representing the bar length. Gray rectangles show the area within the mean shoulder $x$-edges, as detected in \citetalias{Anderson2022}, within which particles are selected. For simulation SD1S, we adopt the same fraction of the bar length as that of the shoulder extent of SD1 at the same time.

It is interesting to note that for all simulations with shoulders, these start outside the outer Inner Lindblad radius (where $\Omega - \kappa/2=\Omega_P$). This is the region where the (main bar-supporting) $x_1$ orbits can start developing loops at their ends \citep{Contopoulos_1980, BT}, and this is the first hint of the importance of these orbits to shoulders, as we confirm below. On the other hand, this might suggest that the frequency ratio $\OmphiR=1/2$, where $\Omphi$ and $\Omega_R$ are the azimuthal and radial frequencies for each orbit, is not relevant for the shoulders. Conversely, Fig.~\ref{fig:Omega_vs_R} shows that the shoulders typically appear near the Ultra Harmonic radius (where $\Omega - \kappa/4=\Omega_P$), which might suggest an important role of the resonance $\OmphiR=1/4$. However, our frequency analysis (Sec.~\ref{sec:results}) reveals a fundamental role for the resonance $\OmphiR=1/2$ in supporting the shoulders, and no important role for the resonance $\OmphiR=1/4$.

The resonance $\OmphiR=1/2$ is the analog of the classical Inner Lindblad Resonance, but applies to orbits of arbitrary eccentricity. Keeping in mind that resonances are defined in the space of frequencies (rather than physical space), in this paper we will use actual frequency ratios to define the main resonances. For instance, the resonance $\OmphiR=1/2$ will be called the Inner Lindblad Resonance (ILR), and the resonance $\OmphiR=1/4$ will be called Ultra Harmonic Resonance (UHR). When identifying the approximate locations of these resonances, based on the epicyclic frequencies, we will refer to their radii, e.g. the Inner Lindblad radius and the Ultra Harmonic radius -- see \cite{Athanassoula_2003} for a discussion on the pros and cons of different ILR definitions.

\begin{figure*}
\begin{center}
	\includegraphics[scale=0.4]{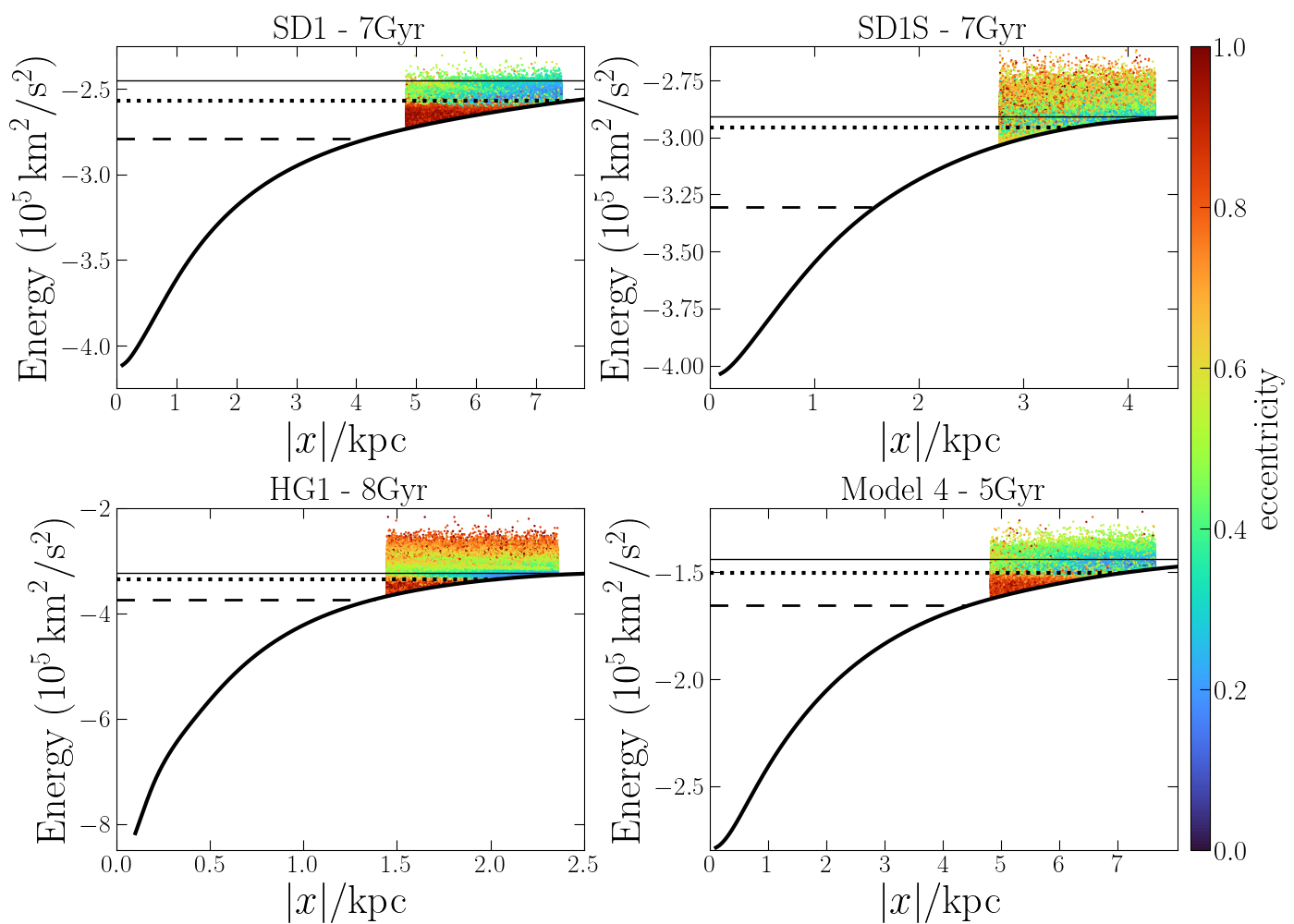}
    \caption{Effective potential along the bar major-axis (thick black solid) and Jacobi integral (points) of the particles selected in the shoulder region from the different simulations, at the selection time. Also shown the effective potential values at L1 (co-rotation, thin solid), at the Ultra Harmonic radius (dotted) and outer Inner Lindblad radius (dashed).}
    \label{fig:pot_eff}
\end{center}
\end{figure*}

With the bar aligned with the $x$-axis, we select star-particles within $|y|/\abar<0.145$ and in the $x$-intervals where shoulders are identified (with no cut in $z$) -- see Figs.~\ref{fig:Sigma_xy_xz_all} and \ref{fig:Omega_vs_R}. Since the detected shoulders are not exactly symmetric around $x=0$, we define the ``shoulder region'' averaging their edges on the $x<0$ and $x>0$ sides. This selection is convenient for identifying orbits supporting the shoulder, but also because, being in the bar outskirts, this region is in the transition to the disc, and particles supporting both structures are selected. Table~\ref{tab:sims} shows the total number of star-particles selected in each model.

Fig. \ref{fig:pot_eff} illustrates the selection (shoulder) region in terms of the effective potential in the rotating frame,
\begin{equation}
    \poteff = \Phi - \frac{1}{2}\OmP^2x^2,
    \label{eq:phi_eff}
\end{equation}
evaluated along the bar major-axis $x$. The points show the Jacobi integral
\begin{equation}
\label{eq:Jacobi}
    \EJ = E - \OmP L_z,
\end{equation}
which is conserved in the frame rotating with angular speed $\OmP$, where $E$ is the energy in the inertial frame and $L_z$ is the angular momentum. These points are color-coded by the eccentricity $\ecc \equiv (r_\mathrm{apo} - r_\mathrm{per})/(r_\mathrm{apo} + r_\mathrm{per})$, where $r_\mathrm{apo}$ and $r_\mathrm{per}$ are the apo- and pericenter radii. These quantities are calculated after the orbit integration explained in Sec.~\ref{sec:orb_int}. The horizontal lines show the effective potential evaluated at the co-rotation radius, $R_\mathrm{RC}$ (solid thin), at the Ultra Harmonic radius (dotted), and at the outer Inner Lindblad radius, $R_\mathrm{ILR}$ (dashed). In agreement with Fig.~\ref{fig:Omega_vs_R}, the selected particles have $E_J> \poteff(R_\mathrm{ILR})$. We note that for the simulations with shoulders the most bound orbits have high eccentricities (thus small $r_\mathrm{per}$) and these orbits extend right up to the Ultra Harmonic radius. Besides that, there are orbits with $E_J> \poteff(R_\mathrm{CR})$ and reasonably large eccentricities, which can visit both the disc and the bar regions -- see Sec.~\ref{sec:orbs_vILR_cloud}.

\section{Orbits}
\label{sec:orbits}
The coordinates of particles selected at each snapshot set the initial conditions for orbit integration in the frozen potential rotating with the corresponding pattern speed \citep[see][]{Athanassoula_2013}. This means that in the rotating frame the potential is static for the whole integration time. We then estimate the frequencies of motion in the three directions (as detailed below), producing a frequency map for each snapshot analysed.

\subsection{Orbit integration}
\label{sec:orb_int}
 For particles selected at a time $t_s$, we integrate orbits in the frozen potentials of five snapshots, $[t_s - 2, t_s - 1, t_s, t_s+1, t_s+2]$ Gyr, with a relative error tolerance of $10^{-15}$ for each coordinate, which is the best we could achieve for all particles at all snapshots. When plotting orbits, we use coordinates in the bar rotating frame, but the frequency analysis uses coordinates in the inertial frame, as detailed below.

We integrate orbits for a (pseudo-)time of $5\Gyr$ (storing $10^5$ points per orbit), which is a good compromise between two limiting time-scales: a lower limit of, at least, tens of orbital periods for precise frequency estimates \citep[][]{Valluri_Merritt_1998}, and
an upper limit given by the predictability horizon for chaotic orbits with finite precision integration \citep[e.g.][]{Roy_1991}. Note that this is not the real time of the simulation, but a parameter in the determination of instantaneous orbital properties in the frozen potential. This integration time 
% Fig.~\ref{fig:Tcirc} shows a histogram of the circular orbital periods for star-particles in the shoulder region of the simulation SD1 at 7 Gyr. The integration time
represents at least $20\times T_\mathrm{circ}(E)$, where $T_\mathrm{circ}(E)$ is the period of a circular orbit with energy $E$, for almost all particles, and this is typically the case for all selections. Furthermore, among the cylindrical coordinates $(R,\varphi,z)$, the azimuthal period is typically the longest, so the integration time is also large enough for the other coordinates.

\subsection{Frequency analysis}
\label{sec:freq_anal}
We employ the Numerical Analysis of Fundamental Frequencies (NAFF) algorithm  originally proposed by \cite{Laskar_1992} and further developed by \cite{Valluri_Merritt_1998, Valluri_2010} to extract the leading frequencies of orbits. Regular orbits are quasi-periodic, so the time-series of each coordinate $q$ can be approximated by
\begin{equation}
    q(t) = \sum_{k=1}^{k_\mathrm{max}} a_k e^{i\omega_k t},
\end{equation}
where $a_k$ is a complex amplitude, and $\omega_k$ is an integer combination of the fundamental frequencies $\Omega_1$, $\Omega_2$ and $\Omega_3$ (time-derivatives of the canonical angle variables). The power spectrum of a Discrete Fourier Transform (DFT) of a regular orbit has discrete peaks at frequencies $\omega_k$ \citep[spectral lines, see][]{Binney_Spergel_1982}. Typically, although not always, the frequency with the largest amplitude (hereafter the {\it leading frequency}) corresponds to the fundamental frequency.

In non-integrable potentials, quasi-periodicity cannot be assumed a priori, which might hinder the extraction of meaningful fundamental frequencies. However, most realistic galactic potentials, although non-integrable, still host a large number of regular or weakly chaotic orbits, a picture (formally) supported by the KAM theorems \cite[][]{Kolmogorov_1954, Arnold_1963, Moser_1962}. Additionally, in non-integrable potentials resonance trapping significantly affects the system's dynamical evolution. Once the fundamental frequencies are identified, resonances stand out in a plot showing ratios of the frequencies, i.e. a frequency map \citep[][]{Laskar_1990, Laskar_1999}, such as the one shown in Fig.~\ref{fig:freq_map_768MR_no_hist}.

We have written a new pure-python implementation of the NAFF algorithm, which we dub \texttt{naif}\footnote{The documentation and installation instructions can be accessed at \url{http://naif.readthedocs.io/en/latest/index.html}.}, based on the \textsc{Fortran} implementation of \cite{Valluri_Merritt_1998}. A summary of the algorithm is presented in Appendix \ref{sec:naff}. Variants of this method have been developed and applied in different contexts \citep[e.g.][]{Binney_Spergel_1982, Carpintero_1998, Valluri_2016, Price-Whelan_2016, BeS_2019, Dodd_2022, Lucey_2022}. The analysis involves important decisions on implementation details, such as the coordinate system (e.g. Cartesian or cylindrical), the use of real or complex time-series, the reference frame, the window function etc. In the study of bars \citep[e.g.][]{Portail_2015, Valluri_2016, Gajda_2016, Parul_2020, Smirnov_2021, Sellwood_2020}, it is common to adopt Cartesian coordinates (or mixed Cartesian and cylindrical) in the bar rotating frame, where the frequency of oscillation along the major axis, $\Omega_x$, is used as a proxy for the azimuthal frequency $\Omphi-\OmP$. Furthermore, spatial coordinates as (real) time-series are often used. See Sec.~\ref{sec:compare_tech} for a discussion on different choices and techniques.

To get some guidance in these decisions, and to test our algorithm, in Appendix \ref{sec:isochrone} we compare our frequency estimates with analytical expressions for orbits in the isochrone potential. We compare the use of real and complex time-series (Appendix~\ref{sec:naff}), and show that the radial frequency $\Omega_r$ is estimated with relative errors $|\delta\Omega_r/\Omega_r| \approx 10^{-6}$ in both cases. For the azimuthal frequency $\Omphi$, we show that the complex time-series $f_\varphi = \sqrt{2|L_z|}(\cos\varphi + i\sin\varphi)$ \citep[][]{Papa_Laskar_1996, Papa_Laskar_1998}, where $L_z$ is the $z$-component of the angular momentum, does offer advantages, compared with $f_\varphi = \varphi$: it improves the accuracy and it does not require setting $\sign (\Omphi) = \sign(L_z)$, which is the case for the real time-series. After excluding the precessing frequency $\Omega_r - \Omphi$, which is the leading one for orbits with large apocenter, we estimate $\Omphi$ with relative errors $|\delta\Omphi/\Omphi| \lesssim 10^{-3}$ for almost all orbits.
 
 All frequency estimates in this paper use the Hanning window, i.e. $p=1$ in Eq.~\ref{eq:chi}. In the following analysis, for an easier identification of important known resonances, we use cylindrical coordinates ($R, \varphi, z$) stored in the inertial frame of the galaxy, rather than the bar rotating frame. This allows identification of the co-rotation resonance, which would require an infinite integration time in the rotating frame -- see Appendix \ref{sec:rot_frame}. Based on the discussions above, we use as input in the frequency analysis the complex time-series:
\begin{equation}
    % \begin{cases}
    \begin{array}{l}
      f_R = R + i v_R \\
      f_\varphi = \sqrt{2|L_z|}(\cos\varphi + i\sin\varphi)\\
      f_z = z + i v_z,
    % \end{cases}
    \end{array}
    \label{eq:f_cyl_complex}
\end{equation}
where $v_R$ and $v_z$ are the radial and vertical velocities, respectively.
For the radial and vertical components we take the absolute value of the leading frequency, while we retain the original sign for $\Omphi$. Since our cuts do not select particles with significantly large apocenter distances, we expect the number of orbits with misidentified fundamental frequencies to be negligible.

\section{Results}
\label{sec:results}
\subsection{Frequency maps: overview}
\label{sec:freq_map_overview}

\begin{figure}
	\includegraphics[width=\columnwidth]{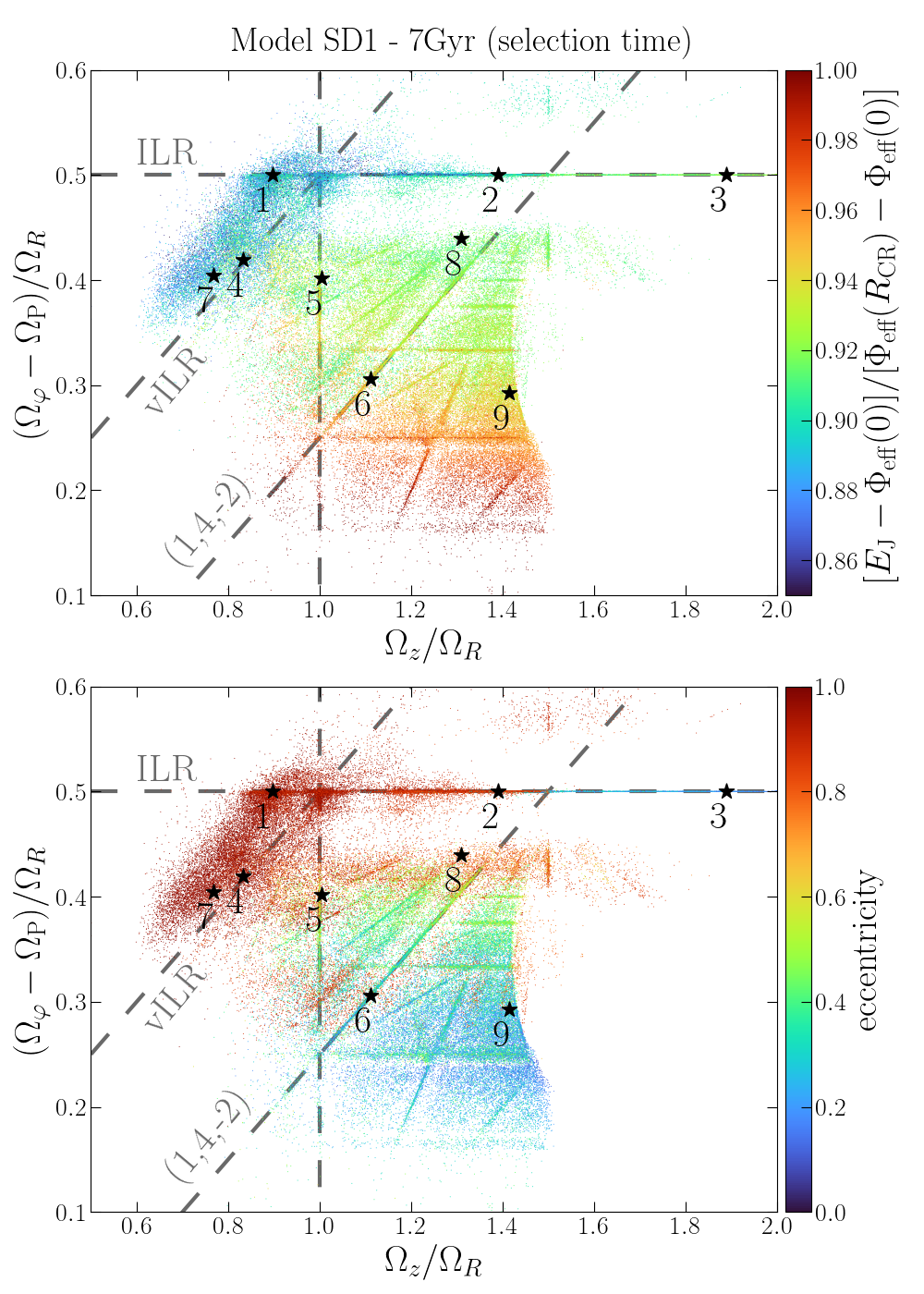}
    \caption{Frequency map for the shoulder region of model SD1 at $7\Gyr$, with dashed lines showing prominent resonances. We indicate the Inner Lindblad Resonance (ILR), the vertical ILR (vILR) and the 3D resonance $\Omega_R + 4(\Omega_\varphi-\OmP)-2\Omega_z = 0$. The upper panel is color-coded by the normalized Jacobi integral, Eq.~\ref{eq:Jacobi_normal}. The upper-left corner is dominated by typical orbits in the central region, and moving towards the lower-right corner, orbits get more disc-like. The {\it vILR cloud} is the high-eccentricity region around the vILR. Numbered stars are the sample orbits shown in Figs.~\ref{fig:orbs_stack_ILR}, \ref{fig:orbs_stack_vILR} and \ref{fig:orbs_stack_other}.}
    \label{fig:freq_map_768MR_no_hist}
\end{figure}

Having demonstrated the accuracy of our numerical procedure, we move on to the analysis of orbits in the frozen potentials of our $N$-body simulations. Fig.~\ref{fig:freq_map_768MR_no_hist} shows the frequency map obtained for star-particles selected in the shoulder region of model SD1 at $7\Gyr$ (see Figs.~\ref{fig:Sigma_xy_xz_all} and \ref{fig:Omega_vs_R}). The upper panel is color-coded by the normalized Jacobi integral (see Fig.~\ref{fig:pot_eff})
\begin{equation}
    \frac{\EJ - \poteff(0)}{\poteff(R_\mathrm{CR})-\poteff(0)},
    \label{eq:Jacobi_normal}
\end{equation}
where $\Phi_\mathrm{eff}$ and $\EJ$ are defined in Eqs.~\eqref{eq:phi_eff} and ~\eqref{eq:Jacobi}, respectively, and $R_\mathrm{CR}$ is the co-rotation radius. The bottom panel is color-coded by the eccentricity, and numbered stars indicate the example orbits presented in Figs.~\ref{fig:orbs_stack_ILR}, \ref{fig:orbs_stack_vILR} and \ref{fig:orbs_stack_other}.

Given the typical shape of rotation curves ({\it cf.} Fig.~\ref{fig:Omega_vs_R}), we can anticipate that orbits with high $\OmphiR$ in Fig.~\ref{fig:freq_map_768MR_no_hist} are typical of the galaxy central regions. Moving down in Fig.~\ref{fig:freq_map_768MR_no_hist} we go towards the end of the bar. For fast bars, this happens near the co-rotation, at $\Omphi - \OmP=0$, and transitioning to the disc we have ${\Omphi - \OmP\lesssim 0}$, corresponding to retrograde stars in the bar frame. Evaluating the functions $\Omega(R)$ and $\kappa(R)$, Eqs.~\eqref{eq:Omega}-\eqref{eq:kappa}, for this model at the average outer shoulder edge (the outermost region of our selection), we obtain $(\Omega - \OmP)/\kappa \approx 0.25$, in broad agreement with the lower values ($\approx 0.15$) on the $y$-axis of Fig.~\ref{fig:freq_map_768MR_no_hist}, and with the fact that the shoulder outer edge in this model matches the Ultra Harmonic radius -- see Fig.~\ref{fig:pot_eff}.

The ratio $\OmzR$ is a proxy for the vertical thinness of the orbit: it is significantly larger than unity for typical orbits in the disc \citep[in the Solar neighborhood, $\OmzR\approx 2$,][]{BT}, and can be $\OmzR \lesssim 1$ in thicker or more spheroidal regions. Thus, the upper-left corner in Fig.~\ref{fig:freq_map_768MR_no_hist} is dominated by typical orbits that can visit the very central regions of the galaxy, and orbits get more disc-like when we move towards the lower-right corner, in agreement with the eccentricity trends seen in the bottom panel. With the radial $\kappa(R)$, Eq.~\ref{eq:kappa}, and vertical $\nu(R) \equiv \sqrt{\partial^2 \Phi/\partial z^2}$ epicyclic frequencies evaluated at the outer shoulder edge, we get $\nu/\kappa \approx 1.70$, also in broad agreement with the rightmost limit of Fig.~\ref{fig:freq_map_768MR_no_hist} (neglecting orbits at the ILR, which are well inside the bar region and thus less well approximated by the linear expressions).

We identify the main regions in the frequency map that we explore in detail below. The horizontal line at $\OmphiR = 1/2$ characterizes the $x_1$ orbits, which are known to be the backbone orbits of bars \citep[][]{Contopoulos_1989, Skokos_2002, Athanassoula_2003}. In principle, $x_2$ orbits, which are aligned with the bar minor-axis, can also be present along this resonance. However, these orbits are only expected within the Inner Lindblad radius \citep[or between two Inner Lindblad radii, see][]{Albada_1982} and, as Fig.~\ref{fig:Omega_vs_R} shows, we select particles outside this region.

As already mentioned in Sec.~\ref{sec:sims_properties}, we refer to the frequency ratio $\OmphiR = 1/2$ as the ILR \citep[see e.g.][]{Athanassoula_2003, Weinberg_2007}. We emphasize that this differs from the classical definition of the ILR based on the epicyclic frequencies, Eqs.~\eqref{eq:Omega}-\eqref{eq:kappa}, or on the existence of $x_2$ orbits \citep[][]{Albada_1982}, but can be seen as their generalizations, in the frequency space, to orbits of arbitrary eccentricities. The UHR, ${\OmphiR = 1/4}$, is also well populated in Fig.~\ref{fig:freq_map_768MR_no_hist}. Also highlighted are the vertical ILR (vILR), where $(\Omphi-\OmP)/\Omega_z = 1/2$; the resonance $\OmzR=1$ and the 3D resonance ${\Omega_R + 4(\Omega_\varphi-\OmP)-2\Omega_z = 0}$, denoted by (1,4,-2). 

We call the cloud of highly eccentric orbits that are to the left and above the vILR the {\it vILR cloud}. We also note two other diffuse populations more clearly identified in the bottom panel: a large blue/green region with low eccentricities and permeated by several resonances (the region of orbits 6 and 9), and an orange stripe below the ILR, with $0.65\lesssim \ecc \lesssim 0.85$ and $0.4 \lesssim \OmphiR\lesssim 0.45$ (the region of orbit 8).

 We now illustrate the orbital morphologies in these different regions, noting that Fig.~\ref{fig:freq_map_768MR_no_hist} does not represent the whole bar region, but only orbits selected at its outermost parts at a particular time. The whole menu of orbits in bars has been extensively studied in other works and is beyond the scope of this paper \citep[see e.g.][]{Athanassoula_1983, Sparke_1987, Combes_1990, Skokos_2002, Patsis_2002, Valluri_2016, Patsis_2019}. For a review, see \cite{Sellwood_1993}.

 \subsubsection{Orbital shapes at the ILR}
 
\begin{figure}
    \centering
	\includegraphics[width=\columnwidth]{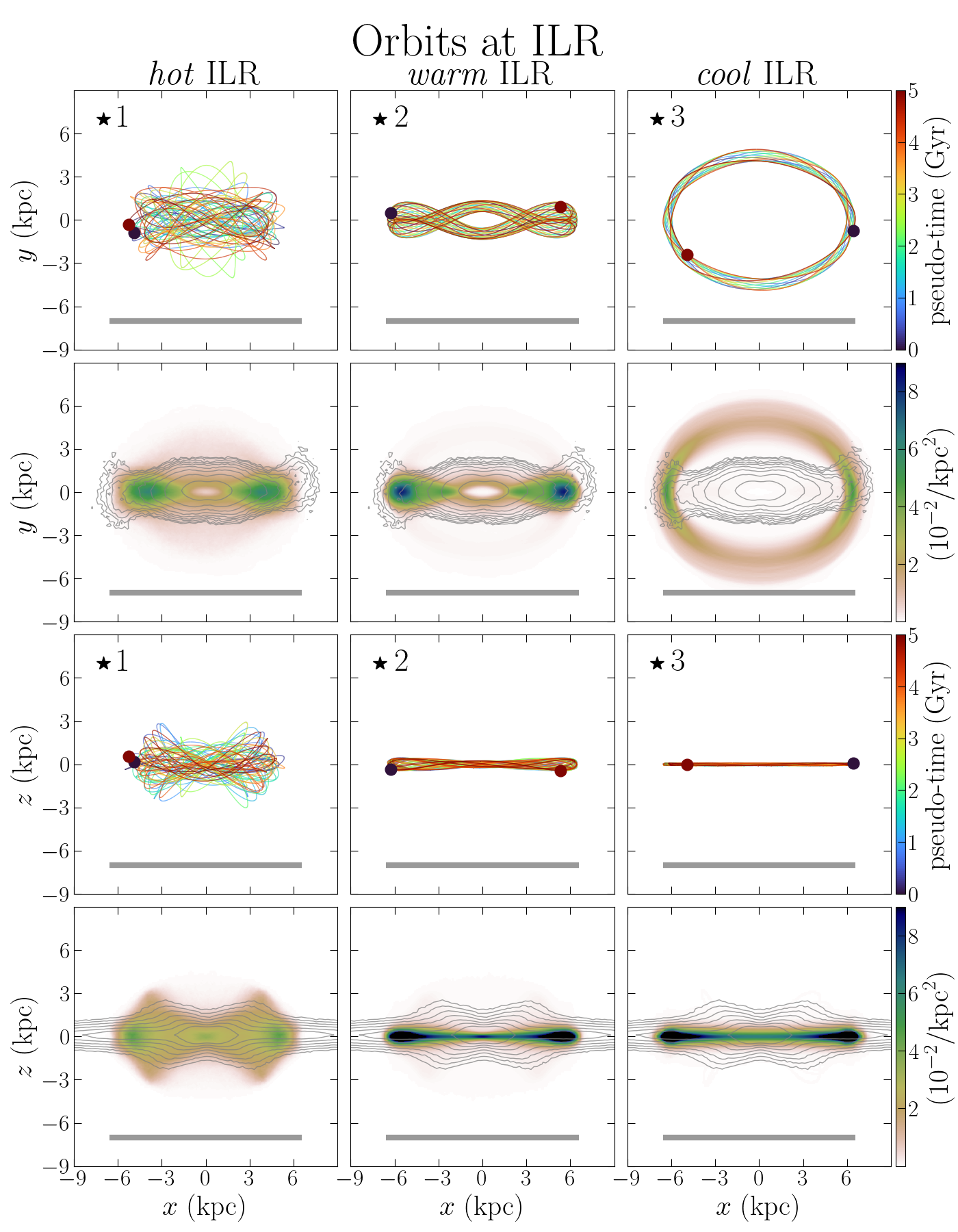}
    \caption{First and third rows: three typical orbits at the ILR selected at different $\Omega_z/\Omega_R$-ranges from simulation SD1 at $7\Gyr$. Orbits are color-coded by time, with dots showing the initial and final locations. Numbered stars identify these orbits in Fig.~\ref{fig:freq_map_768MR_no_hist}. Second and fourth rows: Stacked surface density maps of all orbits at the ILR in the respective $\Omega_z/\Omega_R$ ranges. The orbits and stacked density maps are shown in the bar rotating frame, and the thick gray lines show the bar size. Thin gray lines are the same contours shown in Fig.~\ref{fig:Sigma_xy_xz_all} for the total mass. The vertically warm orbits ${(1 < \OmzR < 3/2)}$ produce a marked density peak in the outskirts of the bar. A similar, but diluted, shape is also produced by the hot orbits $(\OmzR < 1)$.}
    \label{fig:orbs_stack_ILR}
\end{figure}

 We first select orbits at the ILR by requiring $|\OmphiR-1/2| < 0.0025$. By inspection this cut is a good choice to ensure minimal contamination from neighboring orbits. Unless otherwise stated, we use the same tolerance in selections of other resonances. Fig.~\ref{fig:freq_map_768MR_no_hist} suggests grouping the orbits at the ILR into three groups, which we will refer to as vertically {\it cool}, {\it warm} and {\it hot} orbits at the ILR (in some instances we will simply refer to cool, warm and hot orbits for conciseness). Vertically cool orbits are those at the ILR and with $\OmzR > 3/2$. They are the least bound and are generally less eccentric. A typical orbit from this group is shown in the right column (first and third rows) of Fig.~\ref{fig:orbs_stack_ILR}. It has an elliptical shape that is elongated in the direction of the major axis, and it is vertically very thin (see third row).

Transitioning to the middle section of the ILR, the vertically warm orbits have ${1 < \OmzR < 3/2}$. They are more bound (lower \EJ) and eccentric ($\ecc \approx 0.85$). The upper-middle panel in Fig.~\ref{fig:orbs_stack_ILR} shows a typical orbit in this region (orbit 2). These orbits still avoid the center and have characteristic loops at their ends, which have previously been identified in both analytical \citep[e.g.][]{Contopoulos_1978, Papayannopoulos_1983} and numerical studies \citep[e.g.][]{Sparke_1987, Petersen_2021}. In particular, \cite{Contopoulos_1988} found that these orbits typically appear near the Ultra Harmonic radius, which corresponds to our selection region -- see Fig.~\ref{fig:Omega_vs_R}. We refer to these as ``looped'' orbits. 

 Furthermore, for all simulations analyzed in this paper, we find that this transition from elliptical-like to this looped shape occurs at $\OmzR \approx 3/2$, suggesting some kind of influence of vertical resonances. In fact, we note that the crossing of the ILR with $\OmzR = 3/2$ is the convergence point of several 3D resonances in Fig.~\ref{fig:freq_map_768MR_no_hist}. In the next sections we demonstrate the fundamental role of this looped morphology in building the shoulders in the bar major-axis density profiles.

Finally, the vertically hot orbits are those at the ILR and with ${\OmzR < 1}$ (see Fig.~\ref{fig:freq_map_768MR_no_hist}). They are slightly deeper in the potential, and thus have slightly smaller $r_\mathrm{apo}$. Despite this, they are significantly more eccentric ($\ecc \approx 0.98$, see also Fig.~\ref{fig:pot_eff}), because they can get very close to the center, as illustrated in the upper-left panel in Fig.~\ref{fig:orbs_stack_ILR} (orbit 1).

\begin{figure}
\begin{center}
	\includegraphics[scale=0.375]{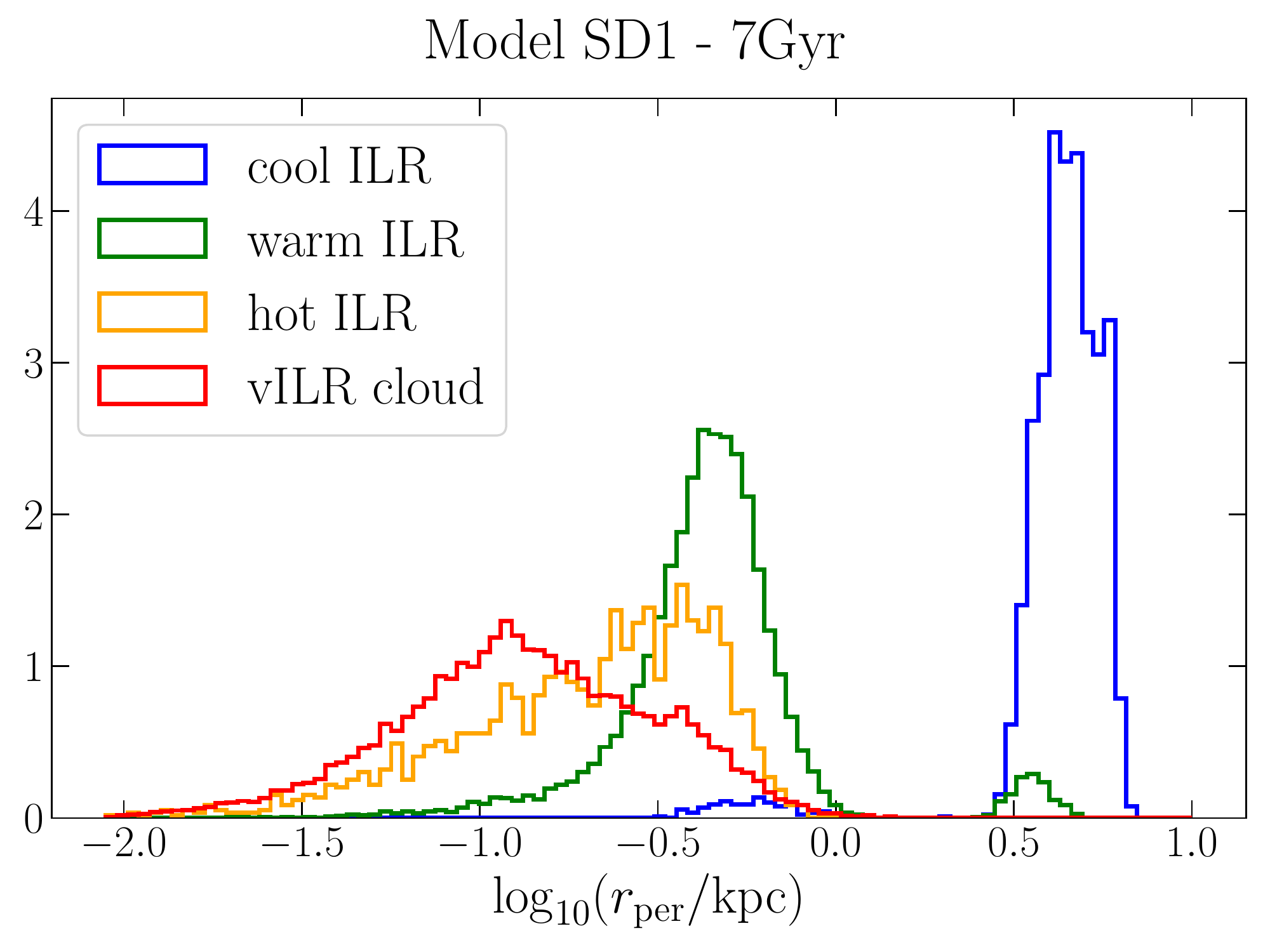}
    \caption{Normalized histogram of $\log_{10} (r_\mathrm{per}/\mathrm{kpc})$ for orbits along the ILR, in the different $\OmzR$ intervals, and for orbits at the vILR cloud -- see main text. A large fraction of the vertically hot orbits at the ILR (and the vILR cloud in general) can get as close as $\sim 10-100$ pc from the center. The small secondary peaks for cool and warm orbits at the ILR are due to cross-contamination of these orbits near the edge $\OmzR=3/2$}.
    \label{fig:hist_rper}
\end{center}
\end{figure}

Fig.~\ref{fig:hist_rper} shows a normalized histogram of $\log_{10} (r_\mathrm{per}/\mathrm{kpc})$ for cool, warm and hot orbits at the ILR of the model SD1 at $7\Gyr$. It is clear that a significant fraction of the hot orbits can get as close as 10-100 pc to the center. Moreover, Fig.~\ref{fig:orbs_stack_ILR} shows that typical orbits in this region do not have the coherent shapes observed before, scattering more in the $x-y$ plane (a signature of their chaotic nature, as demonstrated in Sec.~\ref{sec:freq_map_all_sims}). Finally, their vertical thickness is significantly larger than in orbits 2 and 3.

The second and fourth rows in Fig.~\ref{fig:orbs_stack_ILR} show the orbit-averaged surface densities of orbits at the ILR in each of the three groups in $\OmzR$. These plots are 2D histograms built by stacking all points of a single orbit (a time average) and all orbits (an ensemble average), normalized by the total number of orbits in each region to emphasize characteristic shapes rather than relative contributions of each group. We see that the sample orbits are representative of the populations.

The thin gray lines in the second and fourth rows are the same face-on and edge-on contours shown in Fig.~\ref{fig:Sigma_xy_xz_all} for the whole star-particle distribution. We anticipate the role of the looped morphology of vertically warm orbits at the ILR to produce the density excess in the bar outskirts. A similar morphology is also present for the hot orbits at the ILR, but the density excess appears more diluted -- see Secs.~\ref{sec:orbital_support_shoulder} and \ref{sec:evol_Sigma_stack}. We also note the X-shape in the $x-z$ plane of the orbit-averaged density of the hot orbits, suggesting their contribution to the BP-bulge visible in the contours.

\subsubsection{Orbital shapes at vertical resonances}
\begin{figure}
    \centering
	\includegraphics[width=\columnwidth]{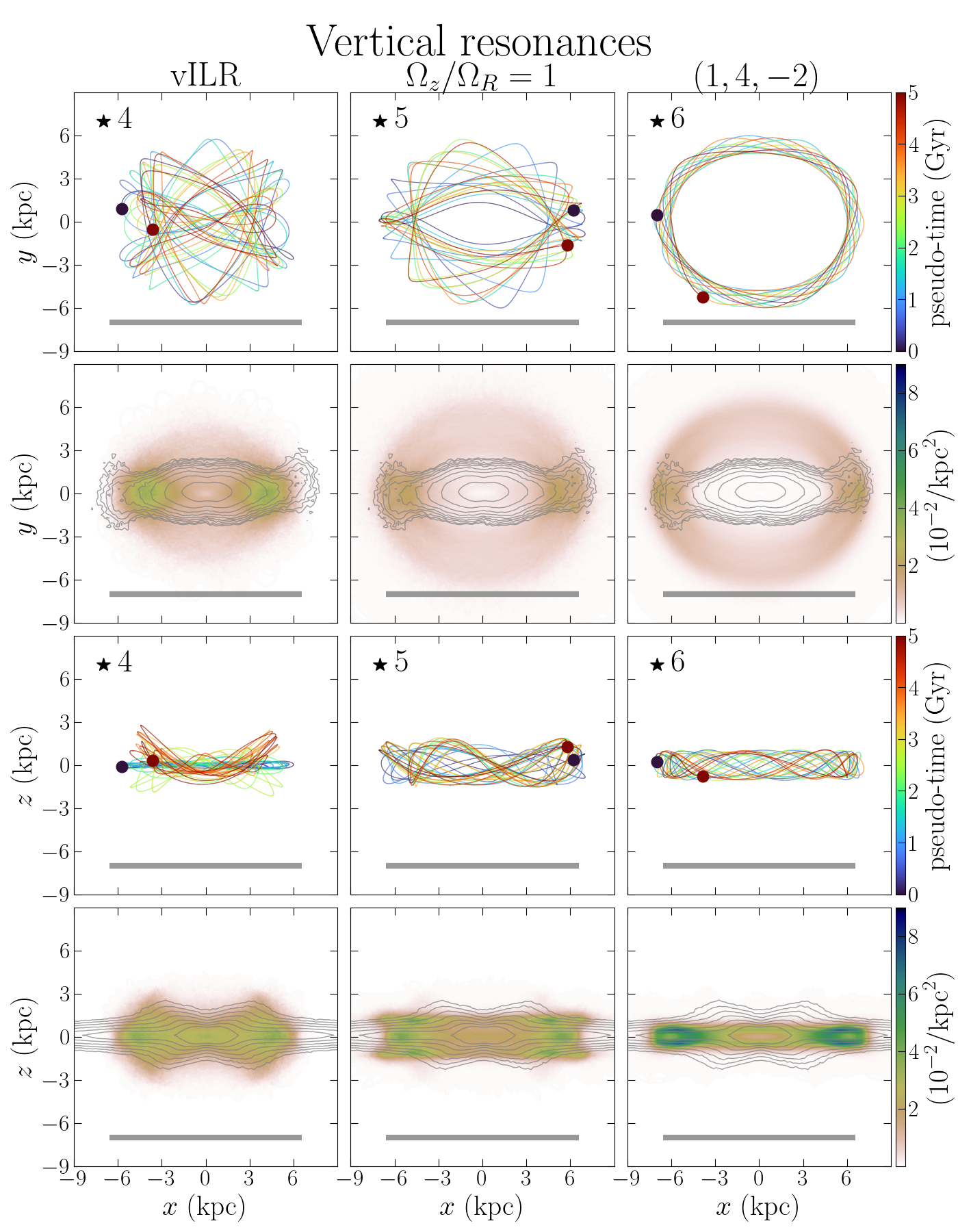}
    \caption{Similar to Fig.~\ref{fig:orbs_stack_ILR}, but for typical orbits selected at three vertical resonances (titles), and requiring $(\Omega_\varphi - \Omega_\mathrm{P})/\Omega_R < 1/2$. The numbered stars indicate these orbits in Fig.~\ref{fig:freq_map_768MR_no_hist}.}
    \label{fig:orbs_stack_vILR}
\end{figure}

We now select orbits at the three vertical resonances highlighted in Fig.~\ref{fig:freq_map_768MR_no_hist} and mentioned before: the vILR, the resonance $\OmzR=1$, and the resonance (1,4,-2). In this selection we exclude the intersections with the ILR and orbits above it by requiring $\OmphiR -1/2 < -0.01$. Typical orbits (orbits 4, 5 and 6 in Fig.~\ref{fig:freq_map_768MR_no_hist}) are shown in the first and third rows of Fig.~\ref{fig:orbs_stack_vILR}, while the panels below show the stacked density maps of all the orbits selected along these resonances. From right to left, we observe a similar trend of orbits becoming more elongated, less regular, and vertically thicker, with orbit 4 reaching all the way to the center. This orbit, at the vILR, has the typical ``banana'' shape in the $x-z$ plane \citep[][]{Pfenniger_1991}. Overall, orbits along these resonances imprint noticeable features in the $x-z$ stacked density maps, but not as distinctly concentrated in the $x-y$ plane as those shown in Fig.~\ref{fig:orbs_stack_ILR}.

\subsubsection{Orbital shapes at the vILR cloud and other regions}
\label{sec:orbs_vILR_cloud}
Finally, we select orbits in three additional regions mentioned before: the vILR cloud (with sample orbit 7); the stripe with intermediate eccentricities (with sample orbit 8), requiring $0.65 < \ecc < 0.85$ and not belonging to any of the previously selected regions; and the low-eccentricity region (with sample orbit 9), requiring $\ecc < 0.65$ and not belonging to any of the previously selected regions.

\begin{figure}
    \centering
	\includegraphics[width=\columnwidth]{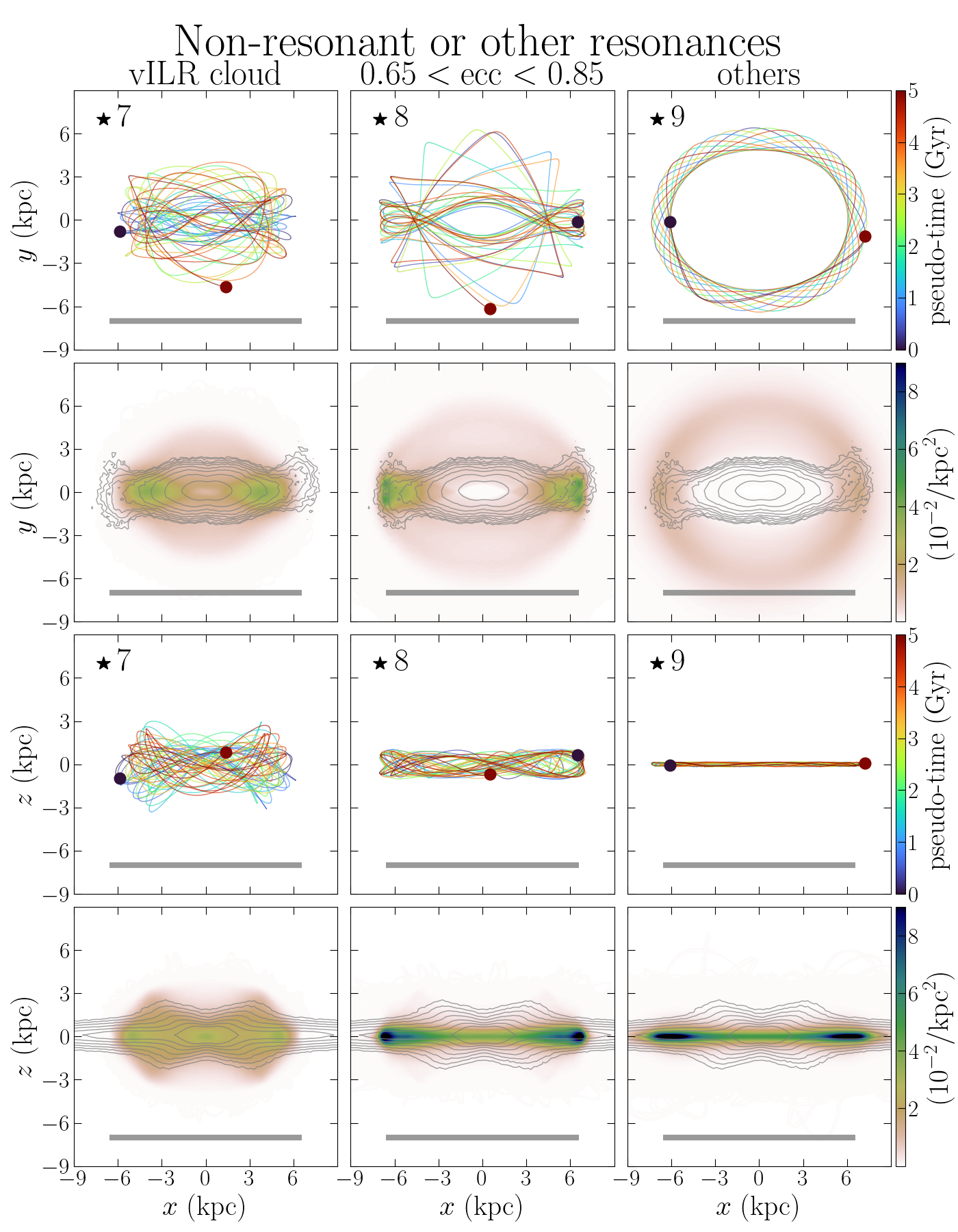}
    \caption{Similar to Figs.~\ref{fig:orbs_stack_ILR}-\ref{fig:orbs_stack_vILR}, but for orbits in the vILR cloud and other locations in the frequency map (see main text). The numbered stars indicate these orbits in Fig.~\ref{fig:freq_map_768MR_no_hist}.}
    \label{fig:orbs_stack_other}
\end{figure}

\begin{figure*}
\begin{center}
	\includegraphics[scale=0.4]{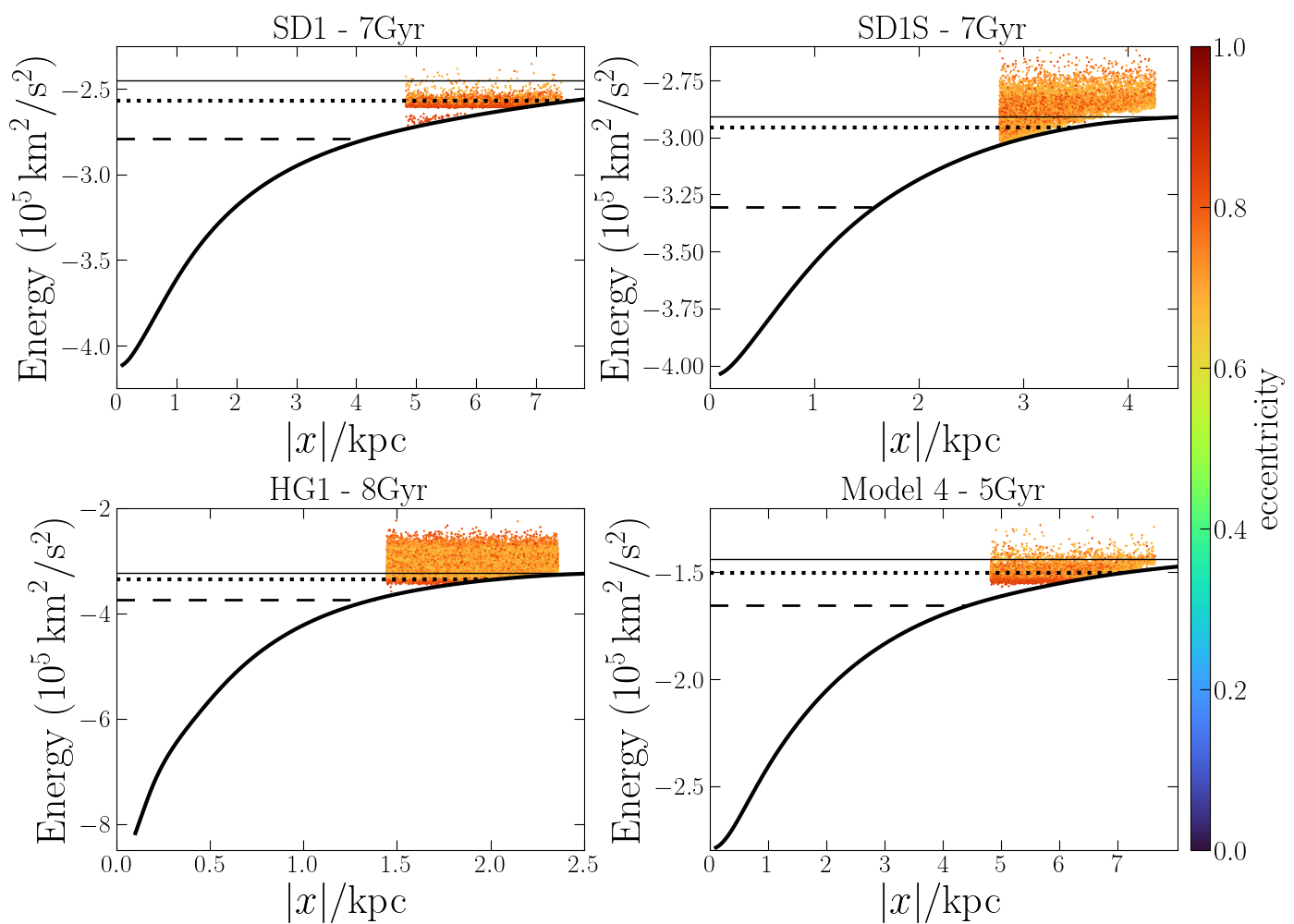}
    \caption{Similar to Fig.~\ref{fig:pot_eff}, but only showing $\EJ$ (points) for the particles selected in the group of orbit 8 in Fig.~\ref{fig:freq_map_768MR_no_hist}. These orbits typically have $0.4 \lesssim \OmphiR \lesssim 0.45$ and in some simulations a significant fraction of these orbits have $\EJ > \poteff(R_\mathrm{RC})$, i.e. can visit both the bar and the disc region, as those identified by \cite{Sparke_1987} and \cite{Pfenniger_1991}}.
    \label{fig:pot_eff_mid_ecc_cloud}
\end{center}
\end{figure*}

Fig.~\ref{fig:orbs_stack_other} shows the sample orbits and the orbit-averaged surface density maps for these regions. From right to left, i.e. in order of decreasing $\OmzR$, we note again the tendency of orbits to become more elongated and less regular, finally approaching the center and becoming significantly thicker. Orbit 9 circulates a little bit further out in the bar (in comparison to orbits 6 and 3). Its orbital group has $\OmphiR \lesssim 0.4$ and these resemble disc orbits in the vicinity of the bar. Orbit 8 has a peculiar shape, with a strong resemblance to the looped orbits at the ILR (e.g. orbit 2), but oscillating between this and a close-to resonant triangular shape. Its orbital group has $0.4 \lesssim \OmphiR \lesssim 0.45$ and it produces an excess in the outskirts of the bar similar to that produced by the looped orbits, but less prominent. These orbits are located near the convergence of several 3D resonances and near the ILR and, as will be  demonstrated in Sec.~\ref{sec:freq_map_all_sims}, are chaotic. These orbits have $0.4 \lesssim \OmphiR \lesssim 0.45$, similar to the location of a peak already observed by \cite{Martinez-Valpuesta_2006, Smirnov_2021}. Some of these orbits (but not all) have $\EJ > \poteff(R_\mathrm{RC})$ and reasonably large eccentricities (see Fig.~\ref{fig:pot_eff}), allowing them to visit both the bar and the disc region, as those identified in $N$-body simulations by \cite{Sparke_1987} and \cite{Pfenniger_1991}. Fig.~\ref{fig:pot_eff_mid_ecc_cloud} is similar to Fig.~\ref{fig:pot_eff}, but only shows points for the orbits in this group. We see that in models SD1S and HG1 a large fraction of these orbits have $\EJ > \poteff(R_\mathrm{RC})$.

An interesting aspect of Figs.~\ref{fig:orbs_stack_ILR}, \ref{fig:orbs_stack_vILR} and \ref{fig:orbs_stack_other} is the similarity of the orbit-averaged density maps in the left column of each figure, especially regarding the X-shape in the $x-z$ projection supporting the BP-bulge (contours). All these orbits are within the vILR cloud, now specifically defined as $\Omphiz -1/2 > - 0.01$, i.e. including hot orbits at the ILR. Fig.~\ref{fig:hist_rper} shows that approximately half of these orbits can get as close as $10-100$ pc to the galactic center.

Having shown the typical spatial distribution of orbits selected in the shoulder region, we now compare the frequency maps of the different simulations.

\subsection{Frequency maps: comparing simulations}
\label{sec:freq_map_all_sims}

\begin{figure}
	\includegraphics[width=\columnwidth]{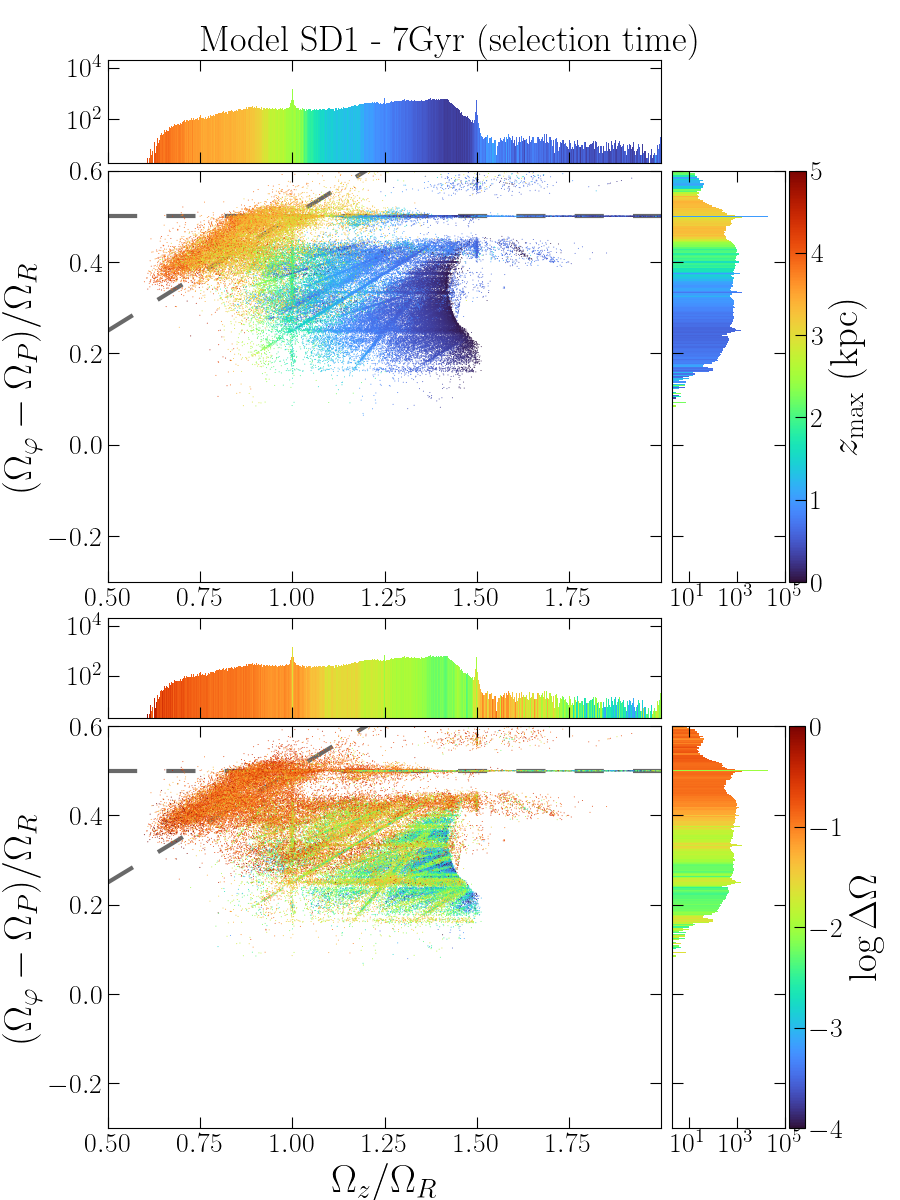}
    \caption{Frequency map for orbits in the shoulder region of simulation SD1 at $7\Gyr$. Smaller panels show histograms (in log-scale) with bin-widths of 0.0025. The upper panel is color-coded by the vertical excursion, and the bottom panel by the frequency drift (high values indicating chaotic orbits). The ILR, $\OmphiR=1/2$, is the most prominent resonance. Orbits in the vILR cloud, to the left and above the vILR $\Omphiz=1/2$, have significantly larger $\zmax$ and are strongly chaotic.}
    \label{fig:freq_map_768MR}
\end{figure}

Here we compare the frequency maps for orbits selected in the shoulder regions for the four simulations (see Sec.~\ref{sec:sims}), at the snapshots where particles are selected ({\it cf.} Fig.~\ref{fig:Omega_vs_R}). Fig.~\ref{fig:freq_map_768MR} shows the results for the model SD1, with the same data as in Fig.~\ref{fig:freq_map_768MR_no_hist}, but with the upper panel color-coded by $\zmax$ and the bottom panel by the frequency-drift $\log \Delta\Omega$, where
\begin{equation}
    \Delta\Omega = \mathrm{max}_{i=1,2,3}\left|\frac{\Omega_i(T_2) - \Omega_i(T_1)}{\Omega_i(T_1)}\right|,
\end{equation}
and $\Omega_i(T_2)$ and $\Omega_i(T_1)$ are the frequencies estimated in two separate halves of the time-series \citep[][]{Valluri_2010}. Since regular orbits conserve frequencies, a large $\log \Delta\Omega$ indicates a chaotic orbit. The side-panels show histograms of the frequency ratios color-coded by the mean $\zmax$ (or $\log\Delta\Omega$) in bins of width 0.0025.

\begin{figure}
	\includegraphics[width=\columnwidth]{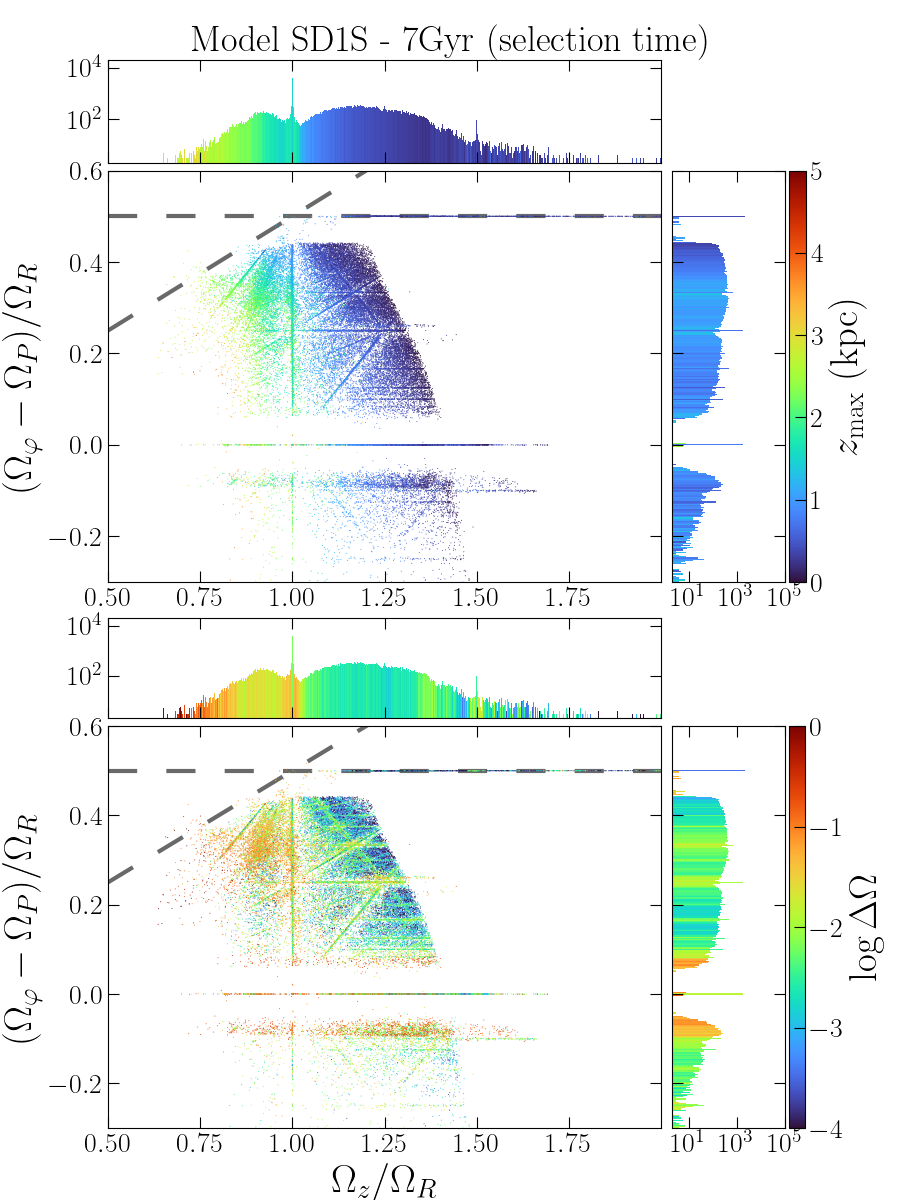}
    \caption{Similar to Fig.~\ref{fig:freq_map_768MR}, but for simulation SD1S (fast bar and no detected shoulders). The most prominent resonances are the ILR, the co-rotation and $\OmzR=1$. The ILR is not as strongly populated as in Fig.~\ref{fig:freq_map_768MR}, and there is no vILR cloud. Orbits with $\OmzR<1$ still have larger $\zmax$ than the rest, but not anywhere as large as in Fig.~\ref{fig:freq_map_768MR}.}
    \label{fig:freq_map_768MRS}
\end{figure}

\begin{figure}
	\includegraphics[width=\columnwidth]{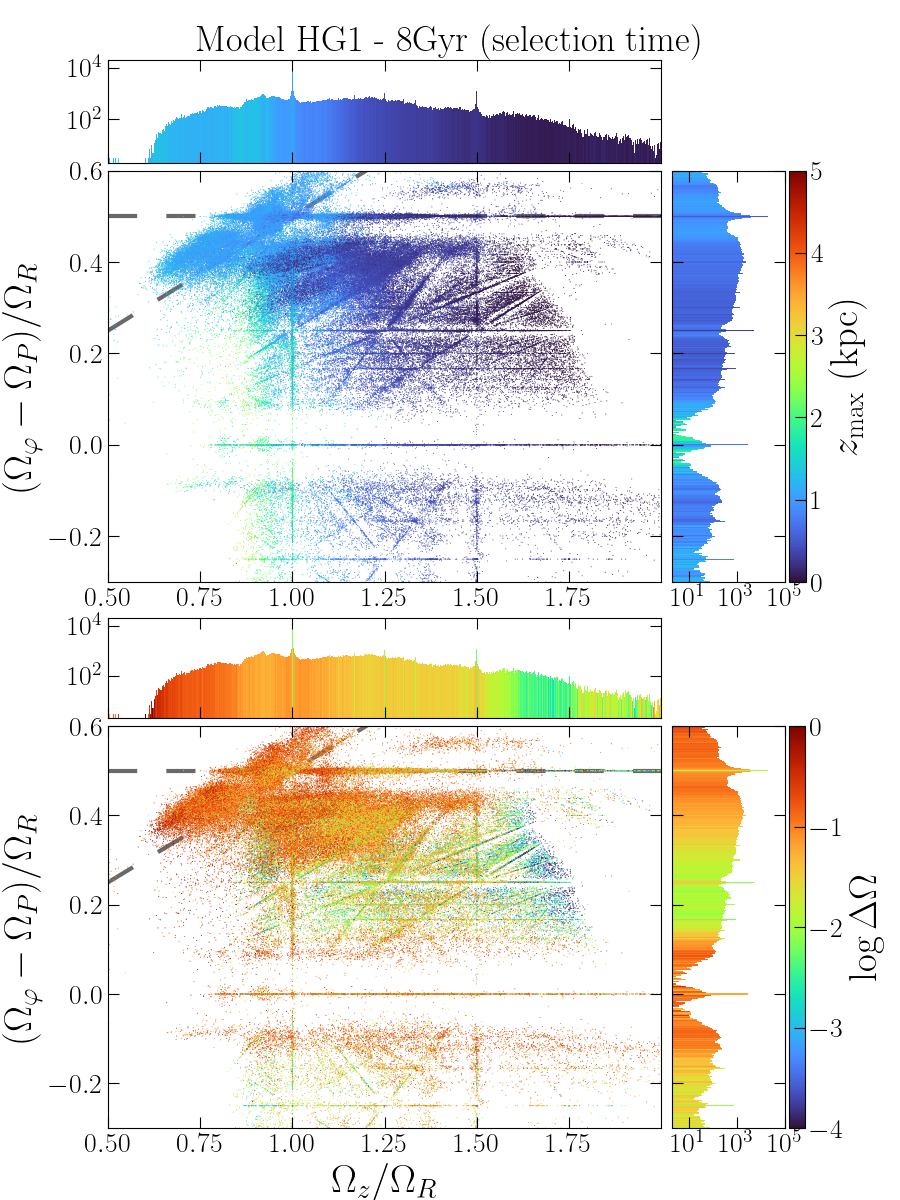}
    \caption{Similar to Fig.~\ref{fig:freq_map_768MR}, but for model HG1, which has gas and a fast bar. We see several well populated resonances, particularly the ILR, the co-rotation and the vertical resonances $\OmzR=1$ and $\OmzR = 3/2$. The vILR cloud is strongly populated with chaotic orbits and relatively high $\zmax$, but not as high as in the model SD1, Fig.~\ref{fig:freq_map_768MR}.}
    \label{fig:freq_map_708main}
\end{figure}

\begin{figure}
	\includegraphics[width=\columnwidth]{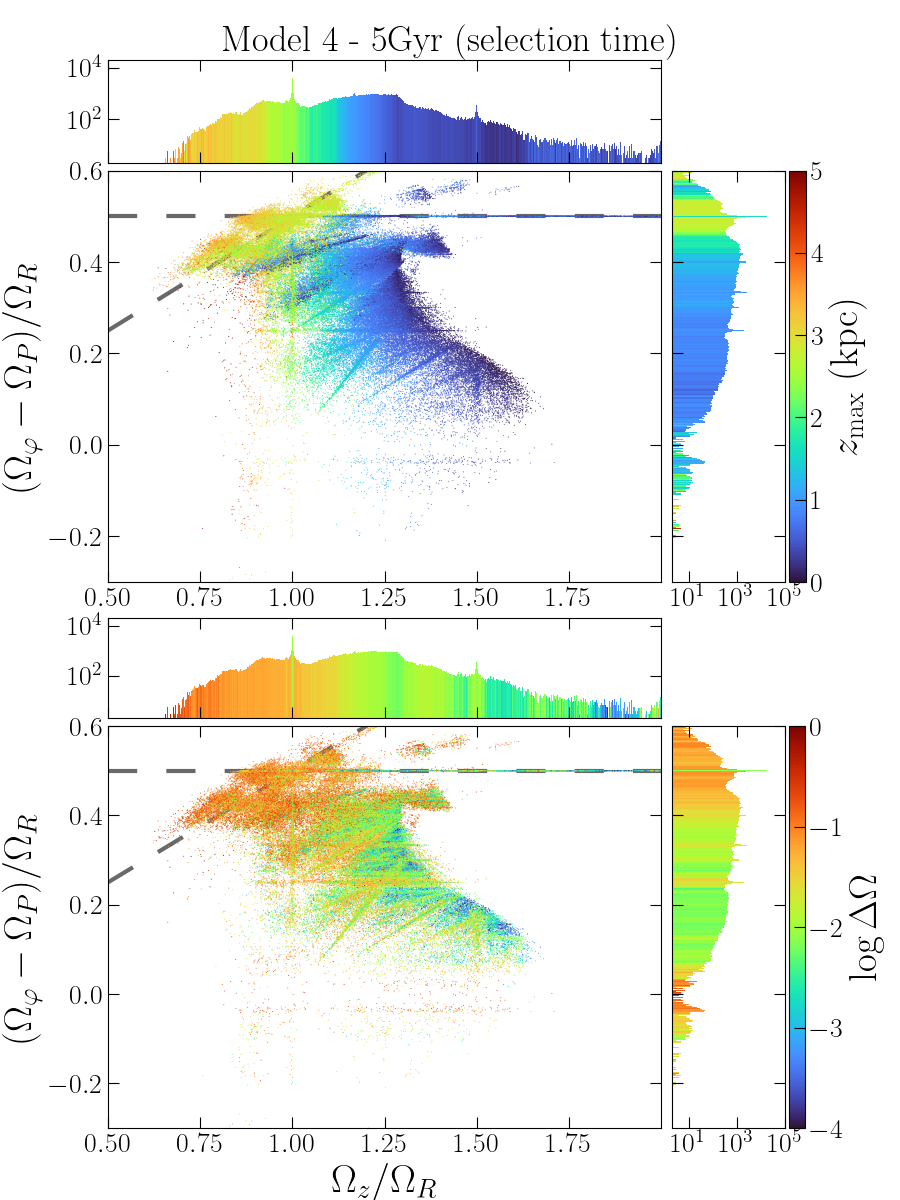}
    \caption{Similar to Fig.~\ref{fig:freq_map_768MR}, but for Model 4, whose bar buckles at $4\Gyr$ and is slow at $5\Gyr$. This figure is similar to that of model SD1, with $\zmax$ values in the vILR cloud intermediate between those in SD1 and HG1, Figs.~\ref{fig:freq_map_768MR} and \ref{fig:freq_map_708main}.}
    \label{fig:freq_map_741D8}
\end{figure}

Figs.~\ref{fig:freq_map_768MRS}-\ref{fig:freq_map_741D8} show the equivalent frequency maps for the models SD1S, HG1 and Model 4. The most noticeable feature in the four models is the well-populated ILR, producing a very prominent peak in the histogram to the right (note the logarithmic scale). This strong prominence of the ILR could be a selection effect since we chose particles which are predominantly part of the bar. However, in a random selection of $10^6$ star-particles in HG1 at $10\Gyr$, we identify more than half of the particles at the ILR; such a strong ILR has also been observed in previous works \citep[e.g.][]{Athanassoula_2003}. In our selection, within $\pm 0.0025$ ($\pm 0.01$) from the ILR we identify $19.8\%$ ($22.2\%$) of the orbits for model SD1, $4.5\%$ ($4.5\%$) for SD1S, $13.5\%$ ($18.3\%$) for HG1, and $15.1\%$ ($18.1\%$) for Model 4, meaning the ILR is well populated in all the models, except for model SD1S.

  In the models SD1S and HG1, we identify the co-rotation resonance at $\OmphiR=0$. This is in agreement with SD1S and HG1 having fast bars, and the selection region, always in the outermost parts of the bar, being close to co-rotation in these models-- see Fig.~\ref{fig:Omega_vs_R}. We emphasize that identifying the co-rotation is only possible when using coordinates in the inertial frame -- see Fig.~\ref{fig:freq_map_768MRS_rot_frame} for a comparison with the bar frame. We also note, in all simulations, big gaps around the co-rotation and ILR, testifying to their influence: orbits that get too close may be trapped or scattered by them.

Another remarkable feature in all models, except SD1S, is the vILR cloud, with significantly large $\zmax$ and $\log\Delta\Omega$. In HG1, which is the only model with gas, the $\zmax$ at the vILR cloud is smaller than in the other models, although $\log\Delta\Omega$ is still comparably large. In a certain sense, the vILR clearly separates the orbits that support the thin and thick parts of the bar. The looped orbits at the (vertically warm) ILR mostly support the thin part, and the orbits at the vILR cloud mostly support the thick part -- see also Fig.~\ref{fig:zmax_vs_OmzR}.

Finally, we see a plethora of additional resonances, particularly in HG1 (Fig.~\ref{fig:freq_map_708main}). Orbits at the main resonances (and away from any crossing resonances) tend to have smaller $\log\Delta\Omega$, i.e. to be regular, as expected since these resonances are stable. On the other hand, orbits in areas with multiple strong resonances, such as the crossing of the ILR and the vILR, or on the floor below the ILR (the group of orbit 8 in Fig.~\ref{fig:freq_map_768MR_no_hist}), tend to be chaotic, as is generally the case at resonance crossings \citep[][]{Chirikov_1979}. These results illustrate the richness of the dynamics in the region between the bar and the disc \citep[see e.g.][]{Contopoulos_1981}, and also the power and precision of the frequency map technique we employ.

\subsection{More on morphologies along the ILR}
\label{sec:zmax_vs_OmzR}

\begin{figure*}
	\includegraphics[width=\textwidth]{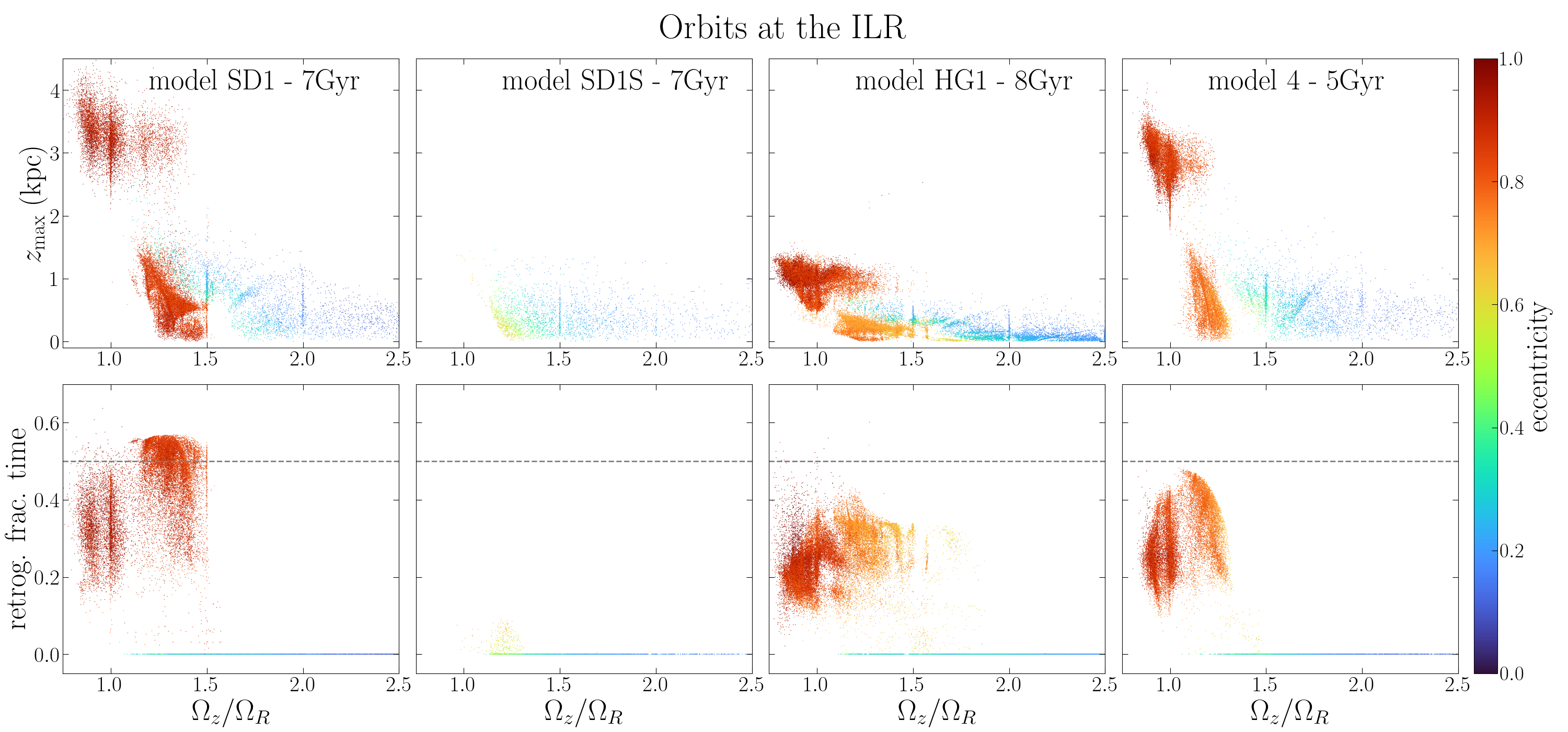}
    \caption{Top: vertical excursion versus $\OmzR$ for particles selected in the shoulder region and at the ILR for all models. Bottom: the fraction of time spent in retrograde motion (in the rotating frame). The horizontal dashed line is at 0.5. The different morphologies identified in Sec.~\ref{sec:freq_map_overview} are clearly seen here (except for SD1S): cool orbits at the ILR have low $\zmax$, low eccentricity and are almost exclusively prograde; warm orbits at the ILR have slightly larger $\zmax$, relatively high eccentricity and expend a significant time in retrograde motion (a symptom of the strong loops at their ends); hot orbits at the ILR have significantly larger $\zmax$ and eccentricity, and moderate fractional time in retrograde motion. The looped morphology is not detected in the shoulder-less model SD1S. We also note the overall lower $\zmax$ for the SPH model HG1, in comparison to models SD1 and Model 4.}
    \label{fig:zmax_vs_OmzR}
\end{figure*}

We explore in greater depth the different morphologies for orbits along the ILR. Fig.~\ref{fig:zmax_vs_OmzR} shows, for the different models, the vertical excursion $\zmax$ (top panels) versus $\OmzR$ for orbits at the ILR ($|\OmphiR -1/2| < 0.01$) at the selection times, color-coded by the eccentricity. All models (except SD1S) show three distinct groups: the cool orbits ($\OmzR \gtrsim 3/2$) have low $\zmax$ and low-eccentricity (blue cloud) -- the elliptical orbits in Sec.~\ref{sec:freq_map_overview}. The warm orbits ($1 \lesssim \OmzR \lesssim 3/2$) have slightly larger $\zmax$ and relatively high eccentricity (orange cloud) -- the looped orbits. And the hot orbits ($\OmzR \lesssim 1$) have significantly larger $\zmax$ and eccentricity (red cloud) -- these are chaotic orbits within the vILR cloud. We also see some hot orbits at the ILR (orbits with high $\zmax$) contaminating the $\OmzR$-interval of the warm orbits, which is not surprising given the crowd of orbits around the crossing of the ILR and the vILR -- see e.g. Fig.~\ref{fig:freq_map_768MR_no_hist}.

As suggested by Fig.~\ref{fig:orbs_stack_ILR} and demonstrated below, the loops at the ends of warm orbits at the ILR play an important role in producing the shoulders along the bar major-axis density profiles. As a consequence of these loops, these orbits spend a considerable time in retrograde motion in the rotating frame of the bar. The bottom panels of Fig.~\ref{fig:zmax_vs_OmzR} show the fraction of time the orbits at the ILR are in retrograde motion, i.e. with $L_z<\OmP R^2$. We see that this fraction is zero for all cool orbits, except for some contamination of the warm orbits in model HG1. In the range $1 < \OmzR < 3/2$ of the warm orbits, the fraction of time in retrograde motion is typically high ($\approx 0.2-0.55$), a signature of the loops developed by these orbits. Finally, the vertically hot orbits also have large fractions of time in retrograde motion, but slightly smaller than the warm orbits.

Fig.~\ref{fig:zmax_vs_OmzR} demonstrates the common $\OmzR$-dependence of the different morphologies seen in Sec.~\ref{sec:freq_map_overview} for the models with shoulders: vertically warm orbits at the ILR have loops and spend a considerable time in retrograde motion; and hot orbits at the ILR jump to large $\zmax$ and also spend a considerable time in retrograde motion, but less than warm orbits on average. The consequence of this for the shoulders is discussed in Sec.~\ref{sec:evol_Sigma_stack}. Moreover, in agreement with the importance of the looped morphology to the shoulders, as suggested in \citetalias{Anderson2022} and confirmed in this work, we note that this morphology is absent in the shoulder-less model SD1S.

\subsection{The orbital support of shoulders}
\label{sec:orbital_support_shoulder}

Here we quantify the contribution of the main orbital groups to the density profiles in the shoulder region. Since the ILR is the main bar-supporting resonance, and inspired by the morphological study in Secs.~\ref{sec:freq_map_overview} and \ref{sec:zmax_vs_OmzR}, we investigate three groups: the vertically cool and warm orbits at the ILR, and the vILR cloud (which includes the hot orbits at the ILR).

\begin{figure}
	\includegraphics[width=\columnwidth]{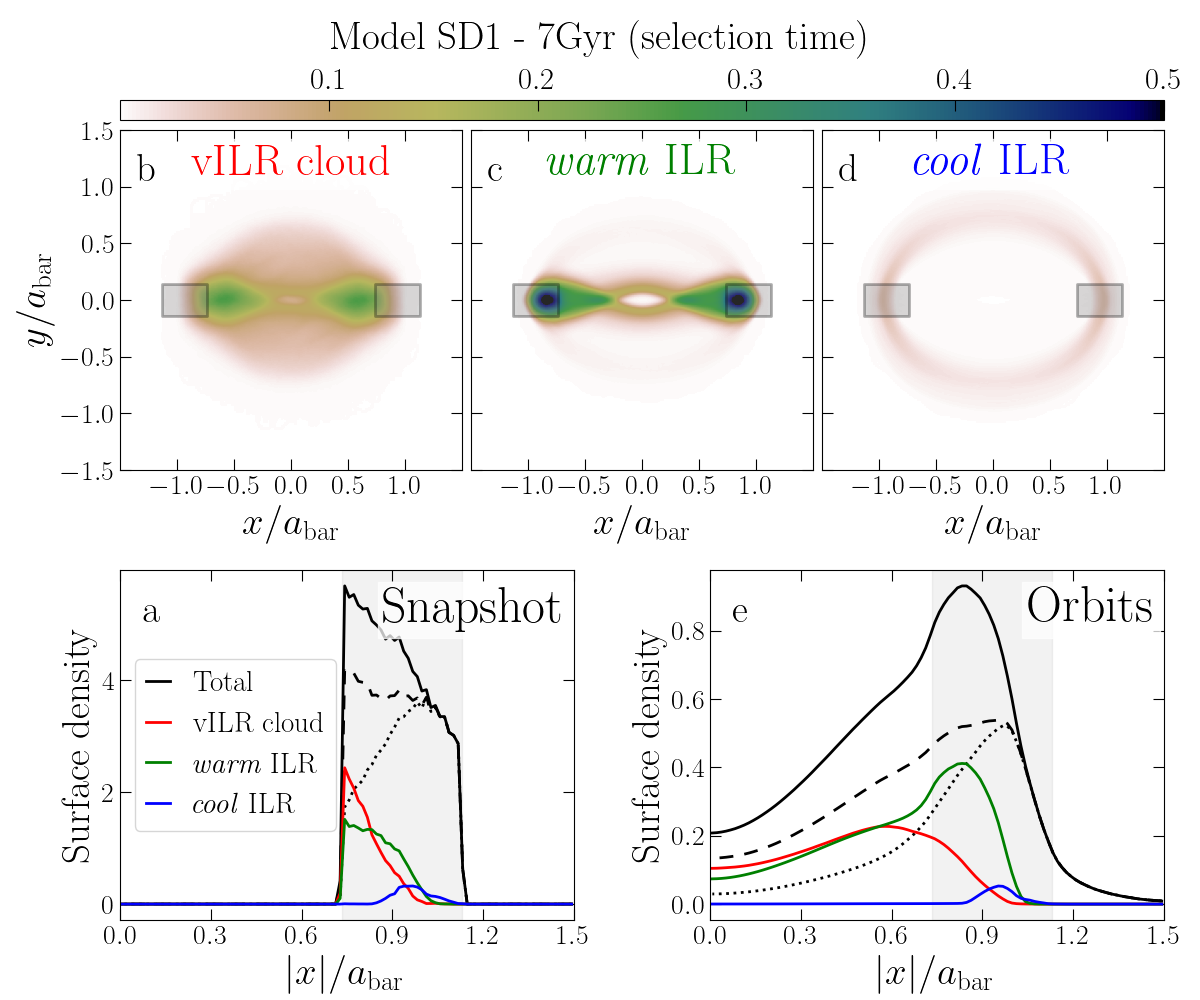}
    \caption{Surface density of particles in the shoulder region (gray shaded areas) of model SD1 at $7 \Gyr$, with coordinates normalized by the bar length at $7 \Gyr$. Panel {\it a}: instantaneous density profile along the major axis ($|y|/\abar<0.145$) in the simulation. The solid black curve shows the total profile, and colored curves show contributions from the groups in panels {\it b}-{\it d} (see main text). The dashed curve is the total profile after subtracting the warm orbits at the ILR ($1<\OmzR<3/2$), and the dotted one results from also subtracting orbits at the vILR cloud. Panels {\it b}, {\it c}, {\it d} show the 2-D orbit-averaged maps for the respective orbital groups, with colors normalized by the total number of particles in the shoulders. Panel {\it e} shows the 1D density profile, similar to panel {\it a}, but for the orbits, and restricting to $|y|/\abar<0.145$. In this panel, the large contribution of warm orbits to the profile in the shoulder region becomes more evident, with a smaller but considerable contribution from the vILR cloud. The contribution from the cool orbits (blue) is negligible.}
    \label{fig:Sigma_1D_stack_ILR_768MR}
\end{figure}

Fig.~\ref{fig:Sigma_1D_stack_ILR_768MR} (panel {\it a}) shows the $N$-body snapshot surface density profile for SD1 at 7 Gyr, for particles in the shoulder region. The density profile is calculated as the number of particles in $(x/\abar)-$bins of width 0.015 and within $|y|/\abar <0.145$, divided by $(\delta x/\abar) \,(\Delta y/\abar)\, N$, where $\Delta y = 2\cdot 0.145$ and $N$ is the total number of selected particles. Since these profiles are almost perfectly symmetric around $x=0$, for better visualization, we present them averaged between the two sides $x<0$ and $x>0$. The gray areas show the average shoulder region, where particles are selected. The black solid curve shows the total density profile, and the colored solid curves show the relative contribution of particles in the different orbital groups, in the same colors as in the titles of panels {\it b}-{\it d}.

In this model the contributions from particles at the vILR cloud (red) and the warm ones at the ILR (green) are similarly large. The black dashed curve shows the total density profile excluding the contribution of the latter. Removing the warm orbits significantly reduces the density profile in the shoulder region, demonstrating their important role in the shoulders. The dotted curve shows the profile after also subtracting particles at the vILR cloud, and we see that they mostly contribute to the innermost parts of the shoulders. Together, these two groups contribute so much to the density profile that their subtraction produces significant deficits in it.

The picture provided by panel {\it a} in Fig.~\ref{fig:Sigma_1D_stack_ILR_768MR} is limited because it shows particles currently in the shoulder region. To evaluate the global profile produced by these particles, one can calculate the profile produced by them at a later time, allowing the particles to phase-mix, as done in \citetalias{Anderson2022}. However, in the mean time the bar pattern speed and the orbits themselves may have evolved into different ones. For a global {\it and instantaneous} view of the density profiles, we stack the orbits integrated for these particles in the frozen potential and plot the surface density in the $x-y$ plane in Fig.~\ref{fig:Sigma_1D_stack_ILR_768MR}, panels {\it b}, {\it c}, {\it d}. To emphasize each group's relative contribution rather than the orbital shapes, we now normalize colors by the total number of particles in the shoulder region.

The orbit-averaged density of the orbits at the vILR cloud (panel {\it b}) resembles that of looped orbits, although it is more diffuse in the $x-y$ plane, which arises because of the chaotic nature of orbits at the vILR cloud -- see bottom panels of Figs.~\ref{fig:freq_map_768MR}, \ref{fig:freq_map_708main}, \ref{fig:freq_map_741D8}. Furthermore, its orbit-averaged density is less extended along the major axis, peaking outside of the shoulder region. Panel {\it c} shows the orbit-averaged density of the warm orbits at the ILR. We see that this profile is more elongated and has a very pronounced peak in the shoulder region. On the other hand, the cool orbits (panel {\it d}) do not represent any visible excess in the shoulder region.

Panel {\it e} shows the density profiles of these stacked orbits (for points within $|y|/\abar<0.145$) along the $x$-axis, analogous to the left panel, and using the same color/style convention. For a comparison with panel {\it a}, these profiles are further normalized by the number of points per orbit. The smaller values, in comparison to panel {\it a}, are due to the ``dilution'' of each particle along its orbit out of the selection region.

In these orbit-averaged profiles, the warm orbits at the ILR contribute much more significantly to the shoulder because: {\it i)} they are more elongated than those at the vILR cloud; {\it ii)} particles stay longer near apocenters; and {\it iii)} the sharply defined loops at their ends make them spend even more time near apocenters. This confirms the important contribution of the looped morphology of the warm orbits at the ILR to the shoulders, but also demonstrates the importance of the vILR cloud (compare the green and red curves in panels {\it a} and {\it e}). Finally, the contribution of the cool orbits at the ILR is negligible also in the orbit-averaged profiles.

\begin{figure}
	\includegraphics[width=\columnwidth]{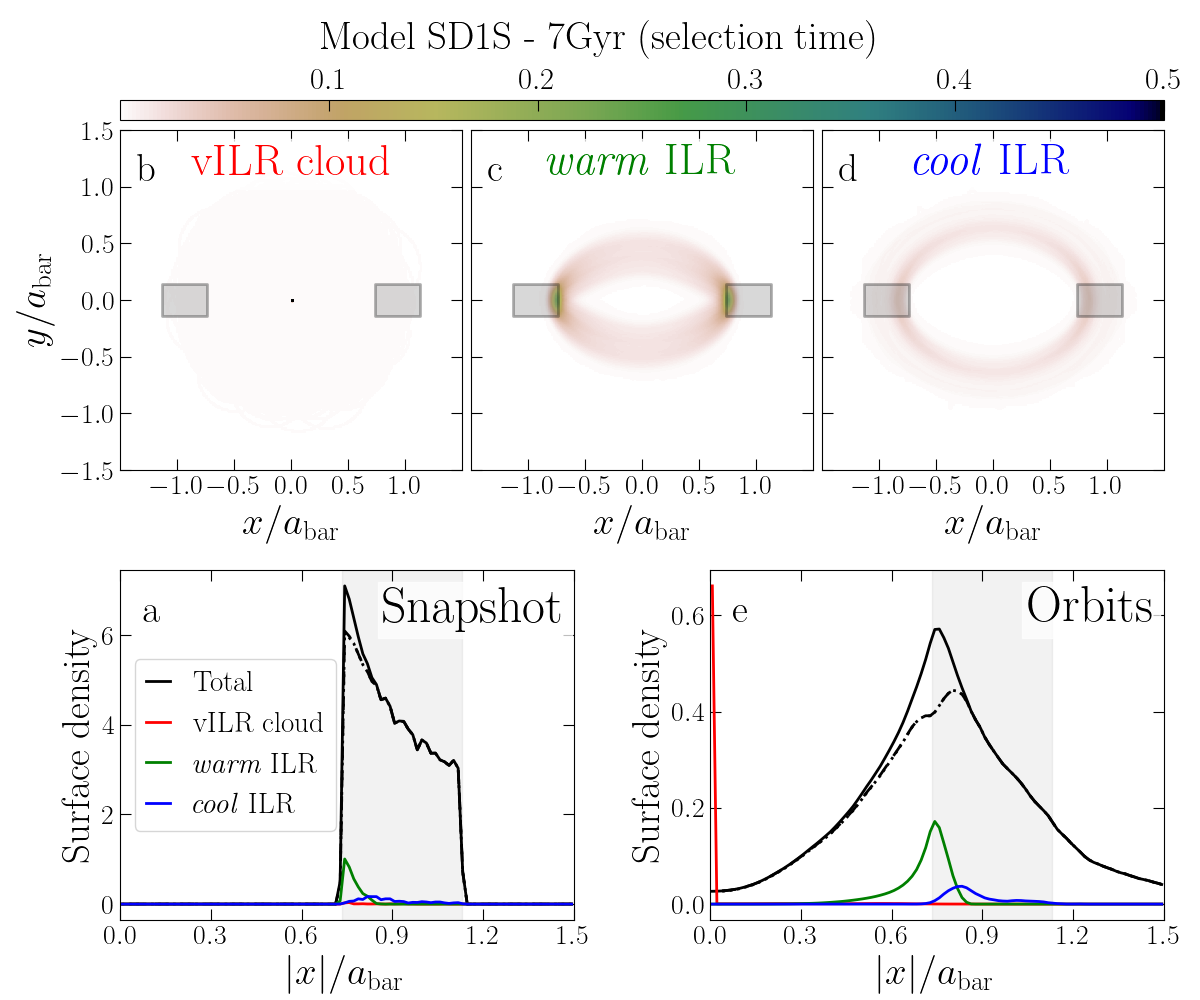}
    \caption{As in Fig.~\ref{fig:Sigma_1D_stack_ILR_768MR}, but for model SD1S, with no detected shoulders. In this case, warm orbits at the ILR do not have loops and their contribution to the density profile (solid green) is much smaller than in the model SD1, Fig.~\ref{fig:Sigma_1D_stack_ILR_768MR}.}
    \label{fig:Sigma_1D_stack_ILR_768MRS}
\end{figure}

\begin{figure}
	\includegraphics[width=\columnwidth]{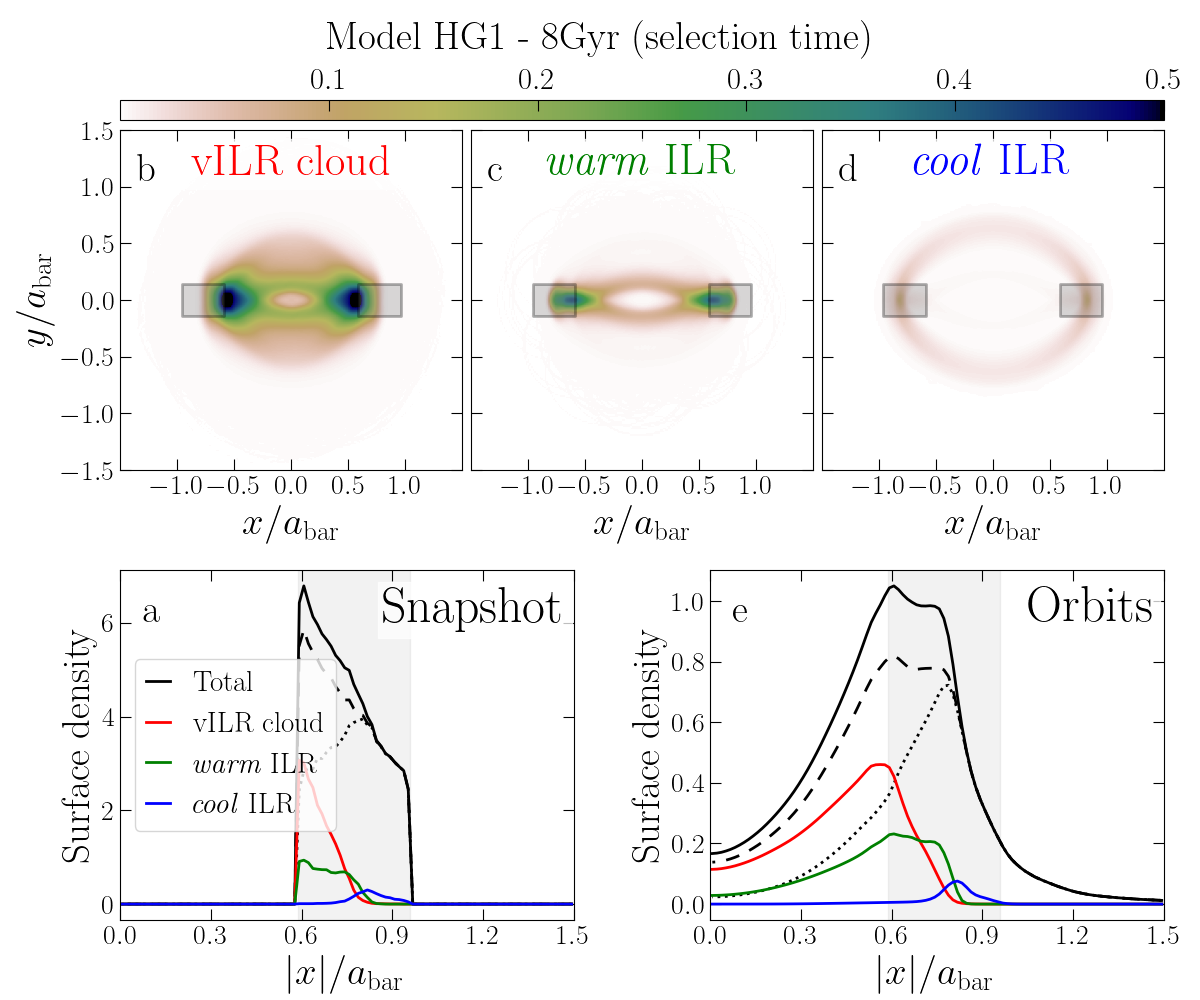}
    \caption{As in Fig.~\ref{fig:Sigma_1D_stack_ILR_768MR}, but for model HG1. The contribution from warm orbits at the ILR (solid green) to the density profile is much smaller than in SD1. However, in this case the number orbits at the vILR cloud is much larger, and they produce an orbit-averaged density in the $x-y$ which resembles the one produced by looped orbits.}
    \label{fig:Sigma_1D_stack_ILR_708main}
\end{figure}

\begin{figure}
	\includegraphics[width=\columnwidth]{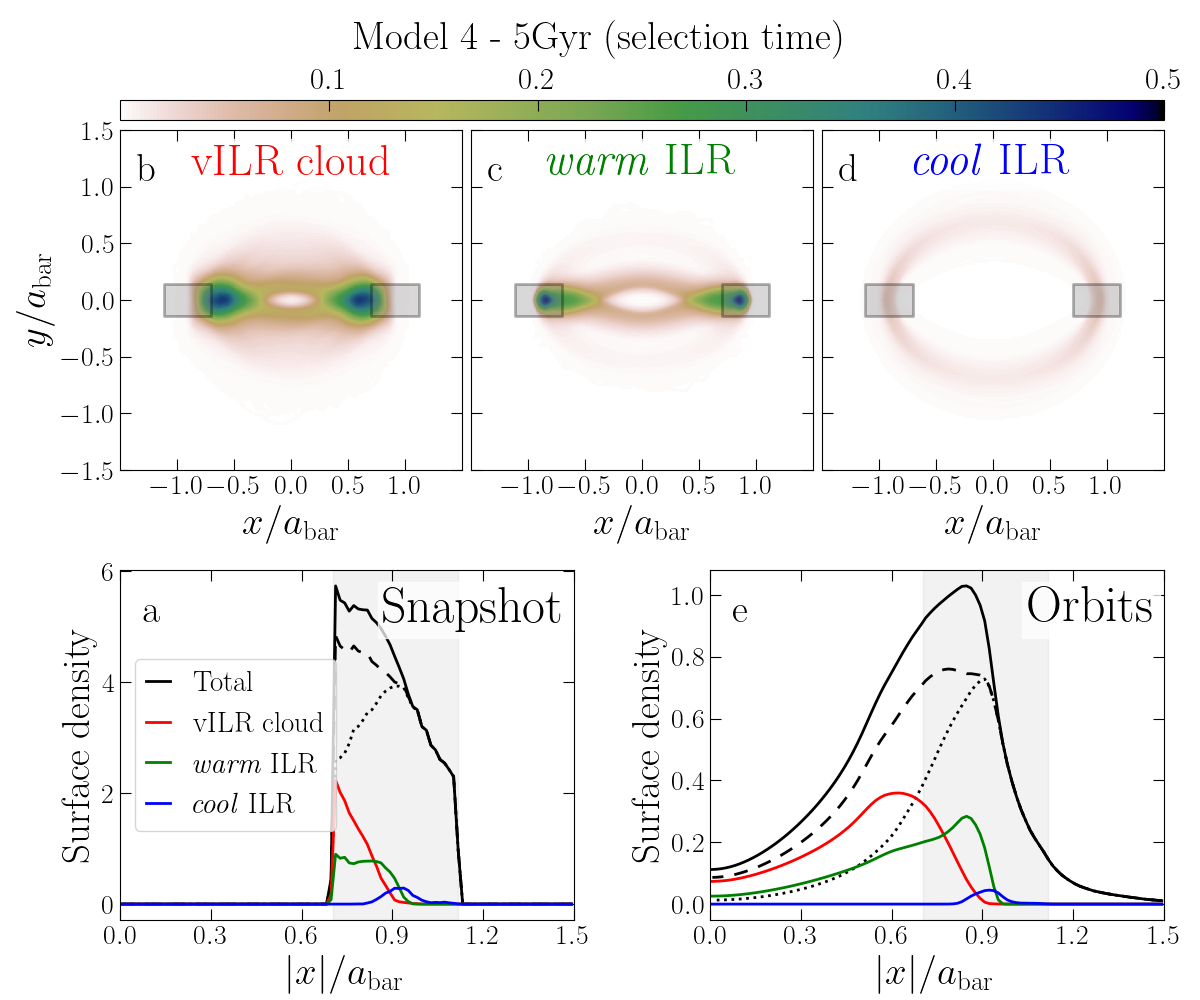}
    \caption{As in Fig.~\ref{fig:Sigma_1D_stack_ILR_768MR}, but for Model 4. The contributions from warm orbits at the ILR (green) and the vILR cloud (red) are comparable, and a looped-like morphology is produced by both groups, as in the model HG1, Fig.~\ref{fig:Sigma_1D_stack_ILR_708main}.}
    \label{fig:Sigma_1D_stack_ILR_741D8}
\end{figure}

Figs.~\ref{fig:Sigma_1D_stack_ILR_768MRS}-\ref{fig:Sigma_1D_stack_ILR_741D8} show the corresponding profiles for models SD1S, HG1 and Model 4, with the same color-normalization as in Fig.~\ref{fig:Sigma_1D_stack_ILR_768MR}. Recall that model SD1S has no shoulders, and that its ``shoulder region'' is set as the same fraction of the bar length as in SD1 at the equivalent snapshot. The warm orbits at the ILR (green) in this model do not develop loops, as already shown in Fig.~\ref{fig:zmax_vs_OmzR}, and these orbits' contribution to the density profile in the shoulder region is negligible, as is the contribution from the vILR cloud. In fact, just a handful of orbits populate the vILR cloud in this model (Fig.~\ref{fig:freq_map_768MRS}).

Fig.~\ref{fig:Sigma_1D_stack_ILR_708main} shows analogous profiles for model HG1, which has weak shoulders -- see \citetalias{Anderson2022}, Fig. 11. In this case, the contribution of the warm orbits is relatively small, and its subtraction produces a profile (dashed) which is close to the total one. On the other hand, in HG1 the vILR cloud orbits are very numerous and 
produce an overall shape in the $x-y$ plane that resembles the looped morphology, contributing significantly to the density in the shoulder region. Fig.~\ref{fig:Sigma_1D_stack_ILR_741D8} shows the equivalent profiles for Model 4. This model is intermediate between SD1 and HG1 in terms of bar length and pattern speed evolution (see Fig.~\ref{fig:Om_p_all_sims}), and so are the relative contributions of warm orbits at the ILR and of the vILR cloud. We also observe significant contributions from the warm orbits and the vILR cloud, with the latter producing a more diluted looped-like shape in the $x-y$ plane, in comparison with the warm orbits.

In summary, we have confirmed the important role of looped orbits to the shoulders, as previously suggested in \citetalias{Anderson2022}, and have shown that this morphology is found in vertically warm orbits at the ILR. We also demonstrate that, although individual orbits in the vILR cloud are chaotic and do not have a well defined looped shape, they still produce an overall shape in the $x-y$ plane that can be seen as a diluted version of the shape produced by the looped orbits. We note that the morphology of vILR cloud orbits in SD1 is the most diffuse in the $x-y$ plane, and also has the largest $\zmax$, among models with shoulders. Conversely, the vILR cloud of HG1 produces the least diffuse shape in the $x-y$ plane, and the smallest $\zmax$, while Model 4 is intermediate between the two. This connection between $\zmax$ and the looped shape is explored via the evolution of the frequency maps presented in the next subsections.

\subsection{Evolution of the frequency maps}
\label{sec:evol_freq_map}

\begin{figure*}
	\includegraphics[width=\textwidth]{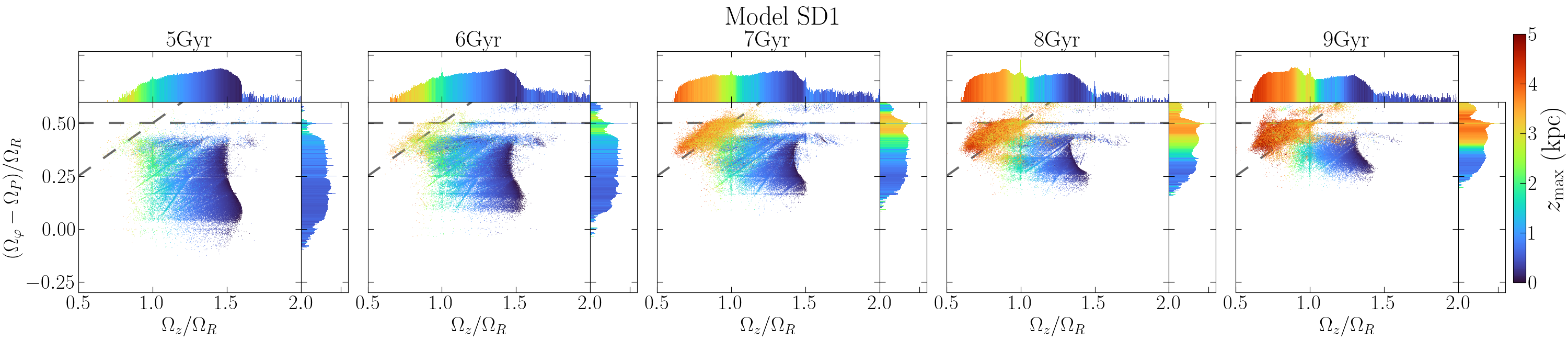}
    \caption{Time evolution of the frequency map for the same set of star-particles selected at $7 \Gyr$ in the shoulder region of model SD1. Points are color-coded by the vertical excursion $\zmax$. We see an overall shift upwards in this plot mainly due to the bar slowing down, i.e. decreasing $\OmP$, and to the left, with orbits becoming vertically thicker over time. Orbits crossing the vILR from right to left have vertical motion dramatically amplified.}
    \label{fig:freq_map_evol_768MR}
\end{figure*}

\begin{figure*}
	\includegraphics[width=\textwidth]{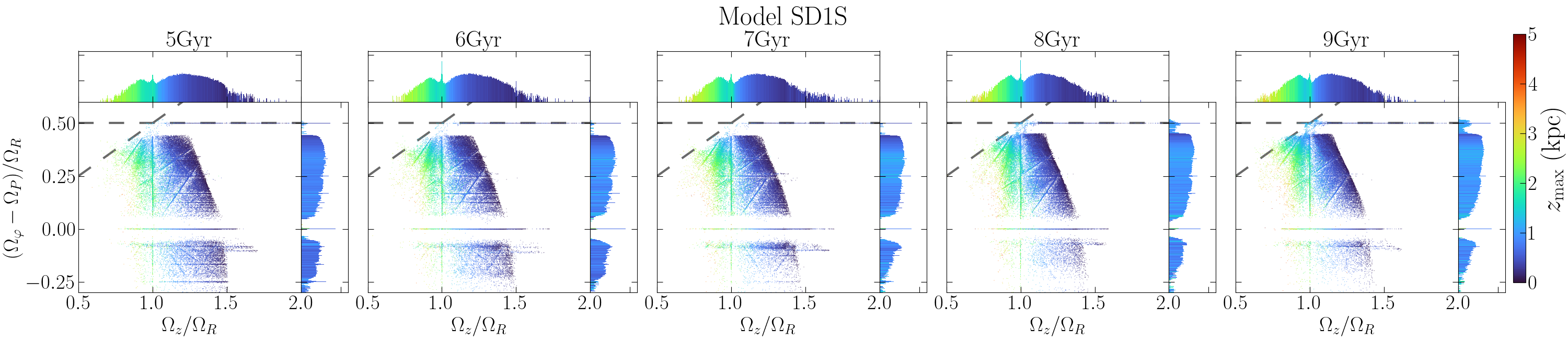}
    \caption{Similar to Fig.~\ref{fig:freq_map_evol_768MR}, but for model SD1S. In this case, we do not observe a significant time evolution of the frequency map, in agreement with the very mild evolution of the bar parameters -- see Fig.~\ref{fig:Om_p_all_sims}.}
    \label{fig:freq_map_evol_768MRS}
\end{figure*}

\begin{figure*}
	\includegraphics[width=\textwidth]{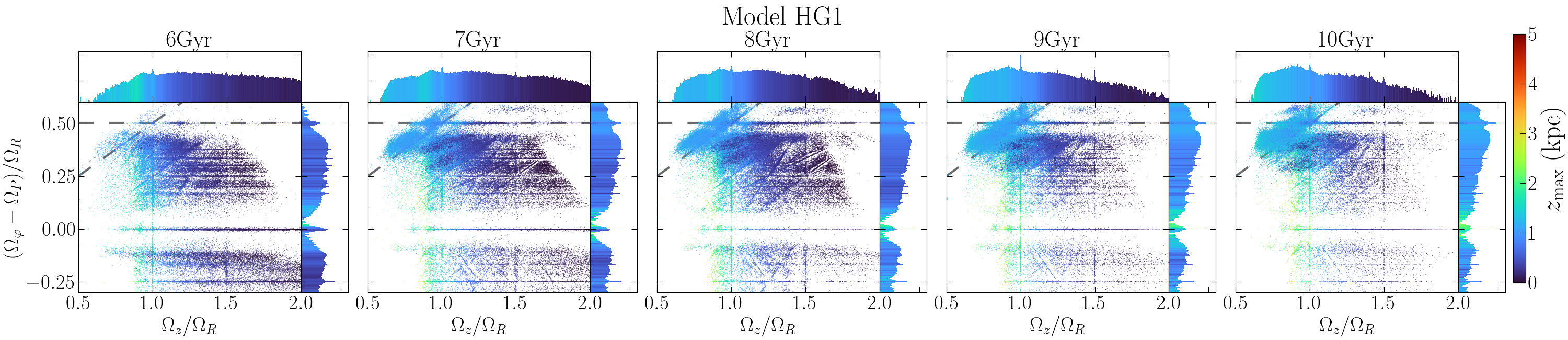}
    \caption{Similar to Fig.~\ref{fig:freq_map_evol_768MR}, but for model HG1. We observe the same trends as in model SD1, i.e. a net shift upwards and to the left. Orbits crossing the vertical resonances $\OmzR=1$ and the vILR have their vertical motion significantly amplified.}
    \label{fig:freq_map_evol_708main}
\end{figure*}

\begin{figure*}
	\includegraphics[width=\textwidth]{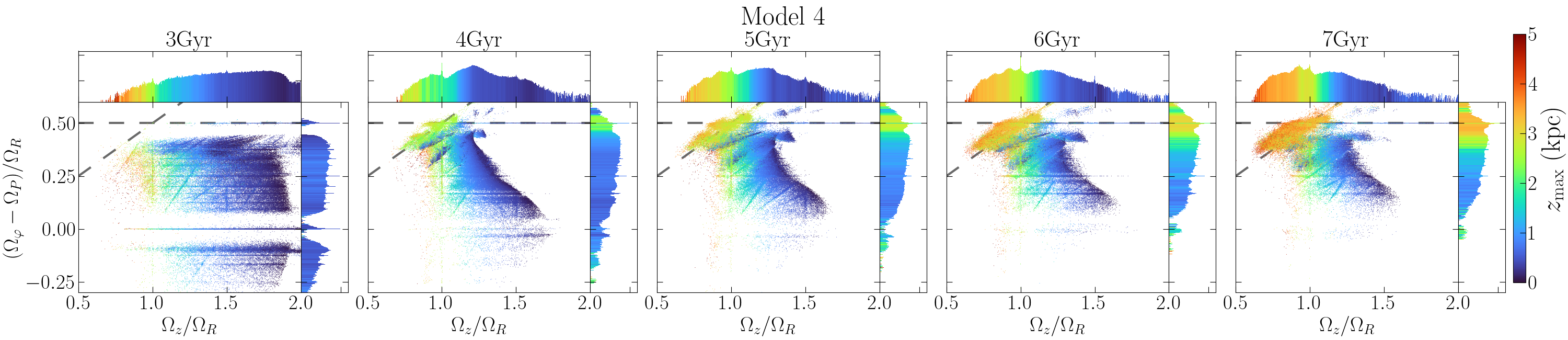}
    \caption{Similar to Fig.~\ref{fig:freq_map_evol_768MR}, but for Model 4. Besides the same trends observed for SD1, we note the abrupt shift from $3 \Gyr$ to $4 \Gyr$ towards smaller $\OmzR$. This is produced by the buckling bar, slightly before $4 \Gyr$, and after which shoulders are detected in the model.}
    \label{fig:freq_map_evol_741D8}
\end{figure*}

Having discussed the orbital support for shoulders, and having shown the frequency maps at the time of selection in Sec.~\ref{sec:freq_map_all_sims}, we now show their time evolution, shedding light on an important mechanism  for the shoulders' evolution. We select the same particles from the model snapshots discussed above and integrate their orbits at four additional snapshots, with the corresponding frozen potentials and bar pattern speeds, and using the particles' coordinates at each snapshot as initial conditions (for HG1, which has star formation, we only use star-particles born before the selection time). The results are shown in Figs.~\ref{fig:freq_map_evol_768MR}-\ref{fig:freq_map_evol_741D8}, with the selection times shown in the central panels.

Fig.~\ref{fig:freq_map_evol_768MR} shows the evolution of the frequency maps for the fiducial model SD1. The main trends, which also occur in the other models with shoulders, are a global shift upwards, i.e. to higher $\OmphiR$, and to the left, i.e. to lower $\OmzR$. One obvious reason for the shift upwards is the decline in $\OmP$ as the bar slows down (see Fig.~\ref{fig:Om_p_all_sims}), with its main resonances sweeping the disc outwards and trapping new particles. Once trapped by the strong ILR, the orbits tend not to shift any longer upwards in the frequency maps.

The ratio $\OmzR$ is a proxy for the vertical thinness of the orbit (Sec.~\ref{sec:freq_map_overview}). Loosely speaking, the net shift to the left in the frequency maps indicates evolution to thicker orbits, which is confirmed by their large $\zmax$ (colors). While the thickness grows gradually for decreasing $\OmzR$, the vertical motion is amplified when orbits reach the vertical resonance $\OmzR=1$, and a more dramatic amplification happens at the vILR. This is particularly important near the crossing point of these two resonances and the ILR. While large vertical excitation by the vILR for orbits close to the ILR [$\OmphiR=1/2$] has been predicted by \cite{Binney_1981} and confirmed by e.g. \cite{Pfenniger_1991}, Fig.~\ref{fig:freq_map_evol_768MR} demonstrates this in a particularly transparent way. A comparison with Fig.~\ref{fig:freq_map_768MR} also suggests that, once orbits cross the vILR, their quasi-periodicity is destroyed, in agreement with \cite{Valluri_2016} who found that ``banana'' orbits are commonly chaotic -- see also \cite{Manos_2022}.

Note that Fig.~\ref{fig:freq_map_evol_768MR} shows the evolution of a fixed set of particles selected in the shoulders at $7\Gyr$, instead of the evolution of the shoulders themselves. In particular, starting from $\approx 4\Gyr$ this simulation persistently manifests shoulders which move outwards following the bar growth -- see Fig. 8 of \citetalias{Anderson2022}.

Fig.~\ref{fig:freq_map_evol_768MRS} shows the frequency map evolution for the model SD1S. In this case, a small overall shift upwards and to the left is barely noticeable, in agreement with the very slow bar evolution in this simulation (Fig.~\ref{fig:Om_p_all_sims}). We also note a significant increase in $\zmax$ for orbits crossing to $\OmzR < 1$, but the vILR seems relevant only very close to the crossing point with the ILR in this model, and there is no vILR cloud.

For HG1, the evolution of the frequency map is shown in Fig.~\ref{fig:freq_map_evol_708main}. Once more, we note a shift upwards and to the left. In this case, the resonance $\OmzR = 1$ also seems more effective in producing large vertical excursions. In the vILR cloud, $\zmax$ is not as significantly amplified as in the model SD1 (Fig.~\ref{fig:freq_map_evol_768MR}). However, the vILR cloud is more populated in this model.

Finally, Fig.~\ref{fig:freq_map_evol_741D8} shows the frequency map evolution for Model 4. The bar in this model buckles slightly before $4\Gyr$, and shoulders are observed right after that -- see Sec.~\ref{sec:sims}. The abrupt shift to lower $\OmzR$ from $3\Gyr$ to $4\Gyr$ is a consequence of this buckling, which vertically thickens the bar region significantly. This interpretation and the frequency map at $4\Gyr$ itself must be considered with caution, since this is too close to the buckling and the gravitational potential may lack the triaxial symmetry we assume in order to estimate it. On the other hand, the main features in the frequency map at $4\Gyr$ are still observed at $5\Gyr$, by which point the potential is more symmetric, and therefore our fit is more reliable, with the same trend of a shift upwards and to the left. This suggests that the frequencies at $4\Gyr$ are reasonably accurate, despite the disturbance from the buckling. In this model, the strong role of the vILR in exciting vertical motion is again observed, producing a well populated vILR cloud with high $\zmax$.
 
\subsection{Evolution of shoulder-supporting orbits}
\label{sec:evol_Sigma_stack}

In Sec.~\ref{sec:evol_freq_map} we showed that orbits selected in the shoulder region were either already trapped by the ILR at previous snapshots or previously had smaller $\OmphiR$, with the latter evolving towards larger $\OmphiR$, until possibly being trapped by the ILR. The orbits also evolve towards smaller $\OmzR$, until possibly crossing the vILR, thereby being excited to high $\zmax$ and becoming chaotic (e.g. Fig.~\ref{fig:freq_map_768MR}). In Sec.~\ref{sec:orbital_support_shoulder} we demonstrated the important role of vertically warm orbits at the ILR, and of the vILR cloud, in building the shoulders. We now analyse the morphological evolution of these warm orbits.

\begin{figure*}
	\includegraphics[width=\textwidth]{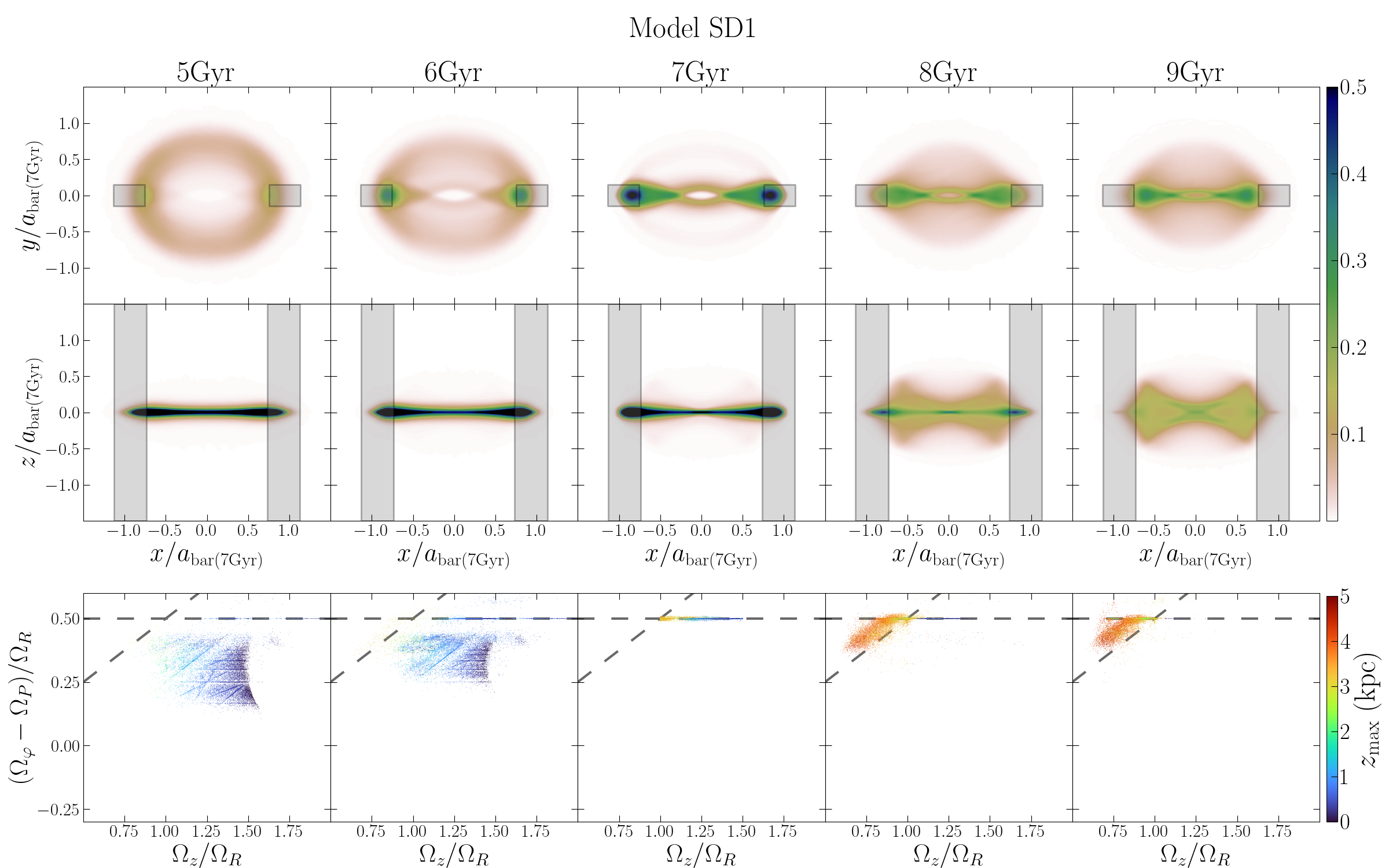}
    \caption{Evolution of the orbit-averaged density maps in the $x-y$ (top) and $x-z$ planes (middle) for star-particles selected to be on warm orbits at the ILR at $7 \Gyr$ in model SD1. The coordinates are normalized by the bar length at the selection time. Vertical shaded gray areas show the shoulder region at the selection time. The bottom row shows the same orbits in the frequency map. Those that at $7 \Gyr$ have the looped-shape, and support the shoulder morphology, evolve from less elongated orbits and typically away from smaller $\OmzR$. Once they cross the vILR, they spread in the vILR cloud, with much larger $\zmax$, diluting the over-densities in the $x-y$ plane.}
    \label{fig:Sigma_stack_xy_ILR_mid_OmzR_768MR}
\end{figure*}

Fig.~\ref{fig:Sigma_stack_xy_ILR_mid_OmzR_768MR} shows, for the model SD1, the orbit-averaged surface density maps at different times for the same star-particles selected to be warm orbits at the ILR at $7\Gyr$, with the central column representing the selection time. The upper (middle) row shows the maps in the $x-y$ ($x-z$) plane, with coordinates normalized by the bar length at the selection time. Here again, colors are normalized by the total number of star-particles in the shoulder region. The shaded areas show the normalized shoulder region at the selection time. The bottom panels show the same orbits in frequency maps.

The upper row shows that the warm orbits at the ILR at $7\Gyr$ mostly come from less elongated orbits which gradually become more elongated, until being trapped by the ILR. A fraction of these orbits had $\OmzR > 3/2$ before the selection time, and either before or after being trapped by the ILR. Before they cross the vILR ($t<7 \Gyr$), these trends are accompanied by negligible evolution in the $x-z$ plane. Once they cross the vILR ($7 - 8 \Gyr$), these orbits occupy the whole vILR cloud region, with large $\zmax$. This is accompanied by a dissolution of the density excess in the $x-y$ plane. Additionally, the orbits start to shrink along the major axis, and become confined to the inner parts at $9\Gyr$, no longer falling in the original ($7\Gyr$) shoulder region. This is closely connected to the abrupt change in the $x-z$ plane from $7\Gyr$ to $8\Gyr$. As shown in Fig.~\ref{fig:freq_map_768MR}, this is associated with the chaotic nature of the orbits in the vILR cloud, which dilutes the density distribution in the $x-y$ plane and weakens their contribution to the shoulders. But perhaps most importantly, we see that, to the extent that the vILR increases the vertical excursion, it decreases the in-plane amplitude, converting in-plane motion to vertical motion, like a pole vaulter.

\begin{figure*}
	\includegraphics[width=\textwidth]{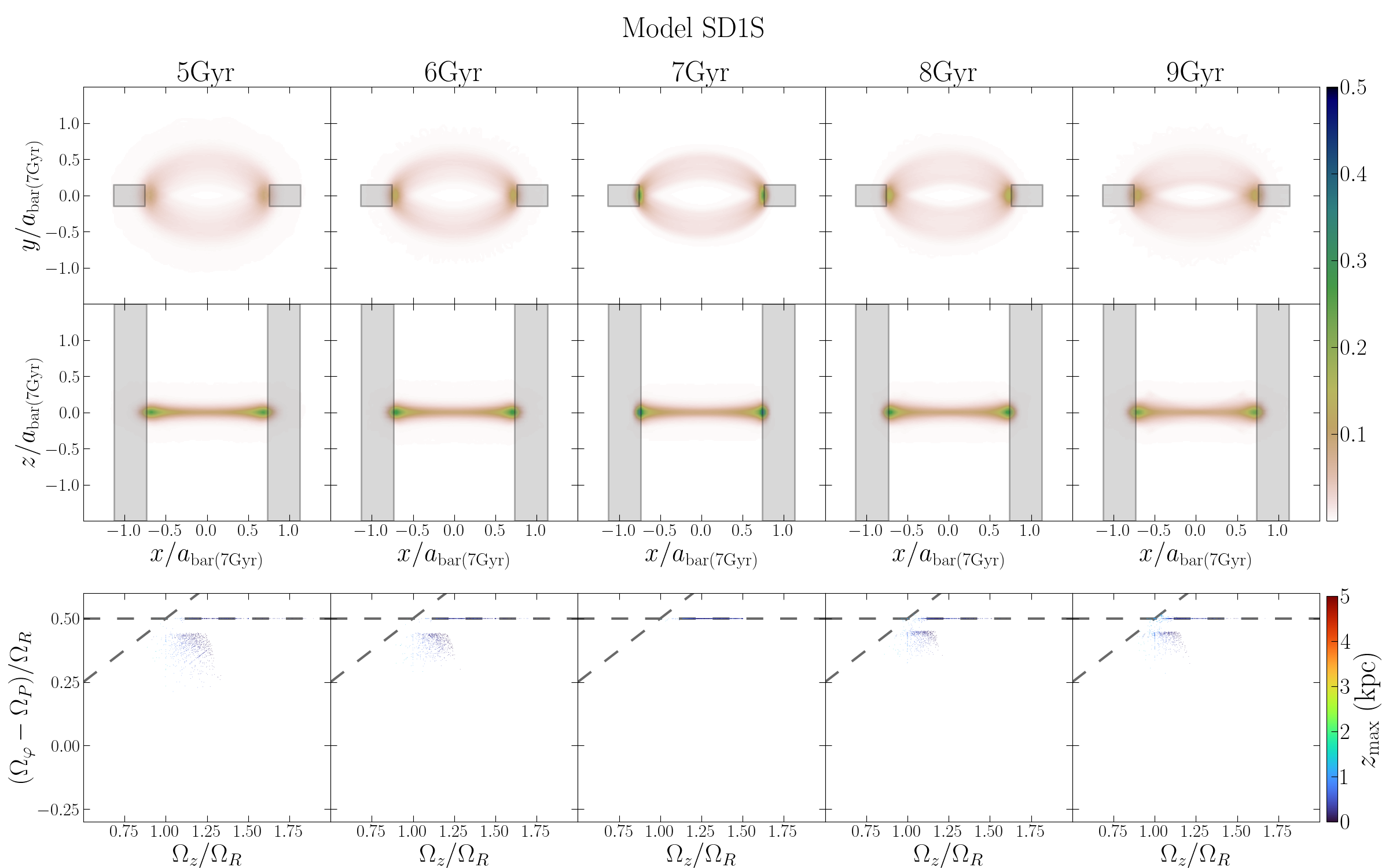}
    \caption{Similar to Fig.~\ref{fig:Sigma_stack_xy_ILR_mid_OmzR_768MR}, but for model SD1S. As the bar barely evolves in this model, the morphology of the warm orbits selected at 7 Gyr at ILR does not change significantly.}
    \label{fig:Sigma_stack_xy_ILR_mid_OmzR_768MRS}
\end{figure*}

For completeness, Fig.~\ref{fig:Sigma_stack_xy_ILR_mid_OmzR_768MRS} shows the corresponding plot for model SD1S. However, since the bar barely evolves in this model (see Fig.~\ref{fig:Om_p_all_sims}), we see no significant evolution in this plot.

\begin{figure*}
	\includegraphics[width=\textwidth]{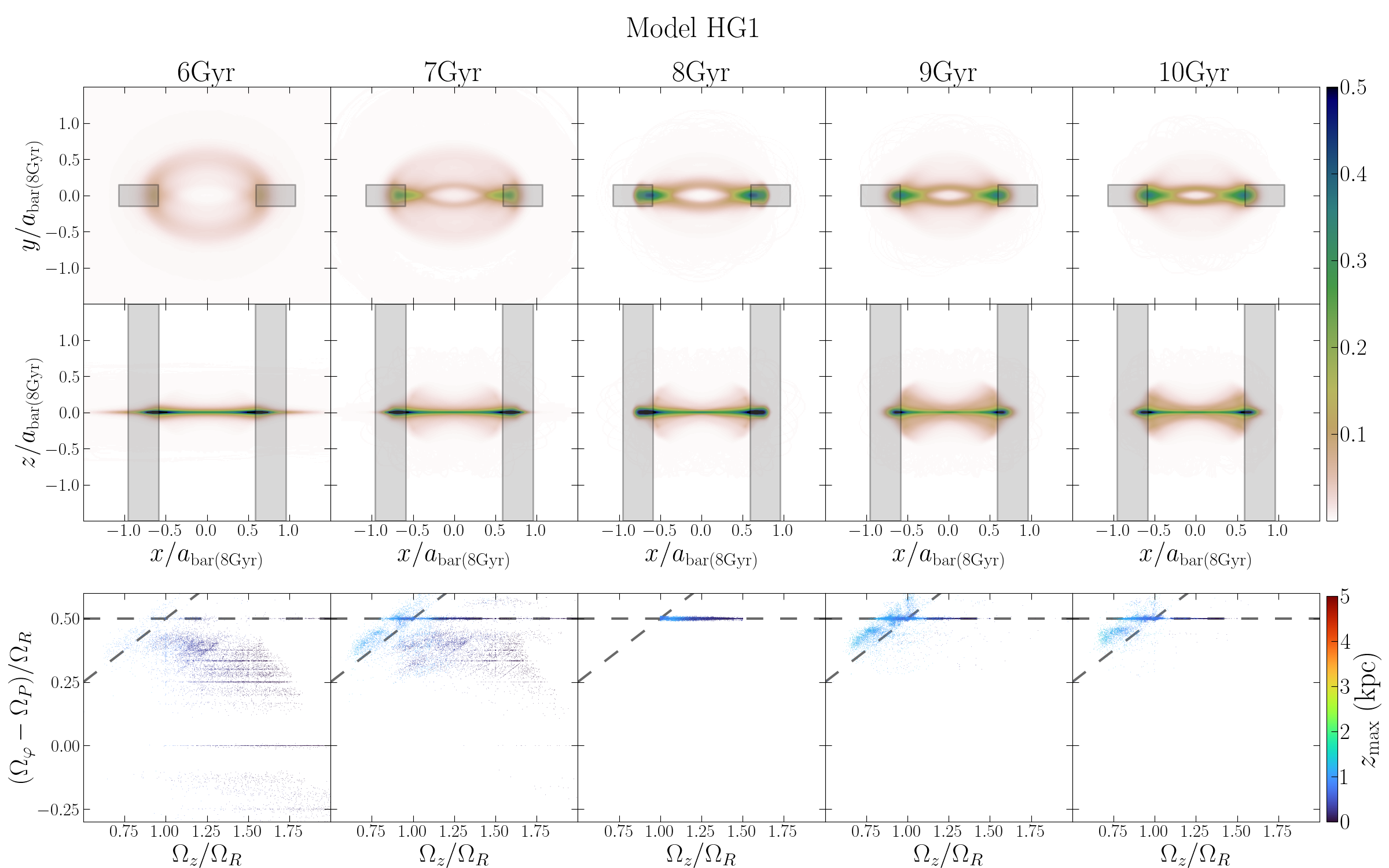}
    \caption{Similar to Fig.~\ref{fig:Sigma_stack_xy_ILR_mid_OmzR_768MR}, but for model HG1. For orbits crossing the vILR, the amplification of $\zmax$ and the dilution of the over-densities in the $x-y$ plane are smaller than in SD1, but we still observe the vertical heating by the vILR.}
    \label{fig:Sigma_stack_xy_ILR_mid_OmzR_708main}
\end{figure*}

\begin{figure*}
	\includegraphics[width=\textwidth]{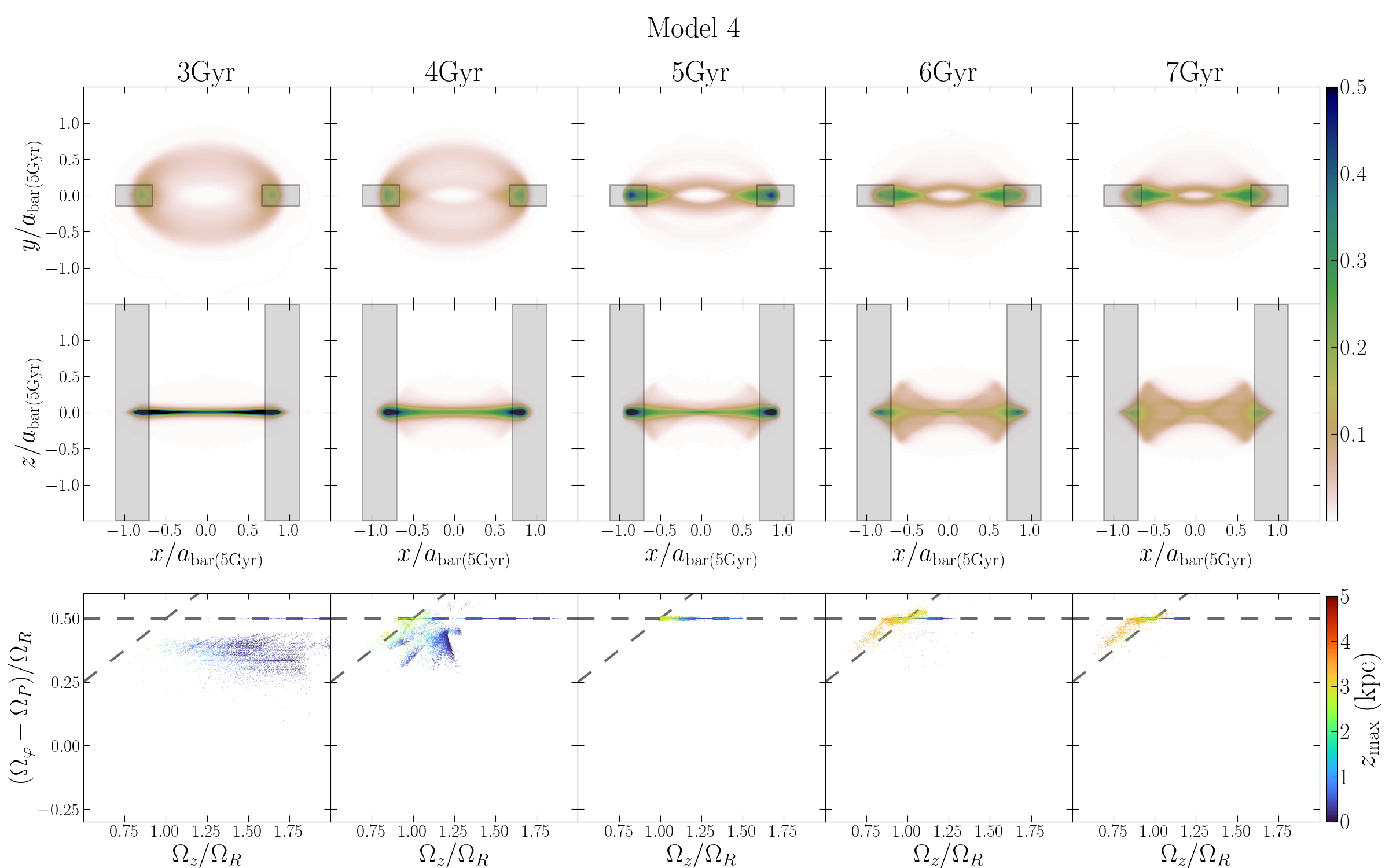}
    \caption{Similarly to Fig.~\ref{fig:Sigma_stack_xy_ILR_mid_OmzR_768MR}, but for Model 4. We see the same trends observed in the models SD1 and HG1, but with the abrupt change from $3 \Gyr$ to $4 \Gyr$ due to the buckling bar.}
    \label{fig:Sigma_stack_xy_ILR_mid_OmzR_741D8}
\end{figure*}

Fig.~\ref{fig:Sigma_stack_xy_ILR_mid_OmzR_708main} shows a corresponding plot for the model HG1, where we observe the same trends. However, in HG1 the vILR is less vigorous in exciting vertical motion and, accordingly, the orbit-averaged density in the $x-y$ plane is less disturbed for particles crossing the vILR to the vILR cloud. Fig.~\ref{fig:Sigma_stack_xy_ILR_mid_OmzR_741D8} shows the equivalent plots for Model 4, and we see the same connection between the dilution of the density excess in the $x-y$ plane and the increase in the vertical motion after the orbits cross the vILR, populating the vILR cloud.

These results demonstrate the role of the vILR in perturbing the shoulder-supporting orbits: once these orbits cross this resonance, their in-plane motion is converted to vertical motion and they become chaotic, contributing to diluting the shoulders.

\section{Discussion}
\label{sec:discussions}

\subsection{Comparison with different techniques}
\label{sec:compare_tech}

We have performed frequency analyses for orbits integrated in frozen potentials of barred galaxy simulations. Frequency analysis of the coordinates in ongoing $N$-body simulations (without introducing frozen potentials) has also been proven possible \citep[e.g.][]{Ceverino_2007, Gajda_2016, Parul_2020, Smirnov_2021}. While this has the apparent advantage of tracking the orbital evolution in a more self-consistent manner, the short time-series (typically $\approx1 \Gyr$), the unavoidable changes in the potential in the considered time interval, and all sources of noise in $N$-body potentials introduce uncertainties in the estimated frequencies. On the other hand, orbit integration in frozen potentials provides {\it instantaneous} orbital properties, and it allows the use of densely sampled time-series (allowing investigation of higher frequencies), with the narrowness of the main resonances in the frequency maps testifying to the high accuracy of our procedure.

The frequency analysis involves the choice of the coordinate system, the reference frame and the use of real or complex time-series. Regarding the last point, \cite{Hunter_2002}, using Cartesian coordinates for a few orbits, reports no improvements in the accuracy of estimated frequencies when using complex time series. Here, we chose cylindrical coordinates for their easier interpretation, given the approximate symmetries of the potential, and complex time-series. In particular, we use the complex Poincaré polar coordinates for the azimuthal angle.

Difficulties in estimating the azimuthal frequency $\Omega_\varphi$ have been reported in several works \citep[e.g.][]{Athanassoula_2002, Athanassoula_2003, Ceverino_2007}, but often not in explicit terms. In Appendix~\ref{sec:isochrone}, exploring the Isochrone Model, we show how to overcome these difficulties, demonstrating that the use of the complex Poincaré polar coordinates improves the accuracy of the azimuthal frequency and facilitates its extraction. The accuracy of estimated frequencies in more generic potentials with (semi-)analytic expressions for the fundamental frequencies deserves further investigation.

Unlike the common practice of performing frequency analysis in the rotating frame of the bar, we opted to store the coordinates in the inertial frame of the galaxy. As demonstrated in Appendix~\ref{sec:rot_frame}, this choice allows a precise identification of the co-rotation resonance, which is not possible, at least with the techniques employed in this work, if the rotating frame of the bar is adopted.

\subsection{Implications for the evolution of shoulder profiles}

\citetalias{Anderson2022} demonstrated that shoulders are present in the density profiles of growing bars. It also suggested the important contribution of looped orbits to the shoulders, which is confirmed with the detailed orbital study performed here (Sec.~\ref{sec:orbital_support_shoulder}). We now return to the questions that motivated this work and that we posed in Sec.~\ref{sec:intro}.

The evolution of the density profile is complex and must depend to some extent on all orbital groups, but we draw a simple picture that is based on the evolution and prevalence of looped orbits along the ILR. In Sec.~\ref{sec:orbital_support_shoulder} we demonstrated that the looped morphology in the $x-y$ plane is very well defined for vertically warm orbits at the ILR (but not in the model SD1S, where orbits are not looped), while cool orbits at the ILR are more elliptical-like and contribute negligibly to the shoulders. Hot orbits at the ILR (and at the vILR cloud) are chaotic and do not share the same looped morphology of warm orbits, but overall still spend a considerable amount of time in retrograde motion and produce a similar, more diluted, looped-like shape in the $x-y$ plane. Their contribution to the shoulders depends on the strength of the vILR in exciting vertical motion.
 
Thus, in order for shoulders to emerge in the density profiles, the ILR needs to trap a significant number of orbits as the bar slows and its resonances sweep through the disc. Additionally, a significant fraction of these orbits need to develop loops at their ends, which typically happens in the warm orbits at the ILR (${1 \lesssim \OmzR \lesssim 3/2}$). Since orbits in the disc and in the bar's thinnest parts typically have $\OmzR > 3/2$, to support shoulders these orbits need to decrease $\OmzR$, either before or after being trapped by the ILR, which occurs when the orbits thicken vertically. While the bar thickens, either via secular processes or via buckling, the shoulder-supporting warm orbits at the ILR can sooner or later be populated. This is compatible with the findings of \citetalias{Anderson2022} that shoulders can appear in either buckling or non-buckling bars and that, in buckling bars, shoulders commonly appear just after a buckling event -- see their Fig. 7.
 
Once shoulders have emerged in the density profile, their supporting orbits can depart from the looped morphology in the $x-y$ plane if they move out of the range $1 < \OmzR < 3/2$. This typically happens with further bar thickening and orbits at the ILR crossing the vILR towards $\OmzR\lesssim 1$ -- see Figs.~\ref{fig:Sigma_stack_xy_ILR_mid_OmzR_768MR}-\ref{fig:Sigma_stack_xy_ILR_mid_OmzR_741D8}. Once more, this thickening can happen either secularly or via a secondary buckling. Thus, if the bar is not replenished with looped orbits, the shoulders can eventually be erased. This explains why \citetalias{Anderson2022} noted several instances of sudden erasure of shoulders after a secondary buckling. It is worth noting that the newly vertically excited orbits will contribute to a thicker potential, thus decreasing the ratio $\OmzR$ of nearby orbits, bringing them towards the vILR, in a positive feedback process.
 
Thus, the common thickening of orbits in the bar region would make shoulders transient features, unless new orbits are trapped at the ILR with the appropriate $\OmzR$ range and develop the looped morphology, a process that seems to happen often for bars evolving secularly -- see Fig. 8 of \citetalias{Anderson2022}. 

The fact that the looped morphology tends to appear in orbits at the ILR and with $1 \lesssim \OmzR \lesssim 3/2$ suggests some connection with the vertical motion. This is intriguing since those orbits were first found in 2D bars \citep[][]{Contopoulos_1988}. An interesting possibility is that this looped morphology might be somehow triggered by vertical resonances -- for instance, comparing Figs.~\ref{fig:freq_map_768MR}-\ref{fig:freq_map_741D8}, we see that the model SD1S, the only one without looped orbits or shoulders, is particularly unpopulated in the region near the crossing of the ILR and ${\OmzR=3/2}$.

 \subsection{Connection with bulges}

Investigating mechanisms for bar thickening, \cite{Sellwood_2020} contrast three scenarios: a bar that buckles, one in which orbits are vertically excited by staying trapped for significant times at the vILR \citep[][]{Quillen_2002}, and another one in which orbits cross the vILR, get vertically heated and remain that way after leaving the resonance \citep[][]{Quillen_2014}. Having found evidence for the long-lasting trapping mechanism only in an artificially vertically symmetrized model, \cite{Sellwood_2020} conclude that the third mechanism is probably more relevant in real galaxies without bar buckling. 

Our results, in particular e.g. Figs.~\ref{fig:freq_map_evol_768MR} and \ref{fig:Sigma_stack_xy_ILR_mid_OmzR_768MR} (bottom row), confirm the rapid crossing of the vILR, as opposed to a long trapping time, and the rapid vertical excitation with a long-lasting X-morphology. It is worth mentioning that the orbits at the vILR cloud, after crossing and being perturbed by the vILR, can get very close to the galactic center (Fig.~\ref{fig:hist_rper}), possibly within the sphere of influence of a supermassive black hole (SMBH). This can potentially represent an important mechanism to persistently replenish the sphere of influence of a SMBH with new stars. This possibility is even more interesting if galactic bars can co-evolve in harmony with black holes, as recently demonstrated by \cite{Wheeler_2023}.
 
In Sec.~\ref{sec:results}, we have shown that orbits crossing the vILR populate what we call the vILR cloud, characterized by chaotic orbits with large $\zmax$ and X-shapes in the $x-z$ plane. This morphology being produced by the vILR agrees with several previous works \citep[e.g.][]{Pfenniger_1991}, where the contribution of the ``banana'' orbits to BP-bulges has been emphasized. On the other hand, \cite{Portail_2015} find that these orbits only contribute in the bulge's outermost regions, while orbits with $\Omega_x/\Omega_z \approx 3/5$ (``brezel'' orbits) are more relevant for the bulge as a whole (in their study, $\Omega_x$ is the frequency of oscillation about the bar major-axis and can be seen as a proxy for our $\Omphi - \OmP$) -- see also \cite{Abbott_2017}.
 
 Secs.~\ref{sec:zmax_vs_OmzR} and \ref{sec:evol_Sigma_stack} demonstrate that after crossing the vILR the banana orbits can also retain, to a certain extent, the loops near apocenters inherited from warm orbits at the ILR. In fact, these loops have already been found in orbits supporting either the thin \citep[][]{Patsis_2014} or the thick \citep[][]{Combes_1990} parts of bars.
 
  Altogether, our results suggest that: {\it i}) the banana orbits can simultaneously contribute to the outskirts of the bulge peanut shape (in the $x-z$ plane) and, depending on the strength of the vertical perturbation, to the shoulders (in the $x-y$ plane) -- see e.g. Fig.~\ref{fig:Sigma_stack_xy_ILR_mid_OmzR_708main} ($9 \Gyr$); {\it ii}) in the study of the orbital support of box-peanut (BP) bulges, the importance of the vILR may be underestimated if we restrict to orbits trapped at it and exclude the vILR cloud, since the latter is populated by orbits crossing the vILR and acquiring a long-term BP-morphology.

  In a detailed observational study with the $S^4G$ survey \citep[][]{Sheth_2010}, \cite{Erwin_2023} find some cases of galaxies with shoulders but no BP-bulge, and no clear case of bars with BP-bulges and no shoulders, suggesting that shoulders appear before BP-bulges. Given the role of the vILR in producing BP-bulge-supporting orbits and the evolution of shoulder-supporting orbits along the ILR, from the vertically warm ones to the vILR cloud after crossing the vILR, the suggestion of \cite{Erwin_2023} emerges as a natural outcome. 
  
  The presence of shoulders has been found to correlate with massive (classical) bulges, or centrally concentrated halos \citep[][]{Kim_2015, Kruk_2018, Athanassoula_Misiriotis_2002}. Furthermore, \cite{Athanassoula_1992} found that looped orbits tend to appear in galaxies with massive bulges (or high central concentration, in general), and with more elongated and slowly rotating bars \citep[looped orbits in a slow bar have also been detected by][]{Chaves-Velasquez_2017}. This is in agreement with the important contribution of looped orbits to shoulders, suggested by \citetalias{Anderson2022} and confirmed here. Additionally, among our three models with shoulders, two (SD1 and Model 4) have slow bars and one (HG1, with weak shoulders) has a fast bar, suggesting that shoulders are not restricted to slow bars, but can also be present, with varying strength, in fast bars. All models have clear signs of BP-bulges, although they are weaker for models SD1S and HG1, which suggests that the correlation shoulder-bulge is not restricted to either BP- or classical bulges.
  
  A possible interpretation of the shoulders-massive bulges correlation might be that massive bulges induce thicker orbits in their outskirts, facilitating occupation of the shoulder-supporting warm orbits at the ILR, while bars in bulgeless galaxies would tend to trap stars in thinner orbits, i.e. with larger $\OmzR$ such as the cool orbits at the ILR, with a morphology that does not support shoulders -- see Fig.~\ref{fig:orbs_stack_ILR}. This line of thought would shift a massive bulge towards being the \textit{cause} for shoulders, which would be in contradiction with the apparent precedence of shoulders with respect to BP-bulges suggested by \cite{Erwin_2023}.
  
  On the other hand, warm orbits at the ILR crossing the vILR are vertically excited and change into chaotic orbits. As we have shown, this is accompanied by these orbits shrinking in the $x-y$ plane and passing close to the center (see e.g. Fig.~\ref{fig:Sigma_1D_stack_ILR_768MR}, panels $b$ and $c$), overall increasing the density of the central regions. This may explain the sudden increase in the central surface density after buckling events observed in simulated galaxies \citep[][]{Raha_1991, Debattista_2006}, since this brings orbits towards the vILR -- see Fig.~\ref{fig:freq_map_evol_741D8}. Therefore, this mass transfer to the center may contribute to the growth of massive bulges which, instead of causing the appearance of shoulders, would be the \textit{consequence} of the evolution of shoulder supporting orbits towards the vILR.

\cite{Kim_2015} suggest that shoulders are indicative of dynamically evolved bars, which should be formed with exponential profiles inherited from the disc. They also suggest that this might explain the correlation with the galaxy's stellar mass and Hubble type, with more massive early-type spirals forming bars earlier. Although the eventual erasure of shoulders found by \citetalias{Anderson2022} complicates this narrative, this would agree with the strong dependence of shoulders on stellar mass found by \cite{Erwin_2023}. Our finding that the most shoulder-supporting morphology appears in orbits at the ILR with intermediate thickness (thus requiring a minimum thickness) is also in line with shoulders being associated with dynamically evolved bars. Additionally, the evolution of these orbits towards vertically thicker orbits supporting bulges also agrees with the suggestion of \cite{Erwin_2023} that shoulders seem to precede BP-bulges.

\subsection{Shoulder definition and possible observational implications}
\begin{figure*}
\begin{center}
	\includegraphics[scale=0.4]{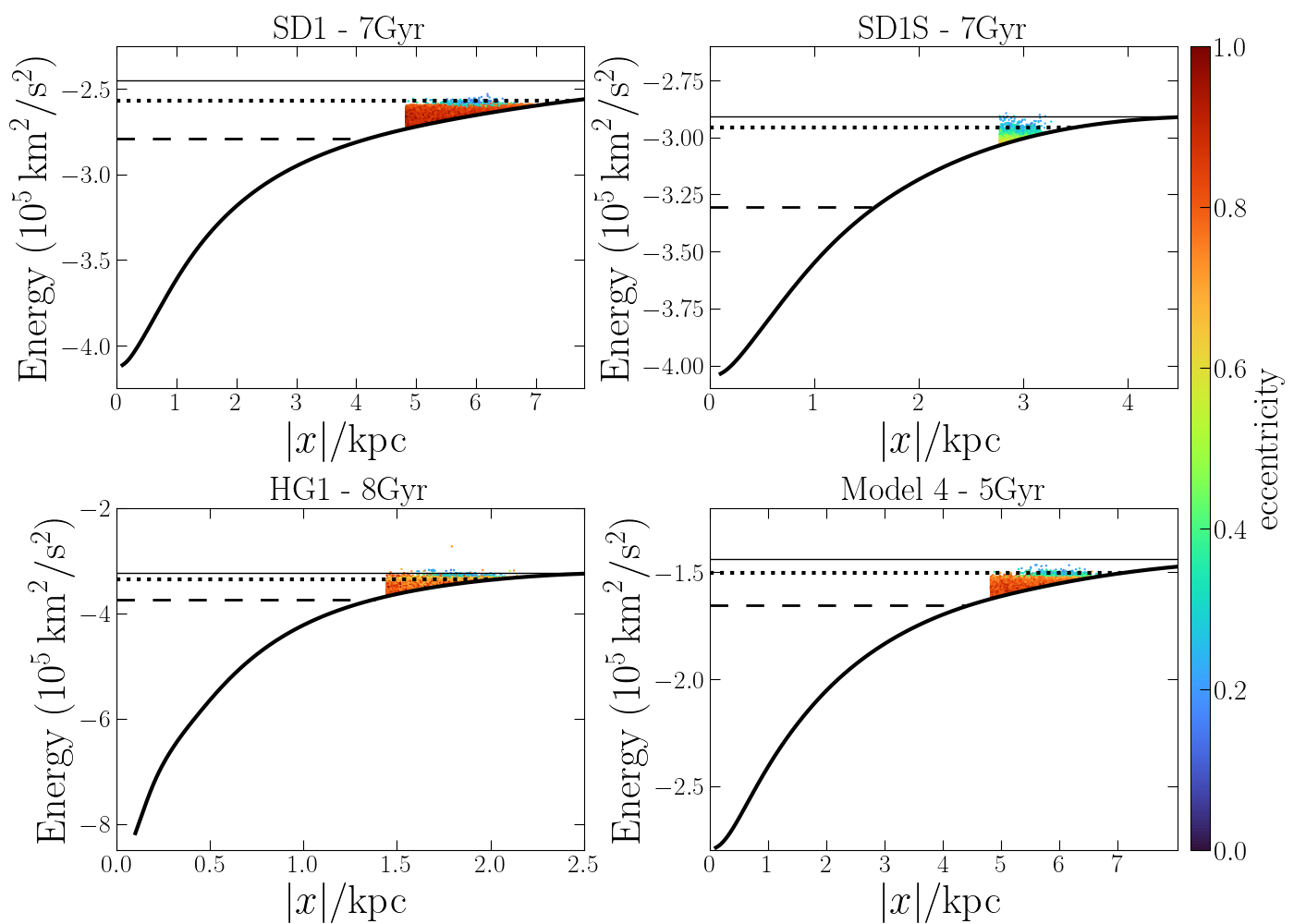}
    \caption{Similar to Fig.~\ref{fig:pot_eff}, but only showing $\EJ$ (points) for warm orbits at the ILR. These orbits are confined to within the UHr.}
    \label{fig:pot_eff_warm_ILR}
\end{center}
\end{figure*}

Our results show that the most relevant orbits to shoulders are the looped ones at the vertically warm ILR. In this way, one can expect shoulders to be restricted to the region where these orbits are present. Similarly to Fig.~\ref{fig:pot_eff}, in Fig.~\ref{fig:pot_eff_warm_ILR} we show the effective potential in the bar-major axis, as well as the approximate locations of the co-rotation, UHR and the ILR. We also show the Jacobi integral, but now restricting to warm orbits at the ILR.

It is clear that these orbits are restricted to the region inside the Ultra Harmonic radius. In Fig.~\ref{fig:pot_eff}, we see that for the model SD1 the outer edge of the shoulder approximately coincides with the Ultra Harmonic radius. However, for the other simulations, the shoulder outer edge (as detected in \citetalias{Anderson2022}) can be significantly outside this radius.

This observation is relevant for two reasons: on one hand, this suggests that identifying the outer edge of the shoulders based on a minimum of the curvature radius of the density profile (as implemented in \citetalias{Anderson2022}) might be overestimating the shoulder extent, if this is to be identified with the region of looped orbits at the ILR. If that is the case, our current selection may be contaminated by a significant number of orbits that are not associated with the shoulders. Thus, the contribution of the looped orbits to shoulders we inferred in Sec.~\ref{sec:orbital_support_shoulder} represents a lower bound to their true importance to shoulders. On the other hand, if we can infer that the shoulders should end at the Ultra Harmonic radius, this can potentially be translated into constraints on the mass distribution and/or bar pattern speed of galaxies where shoulders are detected, or even in the Milky Way.

\section{Summary}
\label{sec:conclusions}

In this paper, we explored three pure $N$-body models and one $N$-body+SPH model which form bars self-consistently. Three of these models, including the one with gas, develop shoulders in the bar-major axis density profile. Among these, one has a bar that buckles. We select star-particles in snapshots and regions where shoulders are identified, and integrate their orbits at different times in frozen potentials rotating with the corresponding bar pattern speeds. We perform frequency analysis of these orbits, using cylindrical coordinates in the inertial frame and selecting the leading frequency, presenting results in the frequency space $\OmphiR$ vs $\OmzR$. We evaluate the contribution of different orbital types to the density in the shoulders region, and follow their evolution. Our main conclusions are:

\begin{itemize}
    \item We confirm the suggestion of \citetalias{Anderson2022} of the importance of orbits with loops at their ends (in the $x-y$ plane) to the shoulders in bar density profiles.
    
    \item We show that this orbital shape typically happens for orbits at the ILR [$\OmphiR=1/2$] and with $1 \lesssim \OmzR \lesssim 3/2$ (vertically warm). Among the orbits at the ILR, those with $\OmzR \gtrsim 3/2$ (cool orbits) have a mildly elliptical shape and contribute negligibly to the shoulders.
    
     \item As the bar slows down, the main resonances sweep the disc and, for a fixed set of particles, the ratio $\OmphiR$ increases, until they are trapped by the ILR. While the bar thickens, the ratio $\OmzR$ tends to decrease, causing a significant number of orbits to cross the vILR (Figs.~\ref{fig:freq_map_evol_768MR}-\ref{fig:freq_map_evol_741D8}).

     \item In agreement with \cite{Quillen_2014} and \cite{Sellwood_2020}, we find that the vILR does not trap stars for long times. However, orbits crossing it are permanently vertically excited, developing an $X$ shape in the $x-z$ plane -- see also \cite{Wheeler_2023}.
    
    \item Warm orbits at the ILR crossing the vILR towards $\OmzR \lesssim 1$ are vertically excited at the expense of in-plane motion, become chaotic and populate a region around the vILR in the frequency map that we dub the {\it vILR cloud}. In comparison to warm orbits at the ILR, those at the vILR cloud are less extended along the $x-$axis and produce a more diffuse net figure in the $x-y$ plane (Figs.~\ref{fig:Sigma_stack_xy_ILR_mid_OmzR_768MR}-\ref{fig:Sigma_stack_xy_ILR_mid_OmzR_741D8}).
    
    \item Orbits at the vILR cloud also contribute significantly to shoulders, but the larger their vertical excursions, the less shoulder-supporting they are.
     
     \item As the bar thickens, the orbital support for shoulders can increase with orbits at the ILR entering the range of  warm orbits, and potentially decrease once they move to $\OmzR \lesssim 1$ (where they cross the vILR). This explains why shoulders can develop right after a buckling event but can also be erased right after a second buckling (as noted in \citetalias{Anderson2022}), since these can suddenly either populate or de-populate those orbits. This simple evolutionary sequence supports the suggestion of \cite{Erwin_2023} that shoulders seem to appear before BP-bulges (with a possible co-formation of shoulders and BP-bulges in buckling bars).

     \item An analysis with the isochrone model (Appendix~\ref{sec:isochrone}) shows that for orbits with large apocenters, the leading frequency in the azimuthal spectrum is the precession one, instead of $\Omphi$. This has possible implications for the analysis of Milky Way halo stars and deserves further investigation. Moreover, frequency estimates with the time-series $f_\varphi = \sqrt{2|L_z|}(\cos\varphi + i\sin\varphi)$ provide higher precision (compared to $f_\varphi = \varphi$) and facilitates its extraction -- see Fig.~\ref{fig:naif_vs_analytic}.
\end{itemize}

In conclusion, bar thickening and vertical resonances can dilute the shoulders in the bar major-axis density profiles, unless the bar keeps trapping new orbits in the appropriate frequency ratios, with the looped morphology. This sheds light on the conclusion of \citetalias{Anderson2022} that long-term shoulders require a growing bar. Our results help to build a coherent picture tying together known analytical results with several recent observational and simulation-based results on the properties and evolution of shoulders in the density profiles of bars.

\section*{Acknowledgements}
We thank Eugene Vasiliev for his help with \textsc{agama} and for always enriching discussions. We also thank Rubens Machado and João Amarante for a careful reading and comments. We are also truly grateful to the anonymous referee, whose comments contributed to improve this paper. V.P.D. and L.B.S. were supported by STFC Consolidated grant ST/R000786/1. L.B.S and M.V. acknowledge the support of NASA-ATP award 80NSSC20K0509 and U.S. National Science Foundation AAG grant AST-2009122. KJD is grateful for support from the Simons Foundation for supporting her sabbatical stay at the CCA, Flatiron Institute. The orbital analysis was carried out on Stardynamics, which was funded from Newton Advanced Fellowship \#~NA150272 awarded by the Royal Society and the Newton Fund. The simulations were run at the High Performance Computing Facility of the University of Central Lancashire and at the DiRAC Shared Memory Processing system at the University of Cambridge, operated by the COSMOS Project at the Department of Applied Mathematics and Theoretical Physics on behalf of the STFC DiRAC HPC Facility (www.dirac.ac.uk). This equipment was funded by BIS National E-infrastructure capital grant ST/J005673/1, STFC capital grant ST/H008586/1 and STFC DiRAC Operations grant ST/K00333X/1. DiRAC is part of the National E-Infrastructure

\software{numpy \citep{numpy},
          scipy \citep{scipy},
          pynbody \citep{Pontzen_2013},
          Agama \citep{Vasiliev_2019},
          naif (this work).}

%%%%%%%%%%%%%%%%% APPENDICES %%%%%%%%%%%%%%%%%%%%%

\appendix
\restartappendixnumbering

\section{Frequency analysis algorithm}
\label{sec:naff}
Here we present the algorithm used to extract frequencies of orbits, which is essentially the same as NAFF, proposed by \cite{Laskar_1992} and further developed by \cite{Valluri_Merritt_1998} -- see also \cite{Price-Whelan_2016}. We refer the reader to these papers for a complete description. We also comment on some details and improvements implemented in the python package used in this work, which we dub \texttt{naif} \citep[][]{naif} -- the documentation is available at \url{http://naif.readthedocs.io/en/latest/index.html}. This preliminary version extracts peak frequencies very accurately, as demonstrated in Appendix~\ref{sec:isochrone}, and will be improved in the future with several tools for frequency analysis.

Assume an orbit is integrated for a total time $T$ and let, for simplicity of notation, the time variable be defined in the symmetric interval $-T/2 \leq t \leq T/2$. In practice, $t$ is an array with $N$ elements $t_n$, and we have a (real or complex) time-series $f(t_n)$ associated with each coordinate. We start by performing a windowed Discrete Fourier Transform
\begin{equation}
\label{eq:DFT}
    F_j = \frac{1}{N}\sum_{n=0}^{N-1} f(t_n) \chi_p (t_n)e^{-2\pi i nj/N},
\end{equation}
where $j=-N/2,\dots, 0 \dots N/2$, and
\begin{equation}
\label{eq:chi}
    \chi_p(t) = \frac{2^p (p!)^2}{(2p)!}\left[ 1 + \cos \left( \frac{2\pi t}{T}\right)\right]^p
\end{equation}
is the window function, with $p \in \mathbb{N}$. In this work we use $p=1$, but other values might improve the estimate \citep[][]{Laskar_1999, Hunter_2002}.

The spectrum is calculated at frequencies $\omega_j=2\pi j/T$, and the minimum frequency that can be probed is $\omega_\mathrm{min} = 2\pi/T$. On the other hand, the maximum (Nyquist) frequency is $\Omega_\mathrm{Ny}=\Omega_{j=N/2} = N\pi/T$. If we conservatively require $\Omega_\mathrm{Ny} \approx 10^3\Gyr^{-1}$, this requires $N\gtrsim (10^3/\pi) (T/\Gyr)$. For an orbit with period of circular motion $T_\mathrm{circ} \approx 0.25$ Gyr, integrated for $T \approx 20T_\mathrm{circ}\approx 5$ Gyr ($T \approx 40T_\mathrm{circ}\approx 10$ Gyr), we have $\omega_\mathrm{min}\approx 1.25\Gyr^{-1}$ ($\omega_\mathrm{min}\approx 0.63\Gyr^{-1}$), and we need $N\gtrsim 1500$ ($N\gtrsim 3000$) points. Using $N\sim 10^4$ points is thus safe for most applications.

A rough estimate of the leading frequency is the spectral line with maximum amplitude $|F_j|$, $\ommax$. Since the spectrum is defined in the discrete set $\omega_j=2\pi j/T$, this estimate's precision is limited by the resolution $\Delta\omega = 2\pi/T$. This estimate is then refined by calculating the projection $\phi(\omega) = \langle f(t), e^{i\omega t}\rangle$ of the time-series $f(t)$ onto the ``frequency vector'' $e^{i\omega t}$, where
\begin{equation}
    \langle f(t), g(t)\rangle \equiv \frac{1}{T}\int_{-T/2}^{T/2}f(t)g^*(t)\chi_p(t)\, \mathrm{d}t.
\end{equation}
 The refined frequency is found by maximizing $|\phi(\omega)|$ around $\ommax$. The maximum is determined with the optimized Brent's algorithm, as implemented in scipy. If the maximum cannot be identified in this first shot, a local maximum is identified via a brute force search. The interval around $\ommax$ used in the original NAFF implementation is $\ommax \pm 2\pi/T$, but we have found that the interval $\ommax \pm 4\pi/T$ is more appropriate to guarantee the Brent's algorithm will find the maximum, and this is the interval used in \texttt{naif}. Having thus identified this first leading frequency $\omega_1$, we calculate the complex amplitude $a_1$ associated with it, where
\begin{equation}
    a_k = \langle f(t), u_k\rangle,
\end{equation}
and $u_k = e^{i\omega_k t}$. If we want to extract additional frequencies, we ``subtract'' this spectral line from the time-series, defining in general
\begin{equation}
    f_k(t) = f_{k-1}(t) - a_k e_k,
\end{equation}
where $e_1 = u_1$, $f_{k}(t)$ is the time-series after extracting the $k$-th frequency $\omega_k$, and $f_0(t)$ is the original one. Here, $e_k$ denotes an orthonormal basis, built from the $u_k$'s via a Gram–Schmidt process:
\begin{equation}
\label{eq:GSO}
    e_k = u_k - \sum_{j=1}^{k-1} \langle u_k, e_j\rangle\, e_j.
\end{equation}
We then iterate through Eqs.~\ref{eq:DFT}-\ref{eq:GSO} to extract as many frequencies as desired.

Compared to the original NAFF algorithm, a small (but relevant) improvement of \texttt{naif} is the following: NAFF first applies a DFT without windowing to look for the rough estimate $\ommax$, and then uses the window function $\chi_p(t)$ in the refinement step when maximizing $|\phi(\omega)|$. This different procedure to look for the same frequency ends up changing the order of the highest amplitudes in some cases. This is later corrected by sorting the frequencies by amplitude and taking the leading one, but requires the extraction of some frequencies even if we are only interested in the leading one. On the other hand, \texttt{naif} performs a windowed DFT, Eq.~\ref{eq:DFT}, already in the first search for $\ommax$, which ends up being always the leading frequency. This removes the need to extract additional frequencies if we are only interested in the leading one.

Finally, \texttt{naif} provides easy access to the spectrum before the extraction of each frequency and also to $|\phi(\omega)|$ around each extracted frequency, which can be useful for detailed investigations of individual orbits.

\section{Numerical frequencies for the isochrone model}
\label{sec:isochrone}

We use \textsc{agama} to generate a self-consistent sample of the isochrone model and integrate each orbit for $\approx 100$ azimuthal periods. A self-consistent ensemble is not strictly necessary, since we simply compare frequencies of orbits individually and do not focus on their relative numbers. However, this guarantees a broad phase-space region for investigation. Since this is a spherical model, we only estimate the radial and azimuthal frequencies, $\Omega_r$ and $\Omphi$, respectively (the transversal frequency $\Omega_\vartheta = \pm \Omphi$). For this model, we set $GM=1$.

\begin{figure}
	\includegraphics[width=\columnwidth]{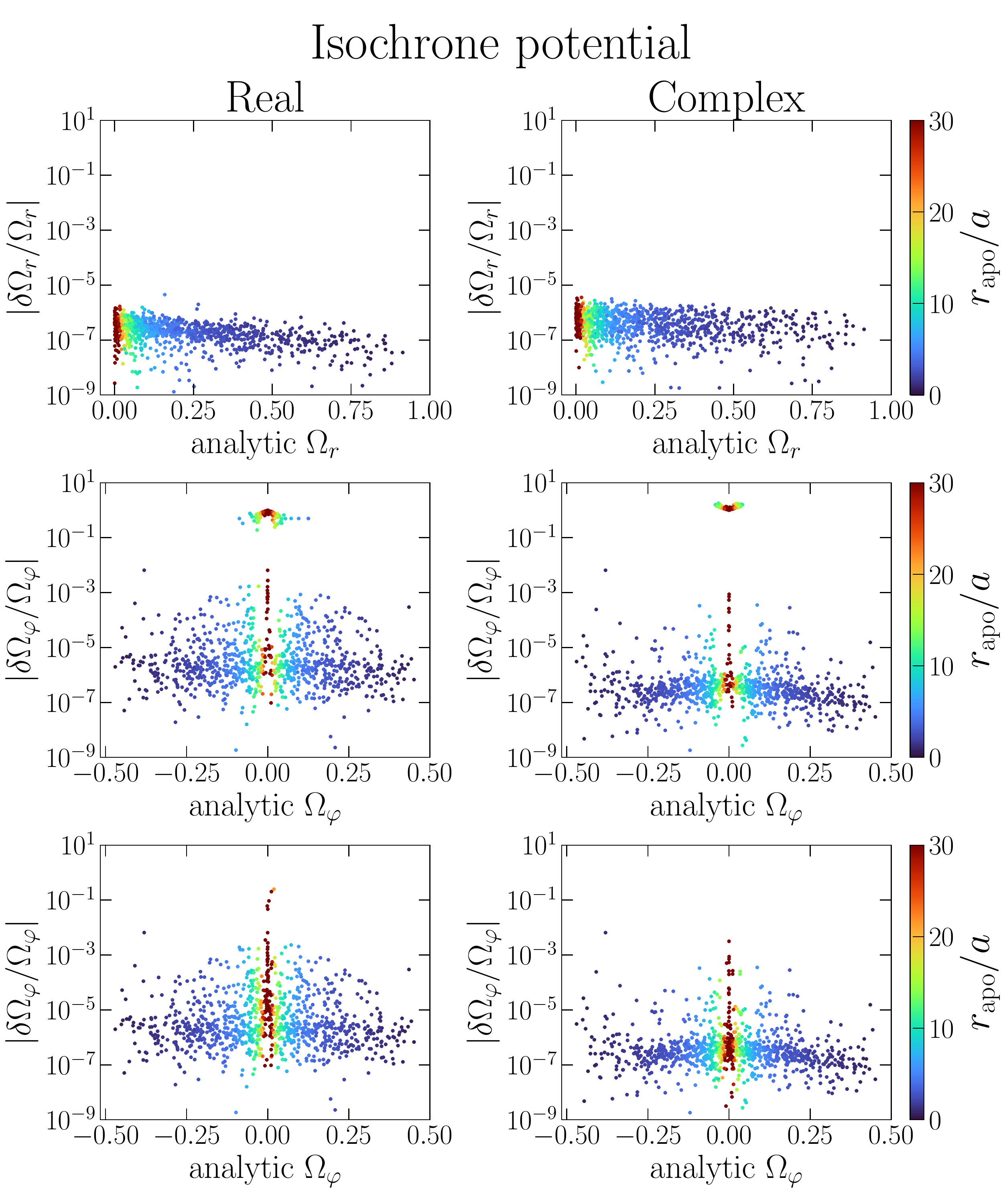}
    \caption{Relative errors in the fundamental frequencies for $10^3$ orbits in the isochrone potential. Left column: real time-series, Eq.~\ref{eq:f_real}. Right column: complex time-series, Eq.~\ref{eq:f_complex}. The radial frequency (top row) is estimated as the absolute value of the leading frequency. Real and complex time-series give similar errors. For $\Omphi$ (middle row), the real time series requires setting $\sign(\Omphi) = \sign(L_z)$, while this is automatically set for the complex time-series. Requiring $\Omphi > \Omega_r/2$ (see main text) and taking the leading frequency, we have $|\delta\Omphi/\Omphi| \lesssim 10^{-3}$ (bottom row). The complex time-series provides $\approx 10\times$ smaller errors for $\Omphi$.}
    \label{fig:naif_vs_analytic}
\end{figure}

Fig. \ref{fig:naif_vs_analytic} shows the relative errors of the estimated frequencies for $10^3$ orbits, in comparison to the analytic expressions for the isochrone model \citep[][]{BT}. Points are color-coded by the apocenter radii (in units of the scale radius). The left column shows results obtained using as input the real time-series
\begin{equation}
    % \begin{cases}
    \begin{array}{l}
      f_r = r \\
      f_\varphi = \varphi,
    % \end{cases}
    \end{array}
    \label{eq:f_real}
\end{equation}
and the right column shows results for the complex time-series \citep[][]{Papa_Laskar_1996, Papa_Laskar_1998}
\begin{equation}
    % \begin{cases}
    \begin{array}{l}
      f_r = r + i v_r \\
      f_\varphi = \sqrt{2|L_z|}(\cos\varphi + i\sin\varphi),
    % \end{cases}
    \end{array}
    \label{eq:f_complex}
\end{equation}
where $v_r = dr/dt$ is the radial velocity and $L_z$ is the $z$-component of the angular momentum. For the radial coordinate (first row), the absolute value of the leading frequency has a relative error $|\delta\Omega_r/\Omega_r| \approx 10^{-6}$, using either the real or the complex time-series.
 
  For the azimuthal angle (second row), using the real time-series ${f_\varphi = \varphi}$ the leading frequencies provide errors $|\delta\Omphi/\Omphi| \lesssim 10^{-3}$ for a fraction of the orbits, but only after imposing ${\sign(\Omphi)=\sign(L_z)}$, while the complex time-series (Eq.~\ref{eq:f_complex}) does not require this intervention. Additionally, there are a few orbits with small analytic $\Omphi$ and $|\delta\Omphi/\Omphi| \approx 1$ for both real and complex time-series. These orbits have large apocenters (${r_\mathrm{apo}/a \gtrsim 15}$), and expend most of the time far from the center, effectively behaving as precessing ellipses in a Kepler-like potential. As a consequence, the leading frequency in the spectrum is $\Omega_r - \Omphi$ (the precessing frequency), although $\Omphi$ is also present, with a smaller amplitude. In a Kepler-like potential, $\Omega_r \approx \Omega_\varphi$, thus the leading frequency is $\approx 0$, which implies $|\delta\Omphi/\Omphi| \approx 1$. In realistic galactic potentials, typical stars have $\Omega \lesssim \kappa \lesssim 2\Omega$. Thus, if we heuristically require $|\Omphi| > |\Omega_r|/2$, and take the leading frequency respecting this restriction, the fundamental $\Omphi$ is recovered with good accuracy ($|\delta\Omphi/\Omphi| \lesssim 10^{-3}$) for almost all orbits (third row). Finally, the complex time-series provides a $\approx 10\times$ better precision (although the estimates are almost identical to those of the real series if orbits are integrated for only $\approx 50$ circular periods).

\section{Inertial vs rotating frame}
\label{sec:rot_frame}
Here we compare the frequencies obtained using coordinates in  the inertial frame and in the bar rotating frame. We use the model SD1S, which has a fast bar at $7 \Gyr$ and the region where star-particles are selected is near the co-rotation radius -- see Fig.~\ref{fig:Omega_vs_R}. Thus, we can expect the co-rotation resonance to be populated in the frequency map, as is the case when we use the inertial frame (Fig.~\ref{fig:freq_map_768MRS}). In this case, the frequency analysis outputs $\Omphi$ and we literally subtract the bar pattern speed $\OmP$ before plotting the frequency map in Fig.~\ref{fig:freq_map_768MRS}.

\begin{figure}
	\includegraphics[width=\columnwidth]{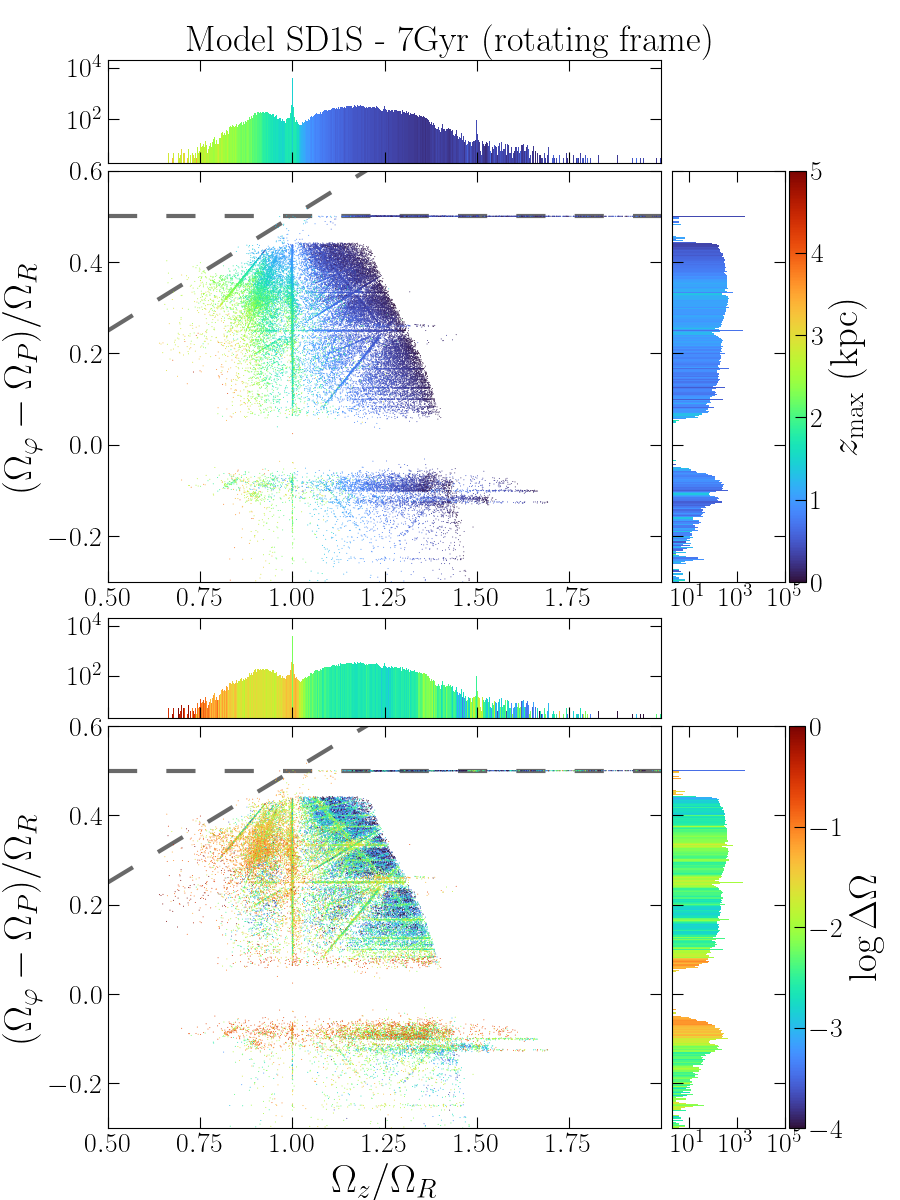}
    \caption{Frequency maps for the same orbits as in Fig.~\ref{fig:freq_map_768MRS}, but here with frequencies estimated using coordinates in the rotating frame. The two frequency maps are almost identical, except for the co-rotation resonance, which is not correcly identified in this case.}
    \label{fig:freq_map_768MRS_rot_frame}
\end{figure}

Fig.~\ref{fig:freq_map_768MRS_rot_frame} shows the frequency map for the same snapshot and the same orbits, but now using coordinates in the bar rotating frame. With $\Omphi$ still representing the azimuthal frequency in the inertial frame, the output of the frequency analysis is now $\Omphi - \OmP$.

Comparing Figs.~\ref{fig:freq_map_768MRS} and \ref{fig:freq_map_768MRS_rot_frame}, we see that the two frequency maps are almost identical, except for the co-rotation ($\Omphi=\OmP$), which is sharply defined in the inertial frame, but is completely lost if we use coordinates in the rotating frame. This is probably due to the leading frequency in this case being the libration around the co-rotation, which might be relevant for detailed investigations of orbits trapped at co-rotation, but it is not of interest for this work.

%% For this sample we use BibTeX plus aasjournals.bst to generate the
%% the bibliography. The sample631.bib file was populated from ADS. To
%% get the citations to show in the compiled file do the following:
%%
%% pdflatex sample631.tex
%% bibtext sample631
%% pdflatex sample631.tex
%% pdflatex sample631.tex

\bibliography{refs}{}
\bibliographystyle{aasjournal}

%% This command is needed to show the entire author+affiliation list when
%% the collaboration and author truncation commands are used.  It has to
%% go at the end of the manuscript.
%\allauthors

%% Include this line if you are using the \added, \replaced, \deleted
%% commands to see a summary list of all changes at the end of the article.
%\listofchanges

\end{document}